\documentclass[prd,nofootinbib,preprint,superscriptaddress]{revtex4}
\pdfoutput=1
\usepackage[T1]{fontenc}
\usepackage{amsmath,amssymb}
\usepackage{epsfig}
\usepackage{subfigure}
\usepackage{float}
\usepackage{graphicx}
\usepackage{bigints}
\usepackage[usenames,dvipsnames]{color}
\usepackage{slashed}
\usepackage{multirow}
\usepackage[colorlinks,citecolor=blue]{hyperref}
\usepackage{pdfpages}
\usepackage{color}
\usepackage{comment}

\newcommand{\mhsqr}[1]{M_{H_1}^2}

\begin{document}
\hfill \preprint~KIAS-P22083
\title{Freeze-in Dark Matter via Light Dirac Neutrino Portal}
\author{Anirban Biswas}
\email{anirban.biswas.sinp@gmail.com}
\affiliation{Department of Physics, Sogang University, Seoul 121-742, South Korea}
\affiliation{Center for Quantum Spacetime, Sogang University, Seoul 121-742, South Korea}
\author{Debasish Borah}
\email{dborah@iitg.ac.in}
\affiliation{Department of Physics, Indian Institute of Technology
Guwahati, Assam 781039, India}

\author{Nayan Das}
\email{nayan.das@iitg.ac.in}
\affiliation{Department of Physics, Indian Institute of Technology
Guwahati, Assam 781039, India}

	\author{Dibyendu Nanda}
	\email{dnanda@kias.re.kr}
\affiliation{School of Physical Sciences, Indian Association for the Cultivation of Science,
2A $\&$ 2B Raja S.C. Mullick Road, Kolkata 700032, India}	
\affiliation{School of Physics, Korea Institute for Advanced Study, Seoul 02455, Korea}

\begin{abstract}
We propose a scenario where dark matter (DM) can be generated non-thermally due to the presence of a light Dirac neutrino portal between the standard model (SM) and dark sector particles. The SM is minimally extended by three right handed neutrinos ($\nu_R$), a Dirac fermion
DM candidate ($\psi$) and a complex scalar ($\phi$), transforming non-trivially under an unbroken $\mathbb{Z}_4$
symmetry while being singlets under the SM gauge group. While DM and $\nu_R$ couplings are considered to be tiny in order to be in the non-thermal or freeze-in regime, $\phi$ can be produced either thermally or non-thermally depending upon the strength of its Higgs portal coupling. We consider both these possibilities and find out the resulting DM abundance via freeze-in mechanism to constrain the model parameters in the light of Planck 2018 data. Since the interactions producing DM also produces relativistic $\nu_R$, we check the enhanced contribution to the effective relativistic degrees of freedom $\Delta {\rm N}_{\rm eff}$ in view of existing bounds as well as future sensitivities. We also check the stringent constraints on free-streaming length of such freeze-in DM from structure formation requirements. Such constraints can rule out DM mass all the way up to $\mathcal{O}(100 \, {\rm keV})$ keeping the $\Delta {\rm N}_{\rm eff} \leq \mathcal{O}(10^{-3})$, out of reach from near future experiments. Possible extensions of this minimal model can lead to observable $\Delta {\rm N}_{\rm eff}$ which can be probed at next generation experiments.
\end{abstract}
\maketitle
\section{Introduction}
\label{sec:Intro}
As suggested by irrefutable evidences from astrophysics and cosmology based experiments gathered over several decades, we live in a universe whose matter content is dominated by a non-baryonic, non-luminous form of matter, known as dark matter (DM) \cite{Zyla:2020zbs, Aghanim:2018eyx}. While it is approximately five times more dominant than ordinary baryonic matter, its total contribution to present universe's energy density is around $26\%$. Present abundance of DM is often quoted in terms of density parameter $\Omega_{\rm DM}$ and reduced Hubble parameter ${\rm h} = \text{Hubble Parameter}/(100 \;\text{km}
~\text{s}^{-1} \text{Mpc}^{-1})$ as \cite{Aghanim:2018eyx}
$\Omega_{\text{DM}} {\rm h}^2 = 0.120\pm 0.001$
at 68\% CL. In spite of so many observational evidences, the particle nature of DM is not yet known. However, it is known for sure that none of the standard model (SM) particles can satisfy the criteria for being a particle DM candidate, leading to several beyond standard model (BSM) proposals in the literature. Among these proposals, the weakly interacting massive particle (WIMP) paradigm is one of the most well studied one. In WIMP paradigm, a particle DM candidate having mass and interaction strength (with SM particles) typically around the electroweak ballpark can give rise to the observed DM abundance after thermal freeze-out, a remarkable coincidence often referred to as the {\it WIMP Miracle} \cite{Kolb:1990vq}. The same interactions responsible for thermal freeze-out of WIMP can also lead to its promising direct detection prospects like observable DM-nucleon scattering. However, the direct detection experiments have not seen any such scattering yet leading to tighter bounds on DM-nucleon couplings. Similar null results have also been reported at indirect detection as well as collider experiments. A recent review on the status of WIMP type DM models can be found in \cite{Arcadi:2017kky}. The null results in WIMP detection have also motivated the particle physics community to look for other viable alternatives like freeze-in or feebly interacting massive particle (FIMP) dark matter
\cite{Hall:2009bx, Blennow:2013jba, Klasen:2013ypa, Elahi:2014fsa, Biswas:2015sva, Biswas:2016bfo, Biswas:2018aib, Borah:2018gjk, Biswas:2019iqm, Barman:2020plp, Borah:2020wyc, Borah:2019bdi, Barman:2021tgt, Belanger:2020npe}
where DM, due to its feeble interactions with SM bath, never enters equilibrium in the early universe. A recent review of such models can be found in \cite{Bernal:2017kxu}. While FIMP offers a viable alternative to WIMP, such models are often difficult to probe due to tiny DM interactions except some special cases \cite{Hambye:2018dpi, Belanger:2018sti, Elor:2021swj}. 

In this work, we propose a FIMP scenario by connecting it to neutrino physics. While the origin of neutrino mass and nature of neutrinos (Dirac versus Majorana) are not yet known, we consider the presence of right handed neutrinos which couple to the left handed ones via tiny SM Higgs couplings resulting in light Dirac neutrinos. The right chiral part of Dirac neutrino, being singlet under the SM gauge symmetry, can act like a portal between the dark and visible sectors. To be more precise, we consider a minimal framework where the SM is extended by three right handed neutrinos, one singlet fermion DM candidate and one additional singlet scalar
to facilitate the coupling of DM with right handed neutrinos. Additional
discrete symmetry $\mathbb{Z}_4$ is imposed in order to forbid unwanted couplings while keeping DM stable. While thermal as well as non-thermal singlet scalar can decay to produce DM as well as right handed neutrinos, the latter can lead to additional relativistic degrees of freedom or dark radiation which can be probed at cosmic microwave background (CMB) experiments. Existing data from CMB experiments like Planck constraints such additional light species by putting limits on the effective
degrees of freedom for neutrinos during the
era of recombination ($z\sim 1100$) as  \cite{Aghanim:2018eyx} 
\begin{eqnarray}
{\rm
N_{eff}= 2.99^{+0.34}_{-0.33}
}
\label{Neff}
\end{eqnarray}
at $2\sigma$ or $95\%$ CL including baryon acoustic oscillation (BAO) data. At $1\sigma$ CL it becomes more stringent to ${\rm N}_{\rm eff} = 2.99 \pm 0.17$. Similar bound also exists from big bang nucleosynthesis (BBN) $2.3 < {\rm N}_{\rm eff} <3.4$ at $95\%$ CL \cite{Cyburt:2015mya}. All these bounds are consistent with SM predictions ${\rm N^{SM}_{eff}}=3.045$ \cite{Mangano:2005cc, Grohs:2015tfy,deSalas:2016ztq}. Future CMB experiment CMB Stage IV (CMB-S4) is expected reach a much better sensitivity of $\Delta {\rm N}_{\rm eff}={\rm N}_{\rm eff}-{\rm N}^{\rm SM}_{\rm eff}
= 0.06$ \cite{Abazajian:2019eic}, taking it closer to the SM prediction. Light Dirac neutrino models often lead to enhanced $\Delta {\rm N}_{\rm eff}$, some recent works on which can be found in \cite{Abazajian:2019oqj, FileviezPerez:2019cyn, Nanda:2019nqy, Han:2020oet, Luo:2020sho, Borah:2020boy, Adshead:2020ekg, Luo:2020fdt, Mahanta:2021plx, Du:2021idh, Biswas:2021kio, Borah:2022obi, Li:2022yna}. While Planck bound on $\Delta N_{\rm eff}$ put moderate constraints on the model parameters, the structure formation bounds on DM free-streaming length turn out to be severe disfavouring DM masses all the way up to $\mathcal{O}(100 \, {\rm keV})$. This also leads to small enhancement $\Delta {\rm N}_{\rm eff} \leq \mathcal{O}(10^{-3})$ which, though safe from Planck bounds, remain out of reach of next generation experiments. Suitable extension of this minimal model can however, lead to enhanced $\Delta {\rm N}_{\rm eff}$ which can be probed in near future.

This paper is organised as follows. In section \ref{sec:model} we discuss our basic setup including the model description and relevant Boltzmann equations required to compute the abundance of DM as well as $\Delta {\rm N}_{\rm eff}$. In section \ref{sec:fsl} we discuss the constraints from structure formation followed by the details of our numerical results related to DM and $\Delta {\rm N}_{\rm eff}$ in section \ref{sec:numeric}. In section \ref{sec:uv} we discuss possible UV completions of our minimal setup and finally conclude in section \ref{sec:conclude}.

\section{The Basic Setup}
\label{sec:model}
There have been several BSM proposals to realise light Dirac neutrinos. In order to keep our framework minimal, we consider only three types of BSM particles sufficient to highlight the interesting phenomenology. They are namely, right handed neutrinos $\nu_R$, fermion singlet DM $\psi$ and a complex scalar singlet $\phi$ transforming non-trivially under an unbroken discrete $\mathbb{Z}_4$ symmetry. The right handed neutrinos couple to left handed lepton doublets via SM Higgs with fine-tuned Dirac Yukawa couplings to generate sub-eV Dirac neutrino masses. All SM leptons as well as $\nu_R$ have $\mathbb{Z}_4$ charge $i$ which keep the Majorana mass terms away. The $\mathbb{Z}_4$ charges of $\psi, \phi$ are chosen to be $-1, i$ respectively which ensures DM has only one tree level coupling of the form  $y_\phi \,\overline{\psi}\,{\nu}_R\, \phi$. On the other hand, $\nu_R, \phi$ can have other couplings as well. For example $\nu_R$ couples to SM lepton doublet $\ell$ and Higgs $H$ as $y_H \, \overline{\ell}\, \tilde{H}\, \nu_R$. On the other hand, the scalar singlet $\phi$ can have quartic interactions with the SM Higgs as $\lambda_{H \phi}\, (H^\dagger H)(\phi^\dagger \phi) $. Thus, the Lagrangian involving the newly introduced fermions can be written as
\begin{eqnarray}
\mathcal{L}_{\rm fermion} = i \,\overline{\nu}_R\, \gamma^\mu \, \partial_\mu \, \nu_R\,
+\,  i \,\overline{\psi}\, \gamma^\mu \, \partial_\mu \, \psi\,
- \, m_{\psi} \overline{\psi} \psi - \left(y_H \, \overline{\ell}\, \tilde{H}\, \nu_R 
+  y_\phi \,\overline{\psi}\,{\nu}_R\, \phi + {\rm h.c.} \right)\,.
\end{eqnarray}
Similarly, the scalar Lagrangian of the model is
\begin{eqnarray}\nonumber
\mathcal{L}_{\rm scalar} &=& (D_{H\mu} H)^\dagger (D_{H}^\mu H)
+ (\partial_{\mu} \phi)^\dagger (\partial^\mu \phi) - 
\Bigg[ - {\mu_{H}^2}\,( H^\dagger H)  + {\lambda_{H}}\, 
( H^\dagger H)^2 + {\mu_{\phi}^2}\,( \phi^\dagger \phi) + \\ &&
\lambda_{\phi} \, ( \phi^\dagger \phi)^2 + \lambda_{H \phi}\,
(H^\dagger H)(\phi^\dagger \phi) + \lambda_{\phi}^\prime 
\left(\phi^4 + (\phi^\dagger)^4\right)\Bigg]\,,
\label{potential}
\label{scalar:pot}
\end{eqnarray}
where, the covariant derivative for $H$ is defined as 
\begin{eqnarray}
D_{H\mu}H & = & \left(\partial_\mu + i \frac{g}{2}\sigma_a W^a_{\mu} 
+ i \frac{g^\prime}{2} B_{\mu} \right) H\,.
\end{eqnarray}
Here, $g$ and $g^\prime$ are the gauge couplings
for $SU(2)_{L}$ and $U(1)_Y$ respectively while the
corresponding gauge bosons are denoted by $W_{\mu}^a$ and $B_{\mu}$. Since $\mathbb{Z}_4$ needs to remain unbroken, the singlet scalar does not acquire any vacuum expectation value (VEV). After the neutral component of the SM Higgs doublet $H$ acquires a VEV $v=246$ GeV, the physical masses of the scalars can be written as
\begin{eqnarray}
m_{h}^2 &=& 2 \lambda _{H}\, v^2\,, \\
m_{\phi}^2 &=& \mu_{\phi}^2 + \frac{1}{2} v^2 \lambda_{H\phi}\,.
\end{eqnarray}
While Dirac Yukawa coupling $y_H$ remains suppressed from neutrino mass criteria, without much relevance to the phenomenology of DM and $\Delta {\rm N}_{\rm eff}$, the two other couplings namely, $y_\phi, \lambda_{H \phi}$ play crucial roles along with the masses of $\phi, \psi$ denoted by $m_{\phi}, m_{\psi}$ respectively. Therefore, the relevant free parameters of this model are the following couplings and the masses,
\begin{equation}
{m_{\phi}\,,m_{\psi}\,,y_\phi, \lambda_{H\phi}}.
\end{equation}

Since both DM and $\nu_R$ will be dominantly produced from $\phi$, it is important to track the evolution of $\phi$ in the early universe. Depending upon coupling of $\phi$ with SM Higgs and its mass $m_\phi$, production of DM, $\nu_R$ can occur while $\phi$ is either in equilibrium or out of equilibrium. In order to discuss the our results in details, we consider three different scenarios and write the corresponding Boltzmann equations as follows. For the detailed derivations of the Boltzmann equations for each of these scenarios, please refer to appendix \ref{appen1}.

\subsection{Case I: $\phi$ in equilibrium}
In this case, $\phi$ remains in equilibrium with the SM bath during DM and $\nu_R$ production from $\phi$ decay. Thus $\phi$ abundance can be considered to be its equilibrium abundance throughout while for the other two species $\psi, \nu_R$, the relevant Boltzmann equations, in terms of comoving number densities of $\phi$ and $\psi$, and comoving energy density of $\nu_{R}$, are given by
\begin{equation} \label{case1_psi_1}
    \frac{dY_{\psi}}{dx} = \frac{ \beta}{x \mathcal{H} } \Gamma_{\phi} \frac{K_1(x)}{K_2(x)} Y_{\phi}^{\rm eq},
\end{equation}
\begin{equation} \label{case1_nuR_1}
    \frac{d \widetilde{Y}}{dx} =    \frac{\beta}{\mathcal{H} s^{1/3} x} \langle E\Gamma \rangle Y_{\phi}^{\rm eq},
\end{equation}
where $x = m_{\phi}/T$ and 
\begin{equation}
    \beta = \left[1 + \frac{T d g_{s}/dT}{3 g_s}\right],
\end{equation}
\begin{equation}
    \langle E \Gamma \rangle = g_{\psi} g_{\nu_R} \frac{\lvert \mathcal{M} \rvert^{'2}_{\phi \to \bar{\nu}_R \psi}}{32 \pi} \frac{(m^2_{\phi} - m^2_{\psi})^2}{m^4_{\phi}}.
\end{equation}
Here $\mathcal{H}$ is the Hubble parameter in radiation dominated universe and $K_i$ is modified Bessel function of i-th order. While the
comoving number density $Y_{\psi}=n_{\psi}/s$, the comoving energy density of $\nu_R$ which remains relativistic during the CMB formation, is defined in terms of its energy density as $\widetilde{Y}=\rho_{\nu_R}/s^{4/3}$.

\subsection{Case II: freeze-out of $\phi$}
For certain choices of model parameters, one can have a scenario where $\phi$ gets thermally produced first followed by its freeze-out and only after that dominant production of DM\footnote{This production mechanism of DM is known as superWIMP formalism, first proposed in \cite{Feng:2003uy}.} and $\nu_R$ take place from decay of $\phi$. Since $\phi$ can no longer be taken to be in equilibrium throughout, we need to track its evolution using the corresponding Boltzmann equation. The system of Boltzmann equations in this case is given by
\begin{equation}
    \frac{dY_{\phi}}{dx} =  \frac{\beta s}{\mathcal{H} x} \left(-\langle \sigma v\rangle _{_{\phi \phi^{\dagger{}} \to X \bar{X}}}\left((Y_{\phi})^2 -  (Y_{\phi}^{\rm eq})^2 \right) - \frac{\Gamma_{\phi}}{s} \frac{K_1(m_{\phi}/T)}{K_2(m_{\phi}/T)} Y_{\phi}\right),
    \label{case2_phi_2}
\end{equation}
\begin{equation}
     \frac{dY_{\psi}}{dx} = \frac{ \beta }{x \mathcal{H} } \Gamma_{\phi} \frac{K_1(x)}{K_2(x)} Y_{\phi},
     \label{case2_psi_2}
\end{equation}
\begin{equation}
       \frac{d \widetilde{Y}}{dx} = \frac{\beta}{\mathcal{H} s^{1/3} x} \langle E\Gamma \rangle  Y_{\phi}.
       \label{case2_nur_2}
\end{equation}
Here $\langle{\sigma v}\rangle _{_{\phi \phi^{\dagger{}} \to X \bar{X}}}$ is the thermally averaged annihilation cross-section \cite{Gondolo:1990dk, Guo:2010hq} of $\phi$ into the SM particles via Higgs portal interactions. These include the contact interaction of $\phi$
with the Higgs ($h$) along with all other Higgs portal
interactions $\phi\phi^\dagger \rightarrow f \bar{f}, V V, hh$, where
$f$ denotes the SM fermions (quarks and leptons) and $V$
denotes the SM gauge bosons. The definition of other parameters
remain same as in case I discussed earlier.

\subsection{Case III: non-thermal $\phi$}
Finally, we consider the remaining possibility where $\phi$ can be out-of-equilibrium throughout due to tiny couplings with the SM Higgs. Thus, the initial abundance of $\phi$ remains negligible, like FIMP DM, and then it starts to populate the universe due to decay or annihilation of SM bath particles. Since $\phi$ has only Higgs portal couplings, the relevant production mechanism is from Higgs decay or Higgs annihilation depending upon $m_\phi$. The distribution function for $\phi$ can be calculated by solving the following equation
\begin{equation}
    \frac{\partial f_{\phi}}{\partial t} - \mathcal{H} p_1 \frac{\partial f_{\phi}}{\partial p_1} = C^{h\to \phi \phi^{\dagger}} + C^{h h \to \phi \phi ^{\dagger} } + C^{\phi \to \bar{\nu}_R \psi},
    \label{case3_phi_3}
\end{equation}
the details of the collision terms on the RHS are given in Appendix \ref{appen1}. Once the distribution function $f_\phi$ is evaluated, it can be used to find the evolution of DM and $\nu_R$ densities by solving the following Boltzmann equations
\begin{align}
    \frac{dY_{\psi}}{dr} = \frac{g_{\phi}\beta}{r \mathcal{H} s} \frac{\Gamma_{\phi} m_{\phi}}{2 \pi^2} \int \frac{\left( \mathcal{A} \frac{m_0}{r} \right)^3 \xi^2 f_{\phi}(\xi,r)}{\sqrt{\left(\xi \mathcal{A} \frac{m_0}{r}\right)^2 + m_{\phi}^2}} d\xi \,,\nonumber \\
    \frac{d \widetilde{Y}}{dr} = \frac{g_{\phi}\beta}{r\mathcal{H} s^{4/3}} \langle E\Gamma \rangle \frac{1}{2 \pi^2} \int_{0}^{\infty} \left(\mathcal{A}\frac{m_0}{r}\right)^3 \xi^2 f_{\phi}(\xi,r) d\xi\,,
    \label{case3_psi_nur_3}
\end{align}
where $r=m_0/T$ with $m_0$ being an arbitrary mass scale and details of $\mathcal{A}, \xi$ are given in Appendix \ref{appen1}.

\section{Structure formation constraints}
\label{sec:fsl}
Fermion DM with mass roughly below a keV is ruled out from galactic phase space arguments \cite{Tremaine:1979we, Boyarsky:2008ju}. This implies that a fermion DM with mass above a keV can still allow, in principle, the formation of structures as we observe in the universe. However, such generic lower bound on fermion DM mass based on phase space arguments, can become more stringent depending upon the production mechanism of DM. Such bounds can be imposed on a particular DM scenario by calculating the free-streaming length (FSL) of DM. While hot DM is already ruled out, warm DM with FSL $\lambda_{\rm FSL} < 0.1$ Mpc is still allowed, and can be favourable over cold DM of FSL $\lambda_{\rm FSL} < 0.01$ Mpc due to the small-scale structure problems associated with the latter \cite{Drewes:2016upu}. Dark matter free-streaming length can be estimated from matter power spectrum inferred from the Lyman-$\alpha$ forest data \cite{Croft:2000hs, Kim:2003qt}. This has been done in several earlier works including \cite{Viel:2005qj}. Quasar data have also been used to for studying free-streaming properties of DM \cite{Hsueh:2019ynk}. For theoretical and simulation based studies of dark matter free-streaming properties, one may refer to \cite{Colombi:1995ze, Boyarsky:2008xj, deVega:2009ku, Schneider:2011yu}. For some recent discussions on structure formation constraints on DM production mechanisms, please see \cite{Merle:2013wta, Decant:2021mhj, Ballesteros:2020adh} and references therein.

The free-streaming length can be defined as 
\begin{eqnarray}\label{FSL1}
\lambda_{\rm FSL} = \bigintss_{T_{\rm prod}}^{T_{\rm eq}}
\dfrac{\langle v(T)\rangle}{a(T)} \dfrac{dt}{dT}\,dT,
\end{eqnarray}
where $T_{\rm eq}$ is the temperature of the universe at the time
of matter-radiation equality while $T_{\rm prod}$ denotes the temperature
during maximum production of DM. The average velocity
of DM ($\langle{v(T)}\rangle$) at a temperature $T$
can be expressed as 
\begin{eqnarray}
\langle{v(T)}\rangle =
\dfrac{\bigintss \frac{p_{1}}{E_1}\,
\frac{d^3 p_1}{(2\pi)^3}\,f_{\psi}(p_1,T)}
{\bigintss \frac{d^3 p_1}{(2\pi)^3}
\,f_{\psi}(p_1,T)}.
\end{eqnarray}
Here $p_1$ is the momentum of DM particle $\psi$ having energy $E_1$.
The above integration over $p_1$ are for all possible values of
the momentum ($p_1$) of $\psi$.  In terms of two new variables
$\xi_{\psi}$ and $r$ as defined in the Appendix\,\,\ref{appen1},
the above definition of $\langle v\rangle$ becomes
\begin{eqnarray}
\langle v(r)\rangle = \dfrac{\mathcal{A}(r)}
{\bigintss \xi^2_{\psi}\,f_{\psi}(\xi_{\psi},r)\,d\xi_{\psi}}\times
{\bigintss \dfrac{\xi^3_{\psi}\,f_{\psi}(\xi_{\psi},r)\,
d\xi_{\psi}}{\sqrt{(\mathcal{A}(r) \xi_{\psi})^2 
+ (\frac{r}{m_0}\,m_{\psi})^2}}}\,.
\end{eqnarray}
The function $\mathcal{A}(r)$ is defined in Appendix \ref{appen1} with $m_0$ being a reference mass scale, considered to be 125 GeV in our analysis. Now, in terms of $r$ the above definition of
FSL takes the following form
\begin{eqnarray}
\lambda_{\rm FSL} = 
\left(\dfrac{11}{43}\right)^{1/3}\,r_0
\int_{r_{\rm prod}}^{r_{\rm eq}}
{\langle{v(r)}\rangle\,g_s^{1/3}}
\frac{\beta}{H(r)} \frac{dr}{r^2}. 
\end{eqnarray}

Therefore, in order to calculate the free-streaming length of dark matter
$\psi$, we first need to find the distribution function
$f_{\psi}(\xi_{\psi},r)$. The non-thermal distribution function
of $\psi$ depends mostly on two factors. One of the factors
is the momentum distribution of the parent particle $\phi$
while the rest is the production mechanism of $\psi$ from
the parent $\phi$. In our case, $\psi$ can be produced
from the decay of $\phi$ as the decay is always kinematically
allowed. The Boltzmann equation for $f_{\psi}$ due to
the process $\phi(K_1) \to \psi(P_1) +
\overline{\nu_{R}}(P_2)$ is given by
\begin{align}
\frac{\partial f_{\psi}}{\partial t} - \mathcal{H} p_1 \frac{\partial f_{\psi}}{\partial p_1} =& \frac{1}{16\pi\,E_{p_1}\,p_1}
\int_{k^{\rm min}_1}^{k^{\rm max}_1} \dfrac{k_1 dk_1}{E_{k_1}}
\rvert \mathcal{M} \rvert^2_{\phi \to \bar{\nu}_R \psi}
\,f_{\phi}(k_1).
\end{align}
where,
\begin{eqnarray}
k^{\rm min}_1 &=& \dfrac{1}{2\,m^2_{\psi}}
\left|-p_1(m^2_{\phi}+m^2_{\psi}) +
\sqrt{p^2_1(m^2_{\phi}+m^2_{\psi})^2-m^2_{\psi}
\left\{4p^2_1\,m^2_{\phi}
-(m^2_{\phi}-m^2_{\psi})^2\right\}}\right|\,,\\
k^{\rm max}_1 &=& \dfrac{1}{2\,m^2_{\psi}}
\left[p_1(m^2_{\phi}+m^2_{\psi}) +
\sqrt{p^2_1(m^2_{\phi}+m^2_{\psi})^2-m^2_{\psi}
\left\{4p^2_1\,m^2_{\phi}
-(m^2_{\phi}-m^2_{\psi})^2\right\}}
\right]\,,
\end{eqnarray}
and when $m_{\phi} >> m_{\psi}$, the above limits on $k_1$ reduce to
the following simplified forms
\begin{eqnarray}
k^{\rm min}_1 &\simeq& \dfrac{m^2_{\phi}}{2\,m^2_{\psi}}
\left(-p_1 +\sqrt{p_1^2 - 4\frac{m^2_{\psi}}{m_{\phi}^2} p_1^2 + m^2_{\psi}}\right)\,,\\
k^{\rm max}_1 &\simeq& \dfrac{m^2_{\phi}}{2\,m^2_{\psi}}
\left(p_1  + \sqrt{p_1^2  - 4\frac{m^2_{\psi}}{m_{\phi}^2} p_1^2 + m^2_{\psi}}\right)\,.
\end{eqnarray}

Now for
\begin{itemize}
    \item Case I: $f_{\phi}(k_1) = e^{-E_{k_1}/T}$
    \item Case II: we can find $f_{\phi}(k_1)$ after the freeze-out of $\phi$ by using -
    \begin{eqnarray}
    \frac{\partial f_{\phi}}{\partial t} - \mathcal{H} k_1 \frac{\partial f_{\phi}}{\partial k_1} =& C^{\phi \to \psi \bar{\nu}_R}
    \end{eqnarray}
    \item Case III: we can find $f_{\phi}(k_1)$ by using -
    \begin{equation}
    \frac{\partial f_{\phi}}{\partial t} - \mathcal{H} k_1 \frac{\partial f_{\phi}}{\partial k_1} = C^{h\to \phi \phi^{\dagger}} + C^{h h \to \phi \phi ^{\dagger} } + C^{\phi \to \bar{\nu}_R \psi}.
    \end{equation}
\end{itemize}

Once we find $f_{\phi}$ from the above equations, we can use that to find $f_{\psi}$ which we can use again to find thermal average velocity and free-streaming length. We can also cross-check the numerical calculations by obtaining $n_{\phi} = g_{\phi}\int \frac{d^3 k_1}{(2\pi)^3} f_{\phi}$ and $n_{\psi}= g_{\psi}\int \frac{d^3 p_1}{(2\pi)^3} f_{\psi}$ and comparing it with the previous section's results. Note that the same expression for $n_{\psi}$ also appear in the denominator for the expression of $\left<v(T)\right>$. 

We will discuss the results for free-streaming length for each case together with DM and $\Delta {\rm N_{eff}}$ results in the upcoming section.

\section{Numerical Results}
\label{sec:numeric}
\begin{figure}[h!]
\includegraphics[height=8cm,width=8.0cm,angle=0]{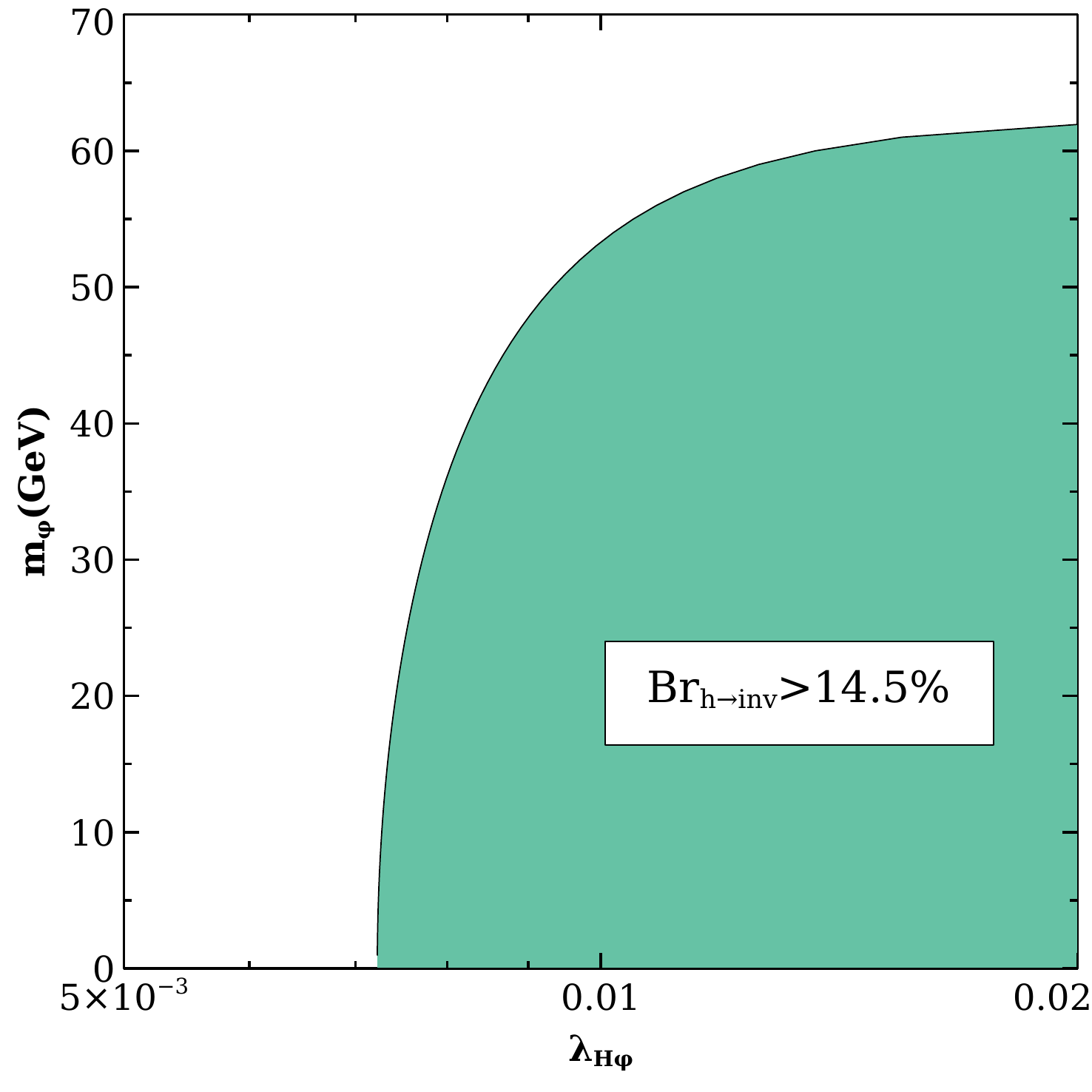}
\includegraphics[height=8cm,width=8.0cm,angle=0]{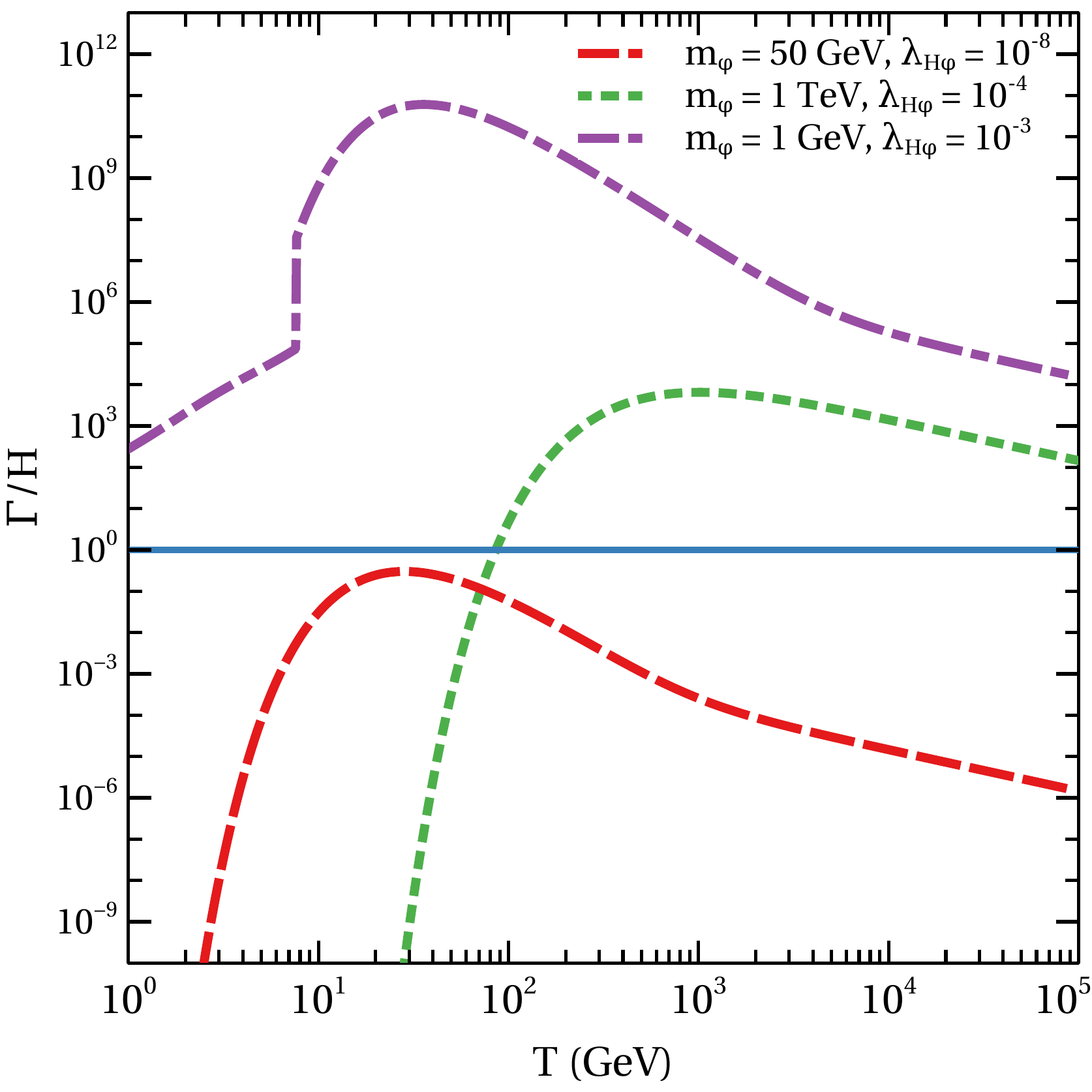}
\caption{Left panel: LHC constraint in $m_\phi-\lambda_{H \phi}$ plane showing the region excluded by upper limit on invisible decay width of the SM Higgs. Right panel: Interaction rates of $\phi$ in comparison to the Hubble expansion rate for benchmark choices of $m_\phi-\lambda_{H \phi}$ used in our analysis.}
\label{fig:phi}
\end{figure}

In this section, we discuss our numerical results for all the three cases mentioned above. After solving the Boltzmann equations for comoving densities of dark sector species, we can find the observable quantities like DM abundance $\Omega_{\rm DM}{\rm h}^2$ and $\Delta {\rm N}_{\rm eff}$ by following the procedure shown in Appendix \ref{appen2}. Since the region of validity for these three cases crucially depends upon the parameters involving complex scalar singlet $\phi$, we first show the parameter space in terms of its mass and Higgs portal couplings in left panel of Fig. \ref{fig:phi} indicating the region excluded by the constraints from the large hadron collider (LHC) on invisible decay of the SM Higgs boson into a pair of $\phi$. The ATLAS and the CMS collaboration have put the limit on invisible Higgs branching ratio as ${\rm BR}_{\rm h \rightarrow {\rm inv}} < 14.6\%$ \cite{ATLAS:2022yvh} and ${\rm BR}_{\rm h \rightarrow {\rm inv}} < 18\%$ \cite{CMS:2022qva} respectively, of which we use the stronger ATLAS bound in the left panel of Fig. \ref{fig:phi}. In the right panel of Fig. \ref{fig:phi}, we show the interaction rate of $\phi$ ($\Gamma$) in comparison to the Hubble expansion rate for three benchmark values of $m_\phi, \lambda_{H \phi}$ to indicate typical Higgs portal couplings required to consider thermal production of $\phi$ in the early universe. Clearly, for Higgs portal coupling $\lambda_{H \phi} \leq 10^{-8}$ validates the non-thermal nature of $\phi$ as we consider while discussing details of case III. In the following, we will choose the benchmark points as well as the scan range while keeping Fig. \ref{fig:phi} in mind.

In addition to bounds on $\Omega_{\rm DM}{\rm h}^2, \Delta {\rm N}_{\rm eff}$ and $(m_\phi, \lambda_{H\phi})$ plane mentioned above, we also note the model independent bounds on DM mass. If DM is very light, it can remain relativistic for a long time after being produced from $\phi$ decay resulting in large free-streaming length. While hot dark matter is ruled out, a warm dark matter (WDM) component is still allowed provided certain bounds are satisfied. Depending upon the details of production mechanism, warm dark matter mass below a few keV is ruled out as shown in several works incorporating different observations \cite{Boyarsky:2008xj, Newton:2020cog, Banik:2019smi}. Coincidentally, similar lower bound exists on fermion DM mass from galactic phase space arguments \cite{Tremaine:1979we, Boyarsky:2008ju}. While these lower bounds can vary slightly depending upon the production scenario and observational constraint imposed, we consider a lower bound of $\mathcal{O}(1)$\,keV in our analysis. We also consider a conservative upper bound on $\phi$ lifetime such that its decay is complete before the BBN epoch $T_{\rm BBN} \sim \mathcal{O}(10)$ MeV. This ensures the production of dark matter as well as dark radiation before the onset of BBN epoch.

\subsection{Case I}
\begin{figure}[h!]
\includegraphics[height=6cm,width=8.0cm,angle=0]{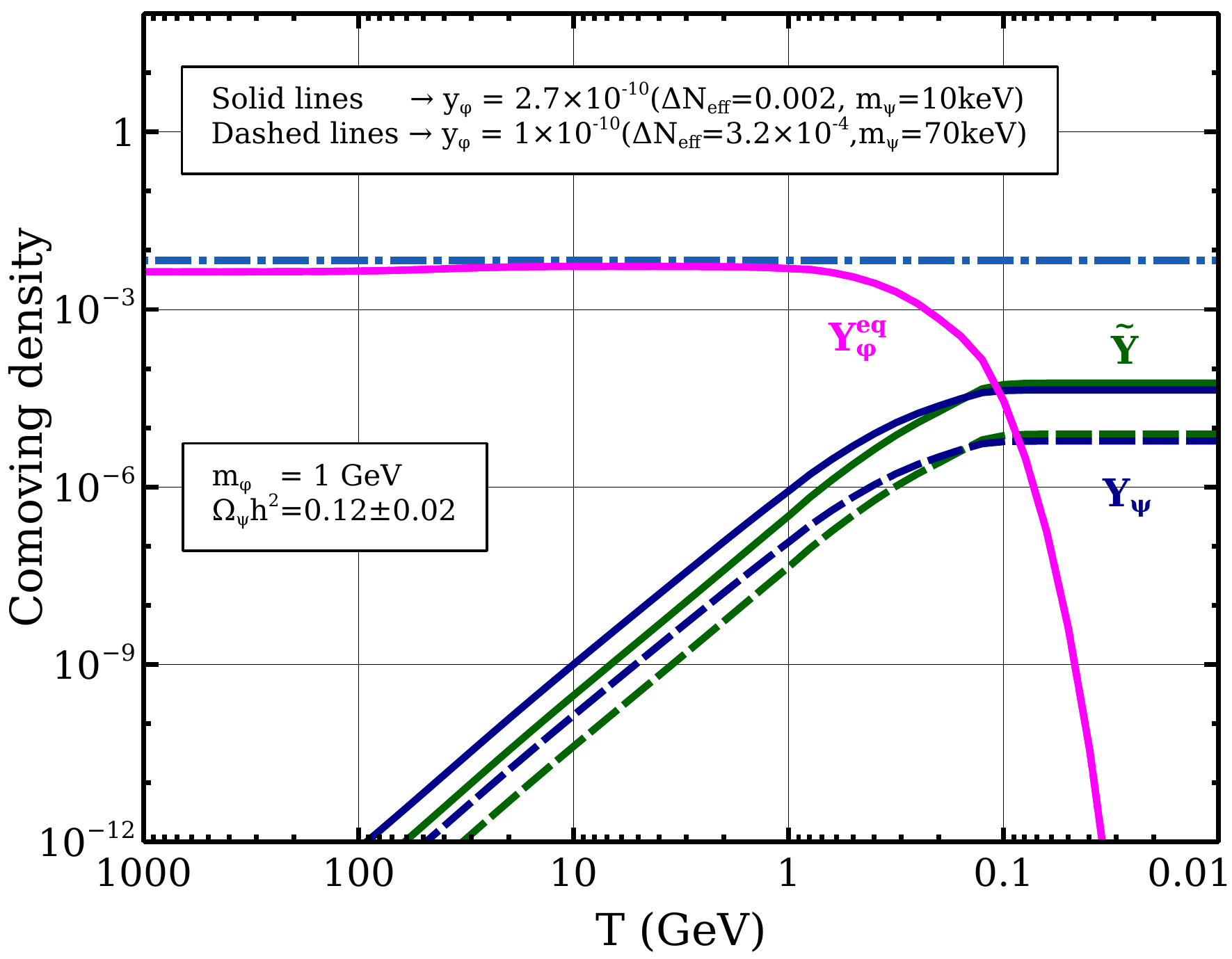}
\includegraphics[height=6cm,width=8.0cm,angle=0]{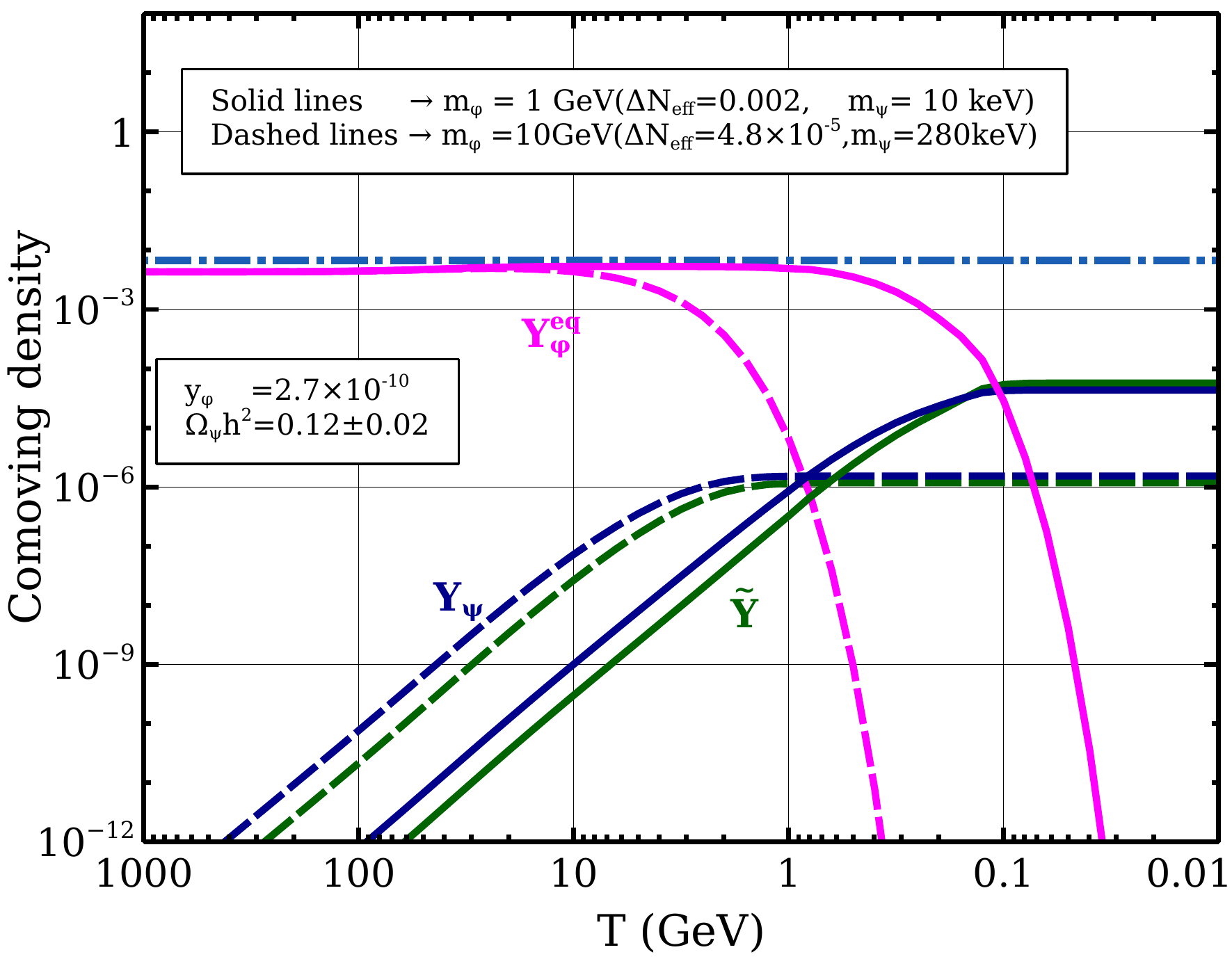}
\caption{Evolution of dark sector particles $(\phi, \psi, \nu_R)$ in case I considering $\phi$ to be in equilibrium throughout. All the lines denote the total comoving number/energy densities of dark sector particles. The left and right panel plots show the change in evolution for two different choices of $y_\phi, m_\phi$ respectively.  Chosen sets of points keep the DM abundance within the Planck limit.}
\label{fig:case1_benchmark}
\end{figure}

In this case, $\phi$ remains in equilibrium while DM and $\nu_R$ production takes place. This is the simplest scenario where we need to solve only two coupled Boltzmann equations for $\psi, \nu_R$ while using equilibrium abundance for $\phi$ throughout. 
Fig.\,\,\ref{fig:case1_benchmark} shows the evolution of dark sector particles as functions of temperature for different sets of parameters. The magenta, blue and green lines correspond to the comoving number densities of $\phi$ (in equilibrium) and $\psi$, and comoving energy density of $\nu_R$ respectively. The three free parameters $m_{\phi}, y_{\phi}$ and $m_{\psi}$ are taken in such a way that DM abundance, $\Omega_{\rm DM} {\rm h}^2$ is always satisfied. While $\phi$ abundance follows the equilibrium abundance as shown by the magenta line, DM and $\nu_R$ freeze in from decay of $\phi$ and get saturated after $\phi$ abundance gets Boltzmann suppressed for $T \lesssim m_\phi$.

Now, let us discuss the phenomenology for this situation with respect to the parameters $m_{\phi}, y_{\phi}$ and $m_{\psi}$ govern by Eqs.\,\,\eqref{case1_psi_1} and \eqref{case1_nuR_1}. The approximate analytical solutions of
these two equations are given in the Appendix C. Equations in \eqref{Appen_C_1} say that for $m_{\phi} \gg m_{\psi}$, both $Y_{\psi}$ and $\widetilde{Y}$ depend on $m_{\phi}$ and $y_{\phi}$ only, making $\Delta {\rm N_{eff}}$ independent of $m_{\psi}$ (from equations in Appendix B). Equation \eqref{Appen_C_4} that gives a relation between $\Delta{\rm N_{eff}}$ and $\Omega_{\rm DM}{\rm h^2}$, carries DM mass as an independent parameter. For correct relic abundance, a minimum value of DM mass will provide a maximum contribution to extra radiation energy density. Keeping this in mind, we plot the solid line in left panel of figure \ref{fig:case1_benchmark}, where we keep DM mass to be $10$ keV. We see that the corresponding $\Delta {\rm N_{eff}}$ value is $0.002$. This is the maximum value of effective relativistic degrees of freedom and it is out of the reach of both Planck 2018 and CMB-S4 limit. An approximate analytical approximation also gives the same value $\Delta {\rm N_{eff}} \approx 0.0016$ (from equation \ref{Appen_C_3}).  For the dashed line in the left panel, we have changed $y_{\phi}$ and observed its effects on $\Delta {\rm N_{eff}}$. In order to satisfy the DM abundance, $m_{\psi}$ has to be increased accordingly for the dashed lines. As expected, $\Delta {\rm N_{eff}}$ is reduced further. The right panel in figure \ref{fig:case1_benchmark} has been plotted for a different value of $m_{\phi}$. Here, due to a larger mass, $\phi$ gets Boltzmann suppressed earlier resulting in a smaller $Y_{\psi}$ and $\widetilde{Y}$. In both the plots, we show a horizontal line denoting the comoving energy density of a single species of right-handed neutrino that corresponds to the $2\sigma$ upper bound from the Planck 2018 data. In conclusion, for this situation when $\phi$ is always in bath, the contribution of dark radiation to effective relativistic degrees of freedom is beyond the reach of future CMB experiments.
\\
\noindent
{\bf Structure formation constraints:} For case I, where the particle $\phi$ is always in thermal equilibrium, we have calculated the free-streaming length for three different benchmark points. Here, we already know the distribution function of $\phi$ using which the distribution function of $\psi$ can be calculated. The Eq. \eqref{FSL1} tells that the free-streaming length is mainly dependent on two factors: the production temperature ($T_{\rm prod}$) and the injected energy to the DM from the decaying particle which will determine the average thermal velocity of the latter. In this section, we will see that if the production temperature is same and injected energy to DM is more, one can expect a larger free-streaming length as the dark matter particle will be relativistic for a longer duration. If the production temperature is high but the injected energy is same, one can expect a smaller FSL due to higher red-shift of DM momentum which will make the DM to be non-relativistic at an earlier epoch.
\begin{figure}[h!]
\includegraphics[height=6cm,width=8.0cm,angle=0]{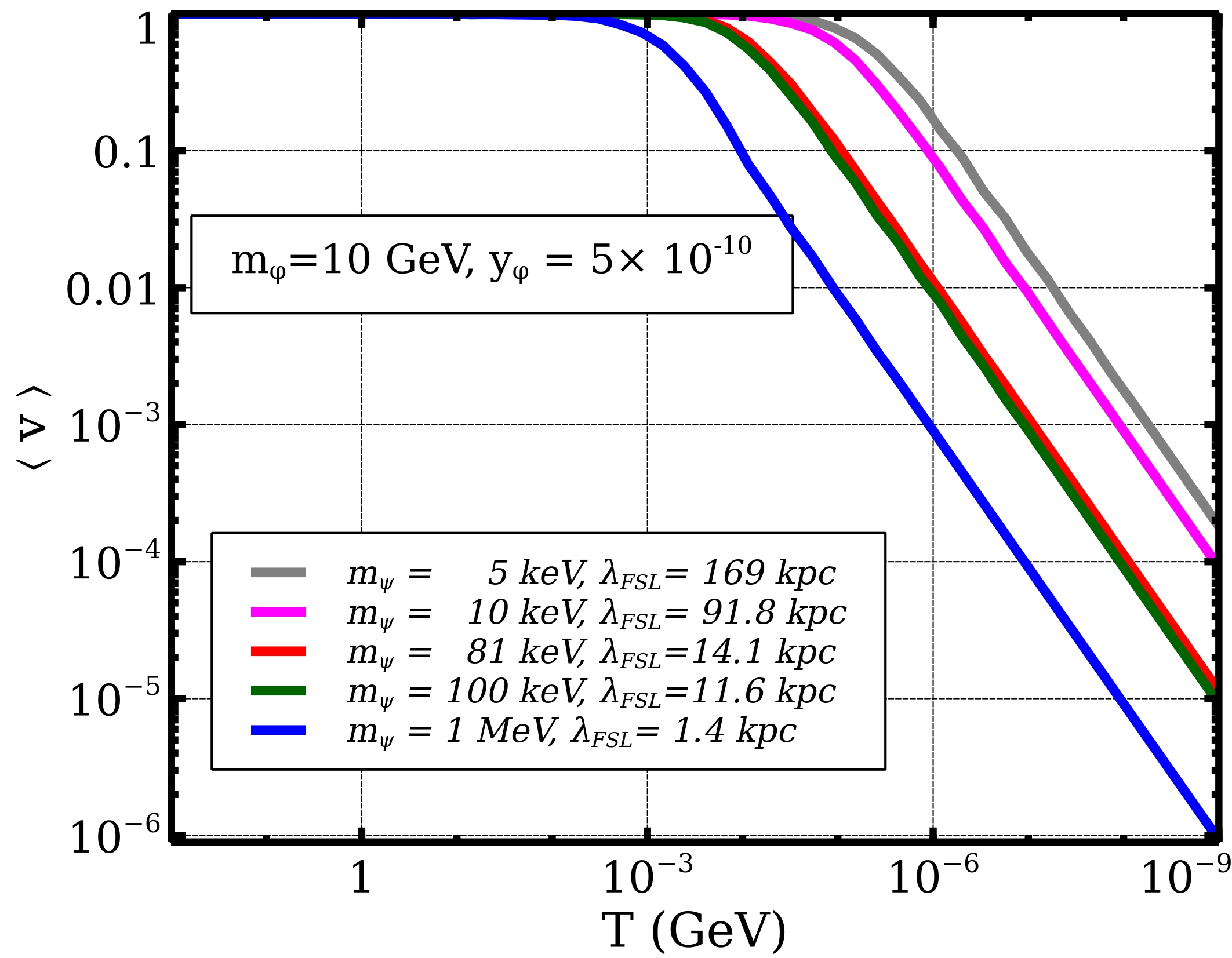}
\includegraphics[height=6cm,width=8.0cm,angle=0]{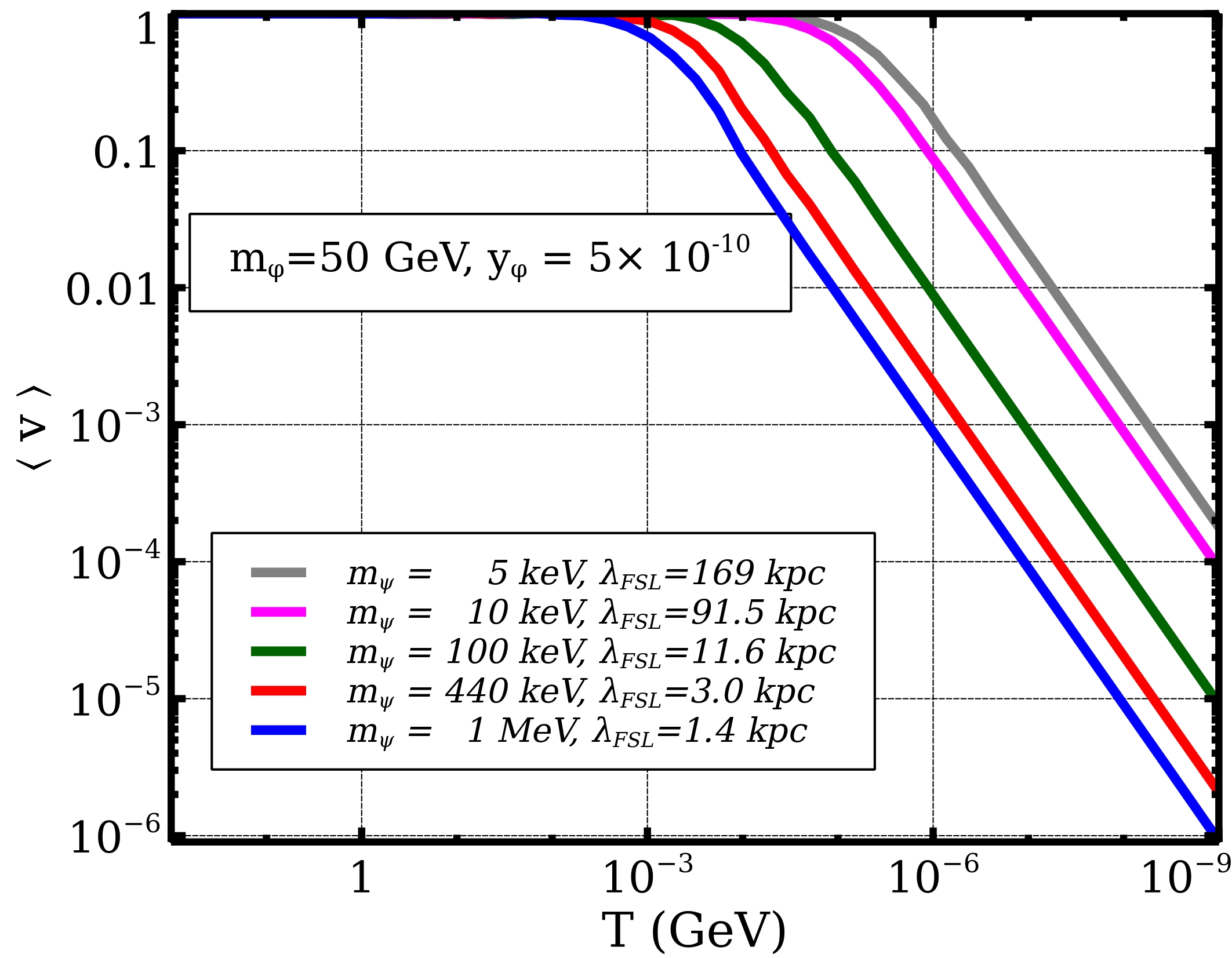}
\includegraphics[height=6cm,width=8.0cm,angle=0]{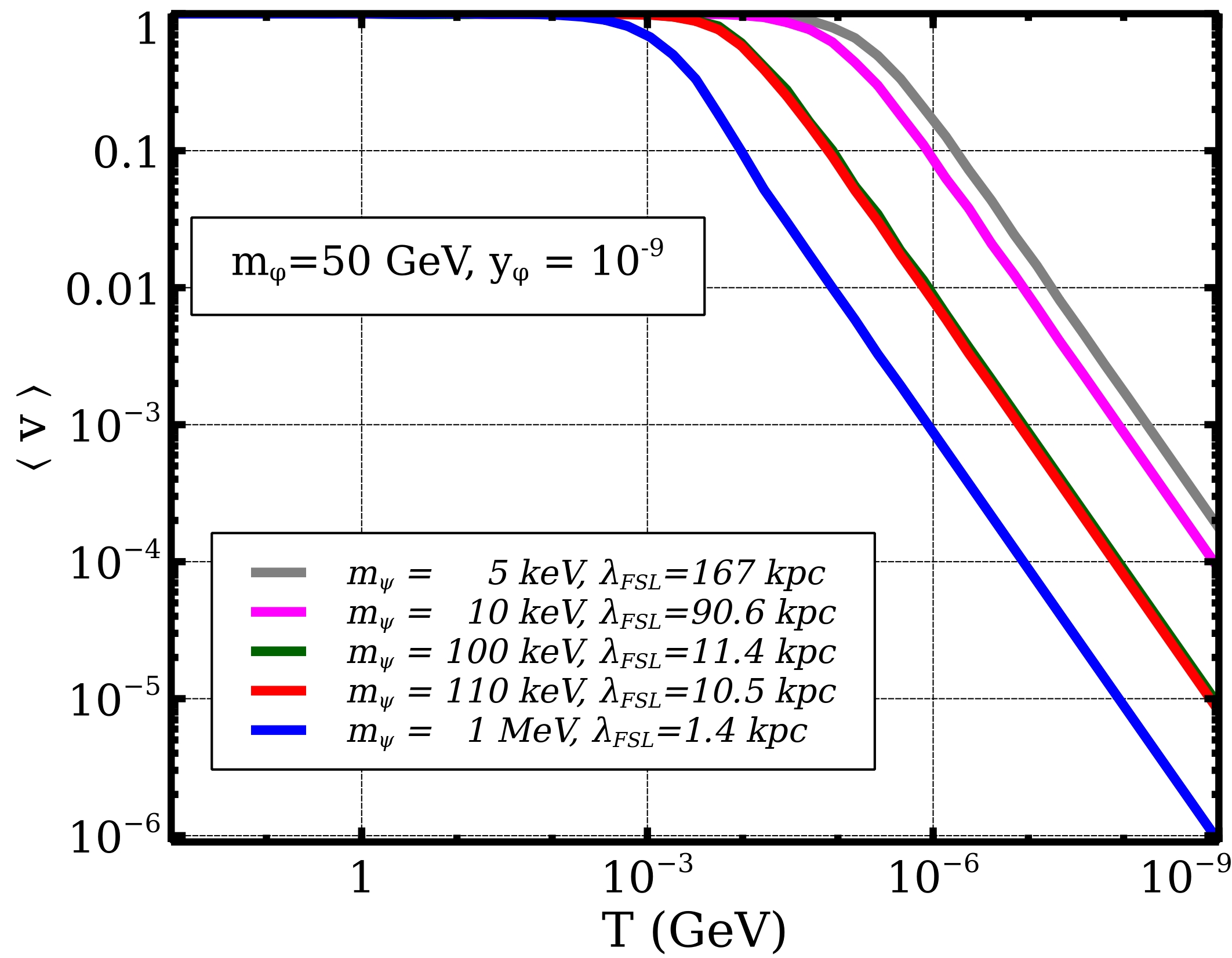}
\caption{Average velocity of DM as a function of temperature in case I for different benchmark combinations of relevant parameters.}
	    \label{fig:case1_FSL}
\end{figure}

In Fig. \ref{fig:case1_FSL}, we have shown the average velocity of DM as a function of temperature for two different values of $\phi$, $m_{\phi}=10$ GeV and $m_{\phi} = 50$ GeV. The values of dark sector coupling, $y_{\phi}$ is same in the upper panel plots of Fig. \ref{fig:case1_FSL}. In all the figures, the red lines show the average velocity of $\psi$ for which $m_{\psi}$ gives the correct DM relic. As can be seen from all the three plots, for $m_{\psi} \gtrapprox 10$ keV, the free-streaming length is less than $0.1$ Mpc, that is, they are in the warm DM region. From the first two plots, we see that for a particular value of $m_{\psi}$ (e.g. $m_{\psi}=1000$ keV), both the plots give very similar values for free-streaming length. This is contrary to the expectation as for higher mass of decaying particle, the injected energy to the DM should be more. The reason why the FSL is still small for higher decaying particle mass is that the production of DM from $\phi$ also occurs at a earlier epoch (see benchmark plot in Fig. \ref{fig:case1_benchmark}). As a result although the DM has higher momentum, its momentum gets red-shifted more. These two different phenomena compete with each other and as a result, we get similar FSL values in both the plots. In the top right panel plot and in the bottom plot, we have kept $\phi$ mass to be same and have changed the dark sector coupling, $y_{\phi}$. Due to same $m_{\phi}$, the initial energy of DM will be the same. Also we have already seen that that changing $y_{\phi}$ does not change $T_{\rm prod}$, the production temperature of dark matter. Hence we can expect same FSL for same DM mass. This is exactly what we can see from the top right plot and the bottom plot. The only difference in these two plots is that the $\psi$ mass satisfying correct DM relic is different. From the above analysis, we have found that the FSL for DM mass corresponding to correct DM relic falls under the warm dark matter region.

\begin{table}[]
    \centering
    \caption{Table for case I}
    \begin{tabular}{|c|c|c|c|c|c|c|}
    \hline
    \multicolumn{3}{|c|}{Parameters} & \multirow{2}{*}{$\Omega_{\rm DM} {\rm h}^2$}& \multirow{2}{*}{$\Delta {\rm N_{eff}}$} & \multirow{2}{*}{FSL(Mpc)}\\ \cline{1-3}  $m_{\phi}$(GeV)& $y_{\phi}$ & $m_{\psi}$(keV)& \multirow{2}{*}{} & \multirow{2}{*}{} & \multirow{2}{*}{}\\ \hline 
    $10$ & $5\times10^{-10}$ &  $81$ & $0.12$ & $1.6\times10^{-4}$ & $0.0141$ \\
    $50$ & $5\times10^{-10}$ &  $440$ & $0.12$ & $2.9\times10^{-5}$ & $0.0030$ \\
    $50$ & $10^{-9}$ &  $110$ & $0.12$ & $1.2\times10^{-4}$ & $0.0105$ \\
    \hline
    \end{tabular}
    \label{tab:case_1}
\end{table}

We summarize our FSL results for case I in table \ref{tab:case_1}, by including only those benchmark points from above analysis which satisfy correct DM relic. Clearly, the constraints on DM mass from FSL criteria can be as severe as $\mathcal{O}(100 \, {\rm keV})$ keeping the $\Delta {\rm N}_{\rm eff} \leq \mathcal{O}(10^{-3})$.

\subsection{Case II}
\begin{figure}[h!]
\includegraphics[height=6cm,width=8.0cm,angle=0]{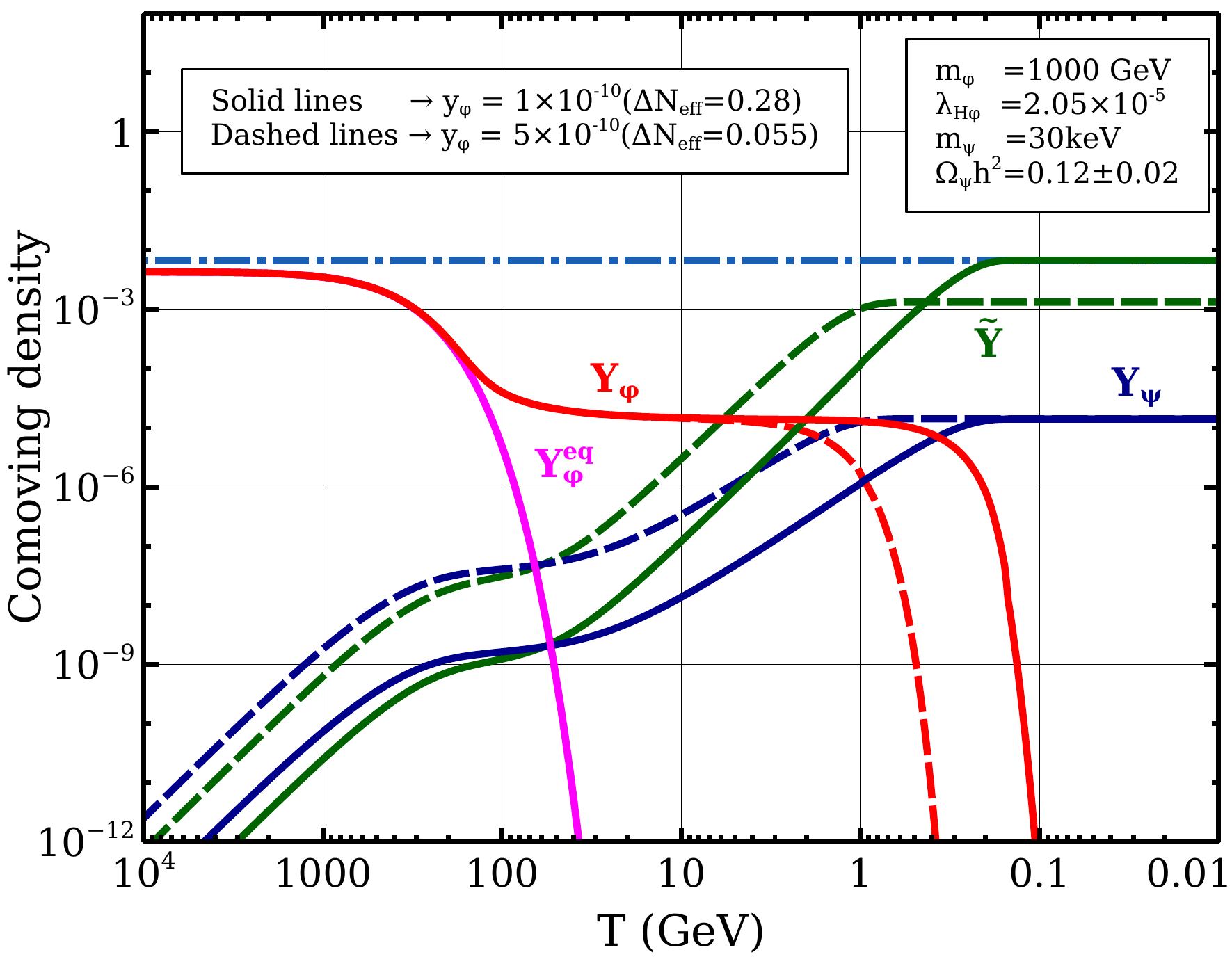}
\includegraphics[height=6cm,width=8.0cm,angle=0]{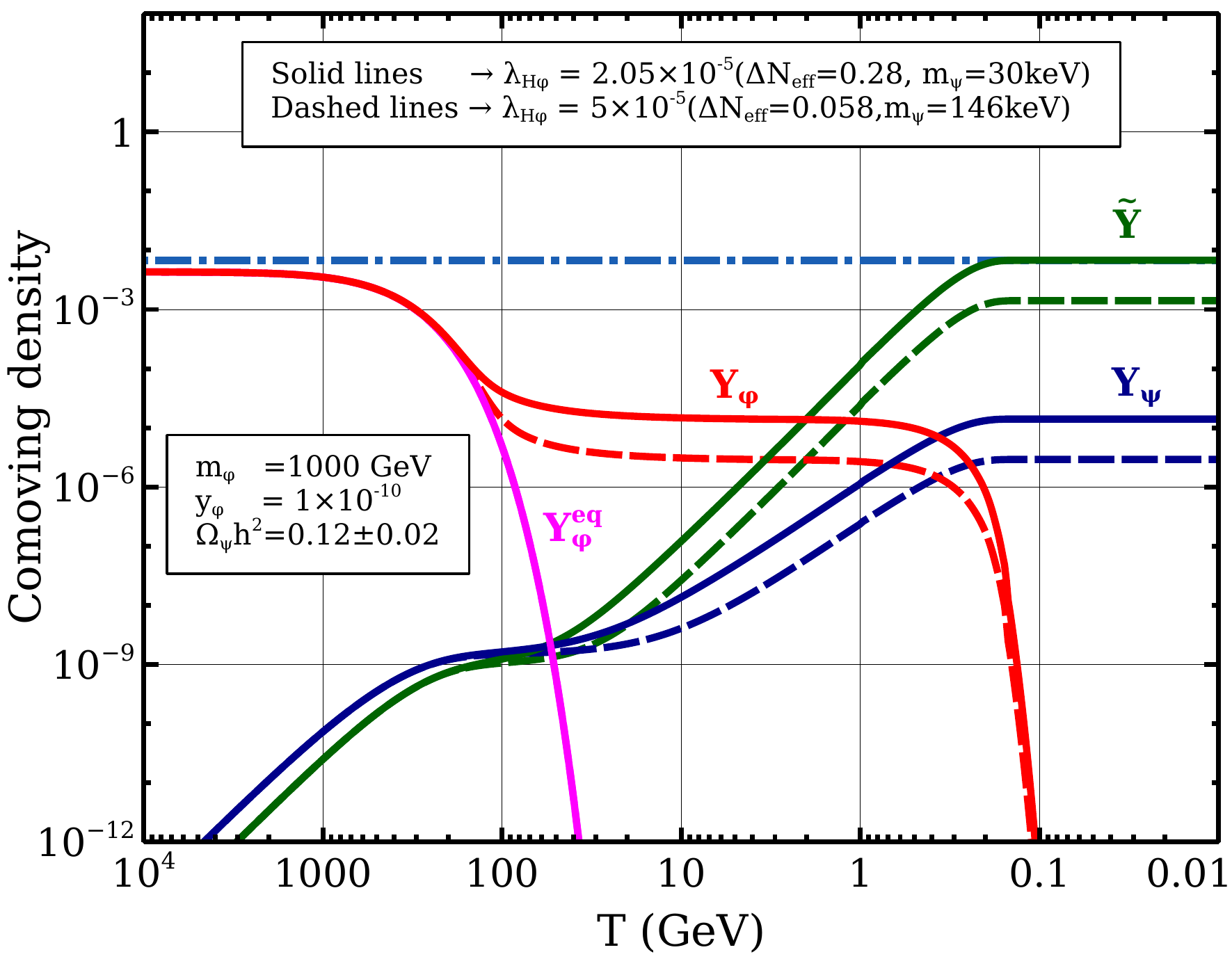}
\includegraphics[height=6cm,width=8.0cm,angle=0]{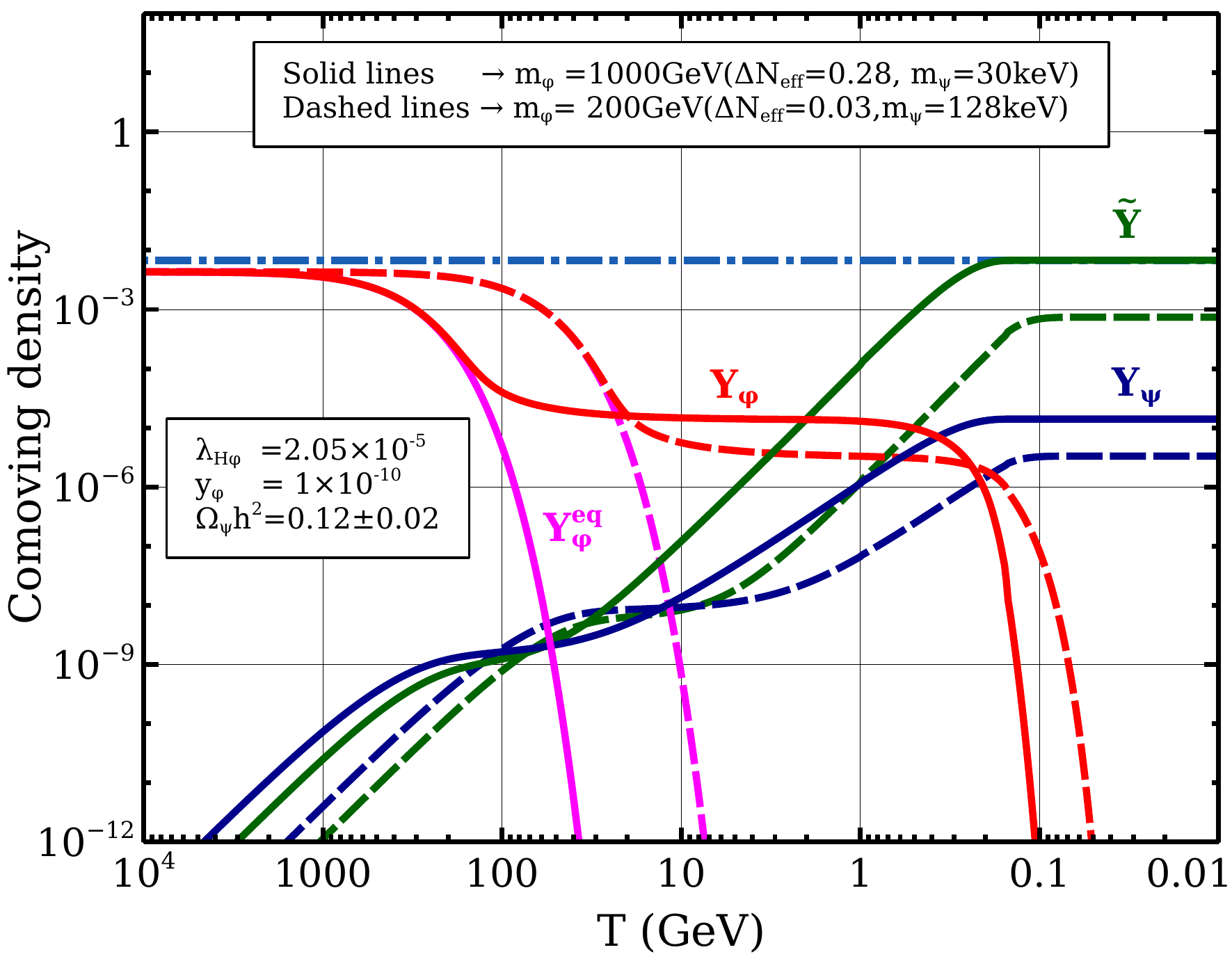}
\caption{Evolution of dark sector particles $(\phi, \psi, \nu_R)$ in case II considering $\phi$ to freeze out from the bath while decaying into $(\psi, \nu_R)$. Top left, top right and bottom panel plots show the change in evolution for two different choices of $y_\phi, \lambda_{H \phi}, m_\phi$ respectively. Chosen sets of points keep the DM abundance within the Planck limit.}
\label{fig:case2_benchmark}
\end{figure}

We now discuss the results for the intermediate scenario where $\phi$ gets produced thermally followed by its freeze-out. This requires solving the Boltzmann equation for $\phi$ as well together with the ones for $\psi, \nu_R$. Therefore, in addition to $m_{\phi}, y_{\phi}, m_{\psi}$, the Higgs portal coupling $\lambda_{H \phi}$ can play crucial role in deciding DM abundance as well as $\Delta {\rm N}_{\rm eff}$. We show the evolution of dark sector particles for case II in Fig. \ref{fig:case2_benchmark}. The top left, top right and bottom panels in this figure show the comparisons for two different choices of $y_\phi, \lambda_{H \phi}, m_\phi$ respectively. Similar to case I, the magenta, blue and green lines correspond to the comoving number densities of $\phi$ (in equilibrium) and $\psi$, and comoving energy density $\nu_R$ respectively. The red line corresponds to the actual comoving number density of $\phi$ which undergoes thermal freeze-out at an intermediate epoch followed by complete decay at later epochs. In all these plots, one can clearly see the production of $\psi, \nu_R$ to be taking place during equilibrium as well as frozen out phases of $\phi$ separated by a kink in between, as seen from the blue and green lines. The Higgs portal coupling of $\phi$ is chosen in such a way that the freeze-out abundance of $\phi$ is non-negligible in order to play substantial role in $\psi, \nu_R$ production. This is clearly visible from the plots shown in Fig. \ref{fig:case2_benchmark}, where the production of $\psi, \nu_R$ from frozen out $\phi$ appear to be significant. Another significant improvement from case I is that mass of DM can satisfy the lower limits discussed earlier even when $\Delta {\rm N}_{\rm eff}$ saturates Planck upper bound.

In the top left panel plot of Fig. \ref{fig:case2_benchmark}, we show the evolution for two different values of $y_{\phi}$ while keeping other parameters fixed. Since $y_\phi$ dictates the decay width of $\phi$, a lower value of $y_\phi$ delays the decay of frozen out $\phi$. Change in $y_\phi$, however, keeps DM density same as the number of $\phi$ gets transferred to number of $\psi$, both of which behave as non-relativistic particles. On the other hand, a lower value of $y_\phi$ or delayed production of $\nu_R$ from frozen out $\phi$ increases the comoving energy density of $\nu_R$ which behaves as radiation with comoving energy density defined as $\widetilde{Y} = \frac{\rho_{\nu_R}}{s^{4/3}}$. This can be understood if we solve the coupled Boltzmann equations given in Eqs. \eqref{case2_phi_2}, \eqref{case2_psi_2}, \eqref{case2_nur_2} analytically after the freeze-out of $\phi$. 
Equations \eqref{Appen_C_5} and \eqref{Appen_C_6} give the approximate analytical expressions for $Y^{\rm fo}_{\phi}, Y_{\psi}$ and $\widetilde{Y}$. As evident from Fig.\,\,\ref{fig:case2_benchmark}, the freeze-out abundance of $\phi$ namely, $Y^{\rm fo}_{\phi}$ gets converted to $Y_{\psi}$; whereas, $\widetilde{Y} \propto \frac{\langle E\Gamma\rangle}{f_{2}} \propto y^{-1}_{\phi}$.

In top right panel plot of Fig. \ref{fig:case2_benchmark}, we show the evolution for two different choices of Higgs portal coupling $\lambda_{H \phi}$. As expected from freeze-out mechanism of WIMP type particles, a larger value of $\lambda_{H \phi}$ leads to smaller freeze-out abundance of $\phi$ and hence smaller yield of $\psi, \nu_R$ at later epochs. On the other hand, for larger benchmark value of $\lambda_{H \phi}$ resulting in smaller yield of $Y_\psi$, we choose a heavier DM mass in order to keep $\Omega_{\rm DM} {\rm h}^2$ within Planck bounds. Finally, in the bottom panel plot of Fig. \ref{fig:case2_benchmark}, we show the evolution of dark sector particles for two different choices of $\phi$ mass. Due to change in Boltzmann suppression, the equilibrium evolution also changes for these two values. Since annihilation cross section decreases with increase in mass, we see larger freeze-out abundance for heavier $\phi$. Naturally, a larger freeze-out abundance for heavier $\phi$ leads to enhancement in comoving densities of DM and $\nu_R$ as well. The benchmark values of $m_\phi, m_\psi$ are chosen in such a way that DM abundance $\Omega_{\rm DM} {\rm h}^2$ remains within Planck limit while heavier (lighter) benchmark of $m_\phi$ keep $\Delta {\rm N}_{\rm eff}$ close to Planck upper bound (CMB-S4 sensitivity). It should also be noted that increasing $\phi$ mass also increases its decay width (for $m_\psi \ll m_\phi$) and hence we notice a delay in production of $\psi, \nu_R$ for lighter $\phi$ mass. Although we noticed enhanced $\widetilde{Y}$ from such delayed production in top left panel plot of Fig. \ref{fig:case2_benchmark}, in bottom panel plot of the same figure, this effect remains sub-dominant. The expected increase in $\widetilde{Y}$ for lighter $m_\phi$ due to delayed production remains subdominant compared to decrease in in $\widetilde{Y}$ for lighter $m_\phi$ due to reduced freeze-out abundance of the latter. Therefore, we only notice an overall increase in $\widetilde{Y}$ for heavier $\phi$ having larger freeze-out abundance. In each of these plots shown in Fig. \ref{fig:case2_benchmark}, the two benchmark parameter values (that is, $y_\phi$ in top left, $\lambda_{H \phi}$ in top right, $m_\phi$ in bottom) are chosen in such a way that one of them leads to $\Delta {\rm N}_{\rm eff}$ close to Planck $2\sigma$ upper limit while the other pushes it close to CMB-S4 sensitivity limit.
 

As seen from the evolution plots of case I and case II discussed above, case II becomes similar to case I if the maximum production of $\psi$ from the decay of $\phi$ happens before
    the freeze-out of latter from the thermal bath. This requires either late freeze-out of $\phi$
    (due large portal coupling $\lambda_{H\phi}$) or a short lived $\phi$ (due to large Yukawa coupling
    $y_{\phi}$). Unless we consider such regimes of couplings, these two cases need to be considered separately, yielding distinct result and phenomenology.

\begin{figure}[h!]
\includegraphics[scale=0.5]{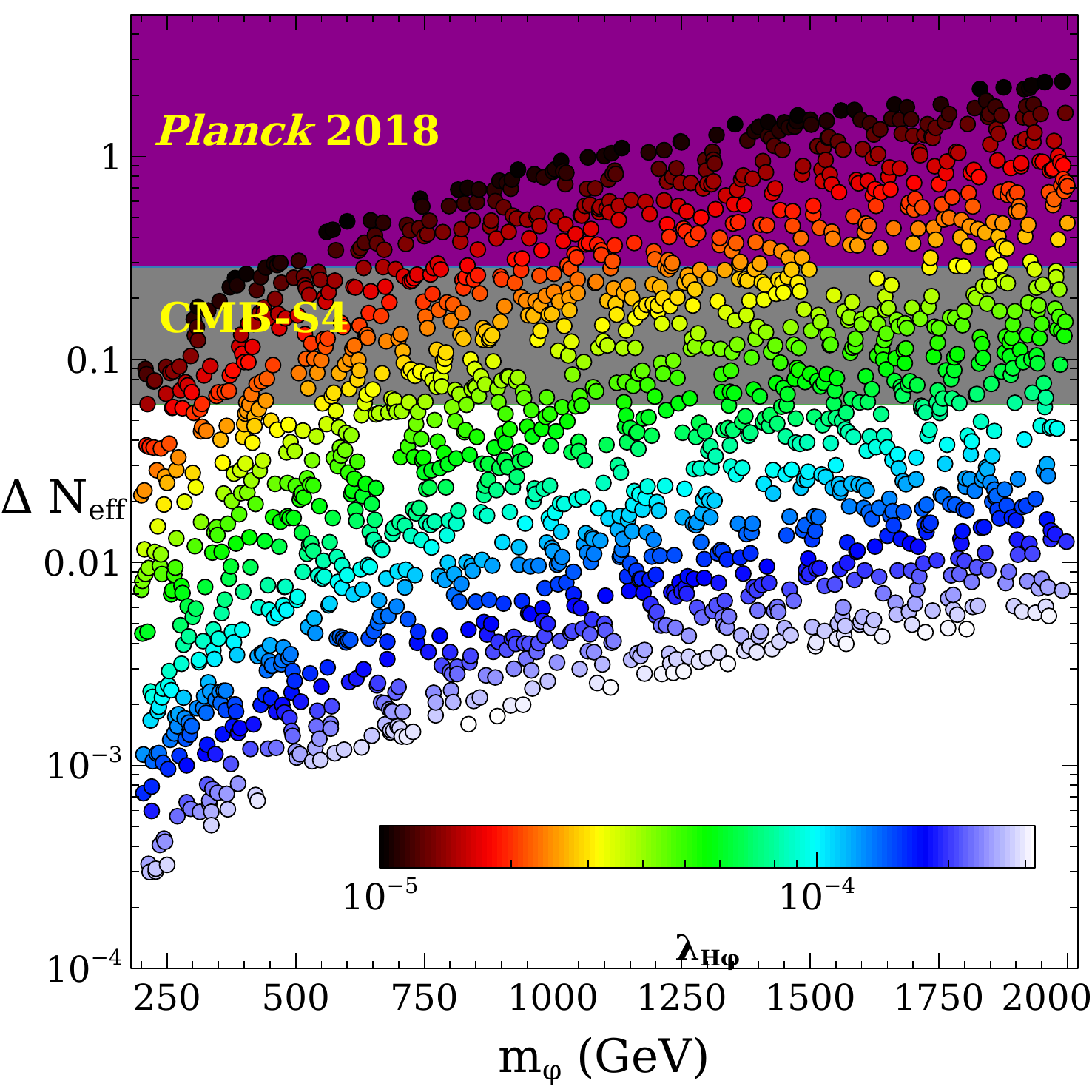}
\includegraphics[scale=0.5]{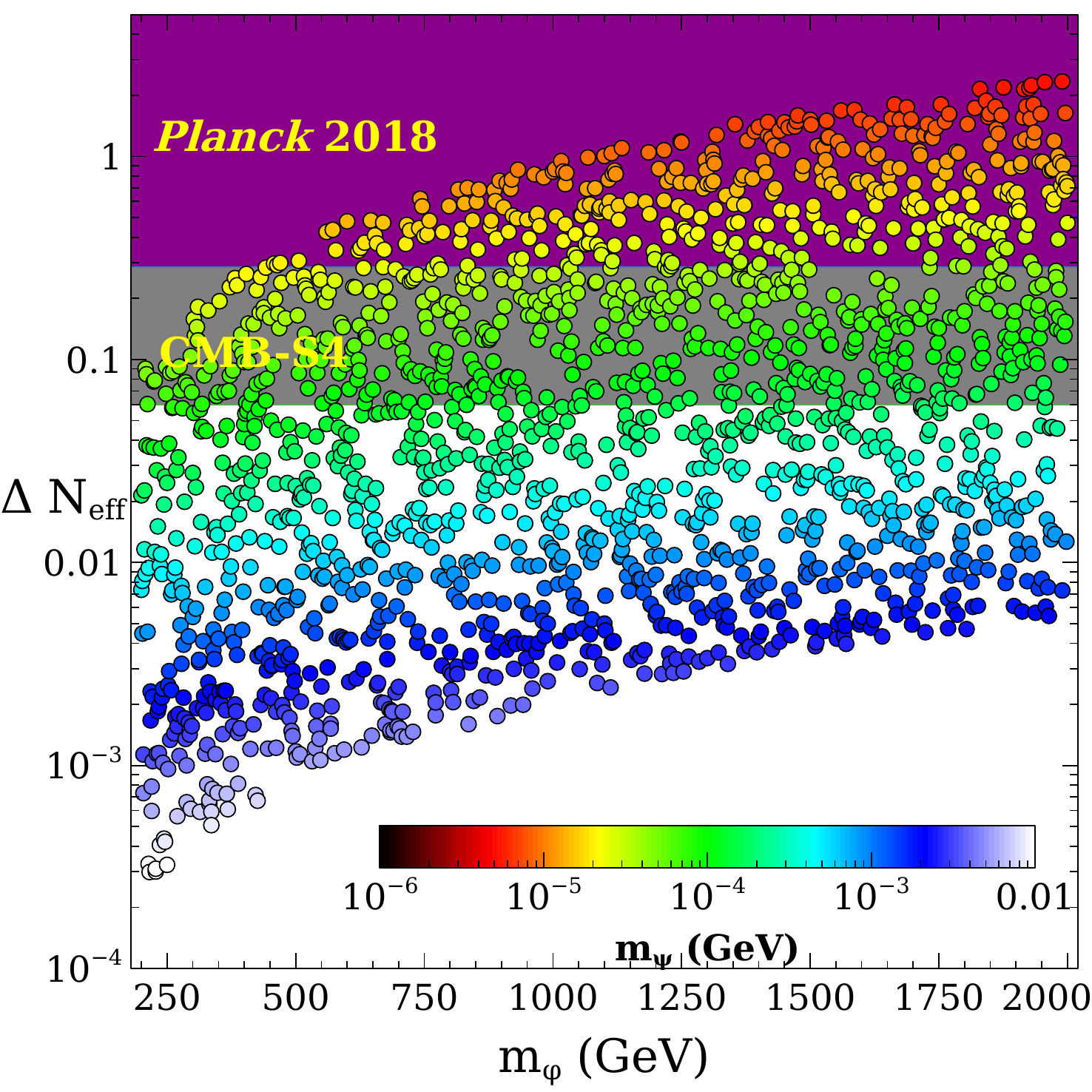}
	\caption{Parameter space plot for case II obtained from numerical scans, shown in terms of $\Delta {\rm N}_{\rm eff}$ vs $m_{\phi}$ while $\lambda_{H\phi}$ (left panel) and $m_{\psi}$ (right panel) are shown in colour code. The other relevant parameter $y_{\phi}$ is kept fixed at $10^{-10}$. The magenta and grey shaded regions indicate the current and future bound on $\Delta {\rm N}_{\rm eff}$ from Planck 2018 ($2\sigma$) and CMB-S4 respectively.}
	\label{fig:case2_scan}
\end{figure}

After highlighting the interesting features of case II with benchmark choices of key parameters, we perform a numerical scan over the parameter space. The relevant parameters are varied in the following range:
\begin{align} 
    200 \, {\rm GeV} \leq & \,m_{\phi} \, \leq 2000 \, {\rm GeV}, \nonumber \\
    10^{-5} \leq & \, \lambda_{H\phi} \, \leq 10^{-3.5}, \nonumber \\
    1 \, {\rm keV} \leq  & \, m_{\psi}  \,\leq 10 \, {\rm MeV}. \nonumber 
\end{align}
The value of $y_{\phi}$ is kept constant and remains fixed at $10^{-10}$, which also ensures that the decay of $\phi$ occurs before the BBN epoch. The resulting parameter space is shown in $\Delta {\rm N}_{\rm eff}$ vs $m_{\phi}$ plane in Fig. \ref{fig:case2_scan}. The colour bar in left and right panel plots show the variation in $\lambda_{H\phi}$ and $m_{\psi}$ respectively. While all the points satisfy the Planck bound on DM relic abundance, the corresponding upper bound on $\Delta {\rm N}_{\rm eff}$ is shown by magenta shaded region. The future sensitivity of CMB-S4 experiment is shown as grey shaded region. From the left panel of Fig. \ref{fig:case2_scan}, we can clearly see that for decrease in $\lambda_{H\phi}$, while keeping $m_{\phi}$ constant, $\Delta {\rm N_{eff}}$ decreases. This is expected as a smaller value of Higgs portal coupling $\lambda_{H\phi}$ leads to a larger freeze-out abundance of $\phi$ followed by enhanced production of $\nu_R$ from $\phi$ decay. Since the same decay also produces DM, we need to choose lower values of DM masses in order to keep its relic abundance within Planck limits. This can be noticed from the right panel plot of Fig. \ref{fig:case2_scan} where the points with large $\Delta {\rm N}_{\rm eff}$ correspond to smaller DM masses. Additionally, for fixed $\lambda_{H\phi}$, if we increase $m_{\phi}$, the corresponding $\Delta {\rm N}_{\rm eff}$ increases. Once again, this is due to larger freeze-out abundance of $\phi$ for heavier masses, as noticed while discussing the evolution plots in Fig. \ref{fig:case2_benchmark}. Accordingly, for heavier $m_\phi$ with fixed $\lambda_{H \phi}$, we need to choose lighter DM masses in order to keep its relic abundance within observed limits, as seen from the right panel plot of Fig. \ref{fig:case2_scan}. Thus, FIMP type DM candidate in our setup with masses all the way up to a few tens of keV can already get disfavoured by Planck 2018 limit $(2\sigma)$ on $\Delta {\rm N}_{\rm eff}$. As we will see in the next section, this lower bound on DM mass gets pushed to hundreds of keV after imposing the structure formation bounds. Accordingly, as these Fig. \ref{fig:case2_scan} suggests, $\Delta {\rm N}_{\rm eff}$ gets pushed down to second or third decimal places.

\begin{figure}[h!]
\includegraphics[height=6cm,width=8.0cm,angle=0]{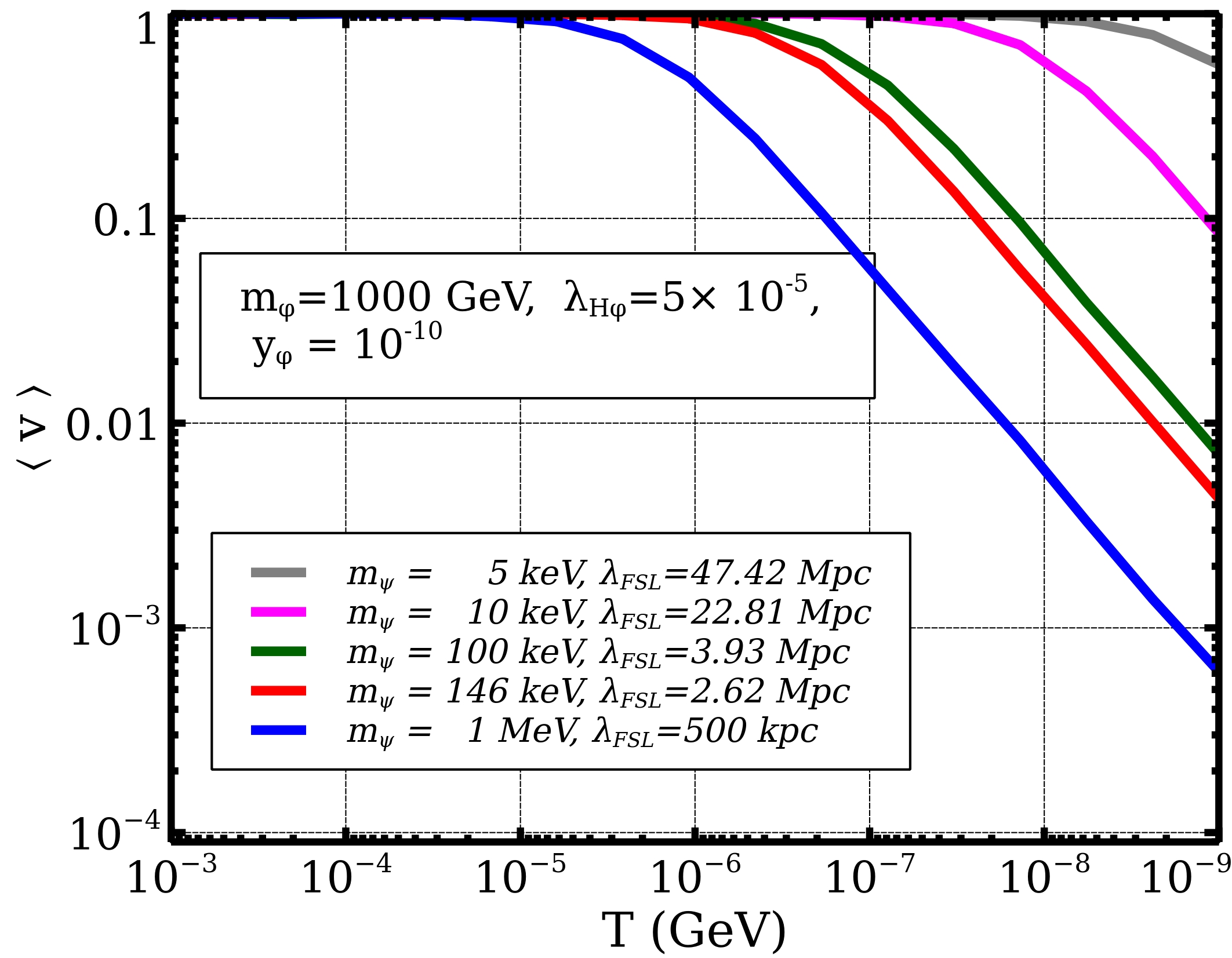}
\includegraphics[height=6cm,width=8.0cm,angle=0]{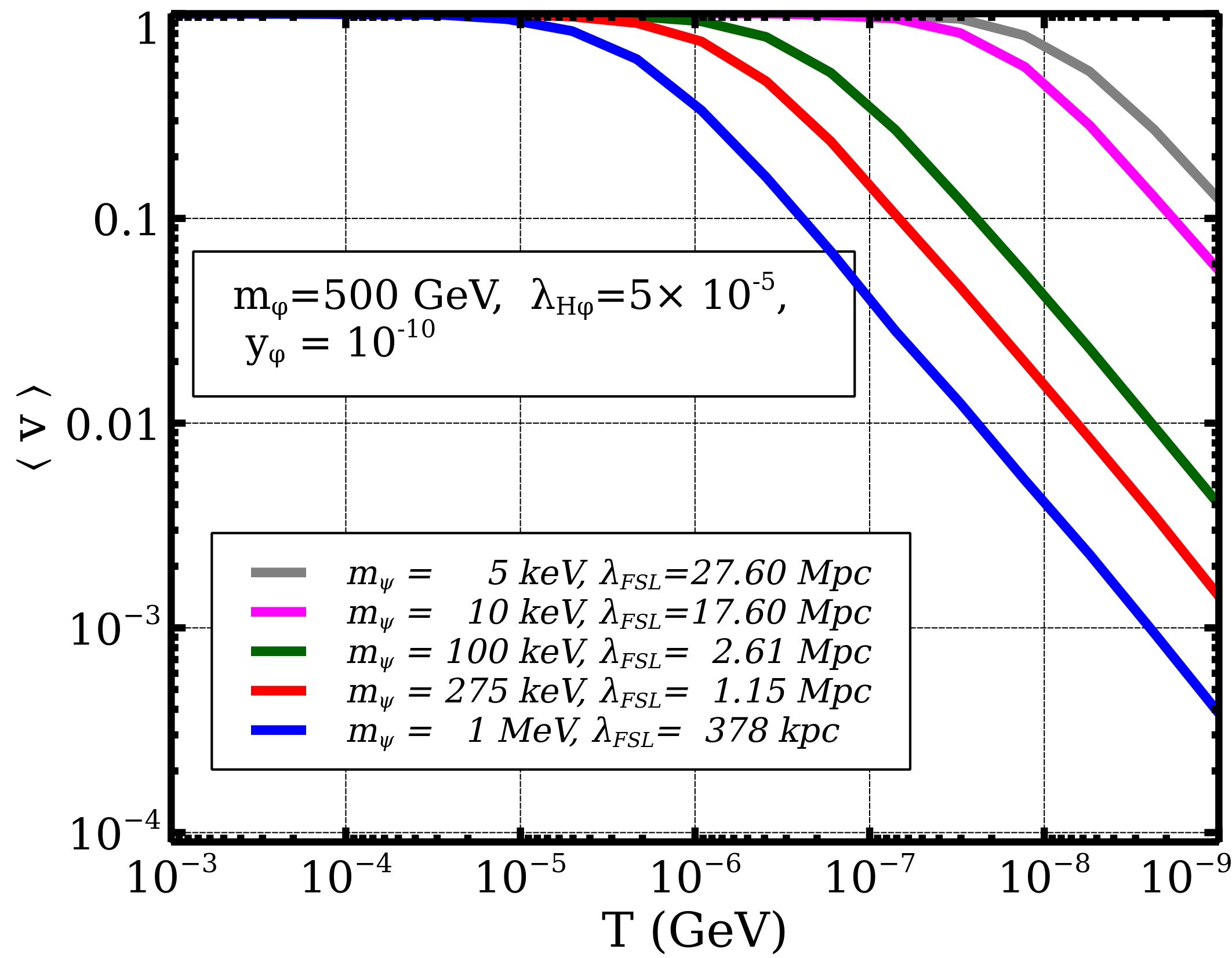}
\includegraphics[height=6cm,width=8.0cm,angle=0]{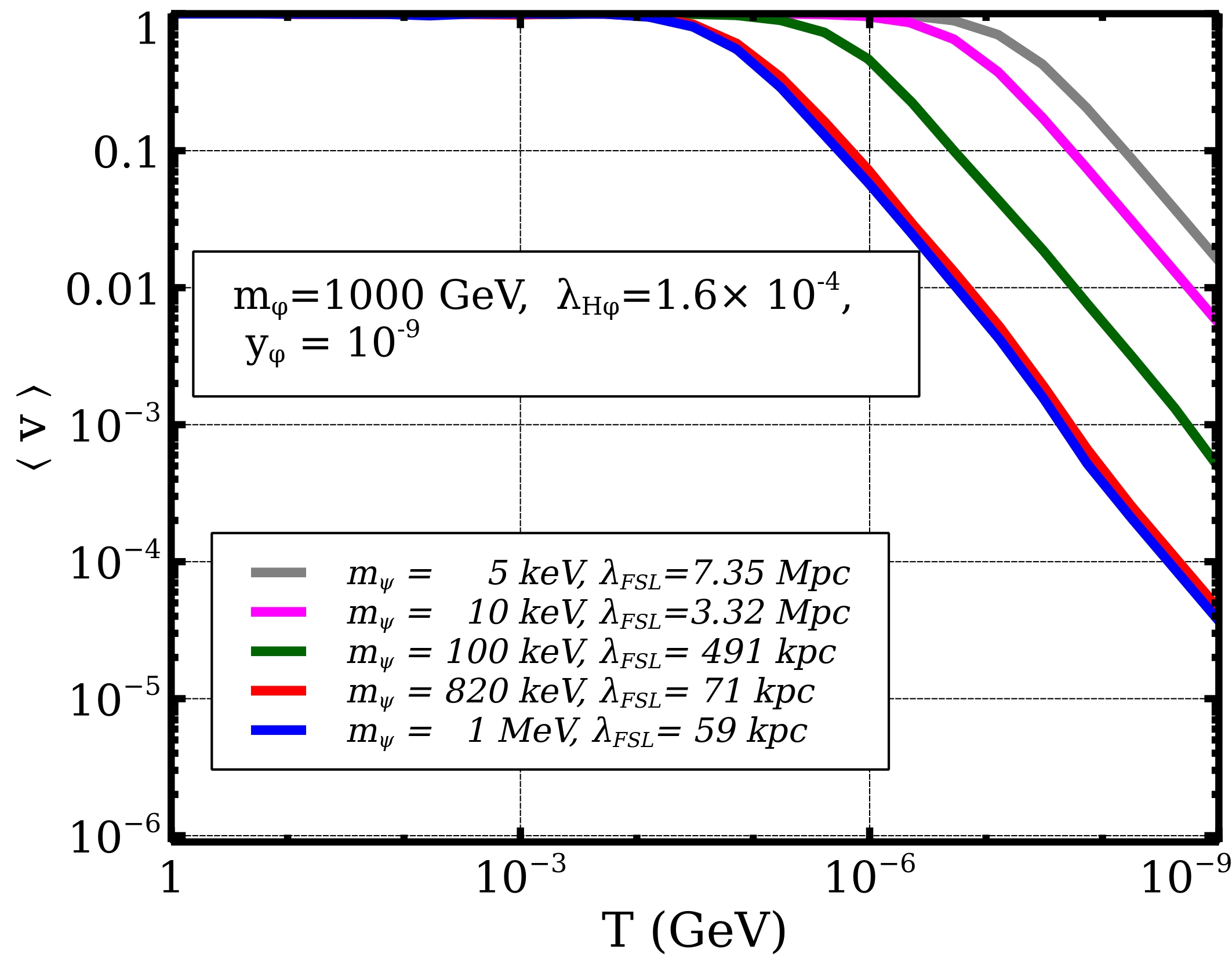}
\includegraphics[height=6cm,width=8.0cm,angle=0]{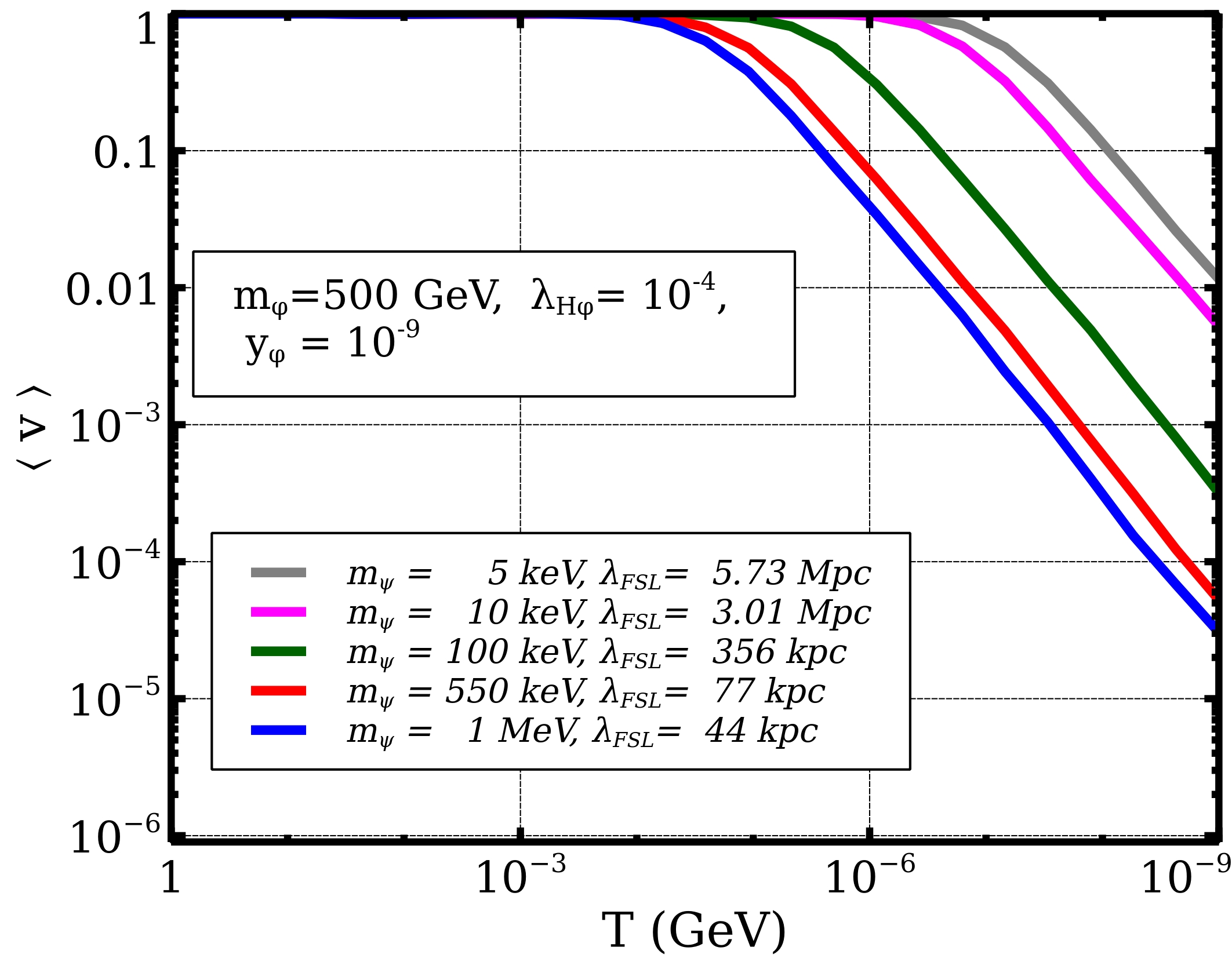}
\caption{Average velocity of DM as a function of temperature in case II for different benchmark combinations of relevant parameters.}
	    \label{fig:case2_FSL_1}
\end{figure}

\noindent 
{\bf Structure formation constraints:} For case II, we have estimated the free-streaming length of dark matter for some benchmark points. The free-streaming length for dark matter when $m_{\phi} = 1000$ GeV, $\lambda_{H\phi}=5\times 10^{-5}$ and $y_{\phi}=10^{-10}$ are shown in the left side of Fig. \ref{fig:case2_FSL_1} for different values of $m_{\psi}$. Except the red colored lines, the other lines do not satisfy the current DM abundance. As expected, for lower mass, the dark matter remains relativistic for a longer period and hence its free-streaming length is higher. Even for the maximum $m_{\psi}$ in the figure, i.e. for $m_{\psi} = 1000$ keV, the free-streaming length is  greater than $0.1$ Mpc, which is roughly the boundary between warm and hot dark matter. Thus for all $m_{\psi}$ in the figure, the free-streaming lengths are found to be higher than $0.1$ Mpc. By decreasing the injected energy to dark matter from the particle $\phi$, the dark matter can be made to become non-relativistic at an earlier epoch. This can be obtained by decreasing $m_{\phi}$. 
The top right panel plot of Fig. \ref{fig:case2_FSL_1} shows the free-streaming length for a smaller $m_{\phi} = 500$ GeV with $\lambda_{H\phi}$ and $y_{\phi}$ having the same value as the top left panel plot. We can see that although the free-streaming length now has a smaller value, but still all the points give hot dark-matter. Another effective way to make dark-matter non-relativistic at an earlier time is to increase the dark sector coupling $y_{\phi}$. This will give a higher decay rate $\Gamma_{\phi}$, leading to a higher dark-matter production temperature. The results can be seen from the bottom plots of Fig. \ref{fig:case2_FSL_1}. The left hand side is for $m_{\phi} = 1000$ GeV and the right hand side is for $m_{\phi} = 500$ GeV. As increasing $\lambda_{H\phi}$ will also increase $m_{\psi}$ for correct DM abundance, the other parameters are tuned in both the figures so that we get DM mass in order of hundred of keV mass, satisfying the relic density constraint. We summarize our FSL results for case II in table \ref{tab:case_2}, by including only those benchmark points from above analysis which satisfy correct DM relic
density.

\begin{table}[]
    \centering
    \caption{Table for case II}
    \begin{tabular}{|c|c|c|c|c|c|c|c|}
    \hline
    \multicolumn{4}{|c|}{Parameters} & \multirow{2}{*}{$\Omega_{\rm DM} {\rm h}^2$} & \multirow{2}{*}{$\Delta {\rm N_{eff}}$} & \multirow{2}{*}{FSL(Mpc)}\\ \cline{1-4}  $m_{\phi}$(GeV)& $\lambda_{H\phi}$ & $y_{\phi}$ & $m_{\psi}$(keV)& \multirow{2}{*}{} & \multirow{2}{*}{} & \multirow{2}{*}{}\\ \hline 
    $1000$ & $5\times10^{-5}$ & $10^{-10}$ & $146$ & $0.12$ & $5.8\times10^{-2}$ & $2.625$ \\
    $500$ & $5\times10^{-5}$ & $10^{-10}$ & $275$ & $0.12$ & $2.2\times10^{-2}$ & $1.146$ \\
    $1000$ & $1.6\times10^{-4}$ & $10^{-9}$ & $820$ & $0.12$ & $7.2\times10^{-4}$ & $0.071$ \\
    $500$ & $10^{-4}$ & $10^{-9}$ & $550$ & $0.12$ & $6.5\times10^{-4}$ & $0.077$ \\
    \hline
    \end{tabular}
    \label{tab:case_2}
\end{table}

\subsection{Case III}
In this subsection, we discuss the results for the last subclass of scenarios mentioned earlier where the mother particle $\phi$ never enters equilibrium due to feeble Higgs portal coupling. In order to simplify the analysis, we consider $\phi$ production to be taking place dominantly from the SM Higgs, either via decay or via annihilation. For $m_{\phi} < m_{h}/2$, the decay process ($ h \rightarrow \phi \phi$) dominates while in the other limit only annihilation ($ h h \rightarrow \phi \phi$) can contribute to $\phi$ production. To show the roles of decay and annihilation separately, we discuss these two limits separately.

\subsubsection{$m_{\phi}<m_{h}/2$}
\begin{figure}[h!]
\includegraphics[height=6cm,width=8.0cm,angle=0]{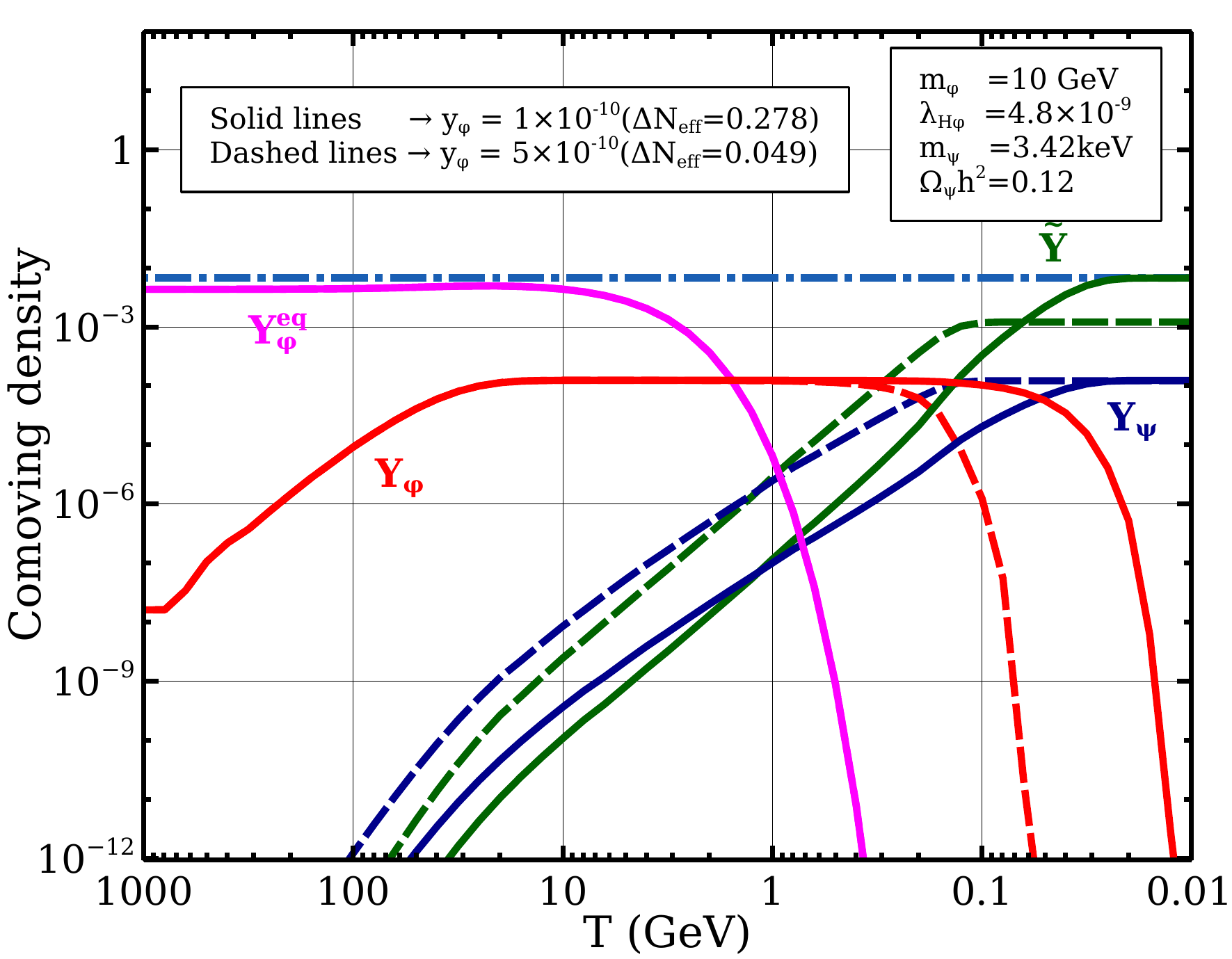}
\includegraphics[height=6cm,width=8.0cm,angle=0]{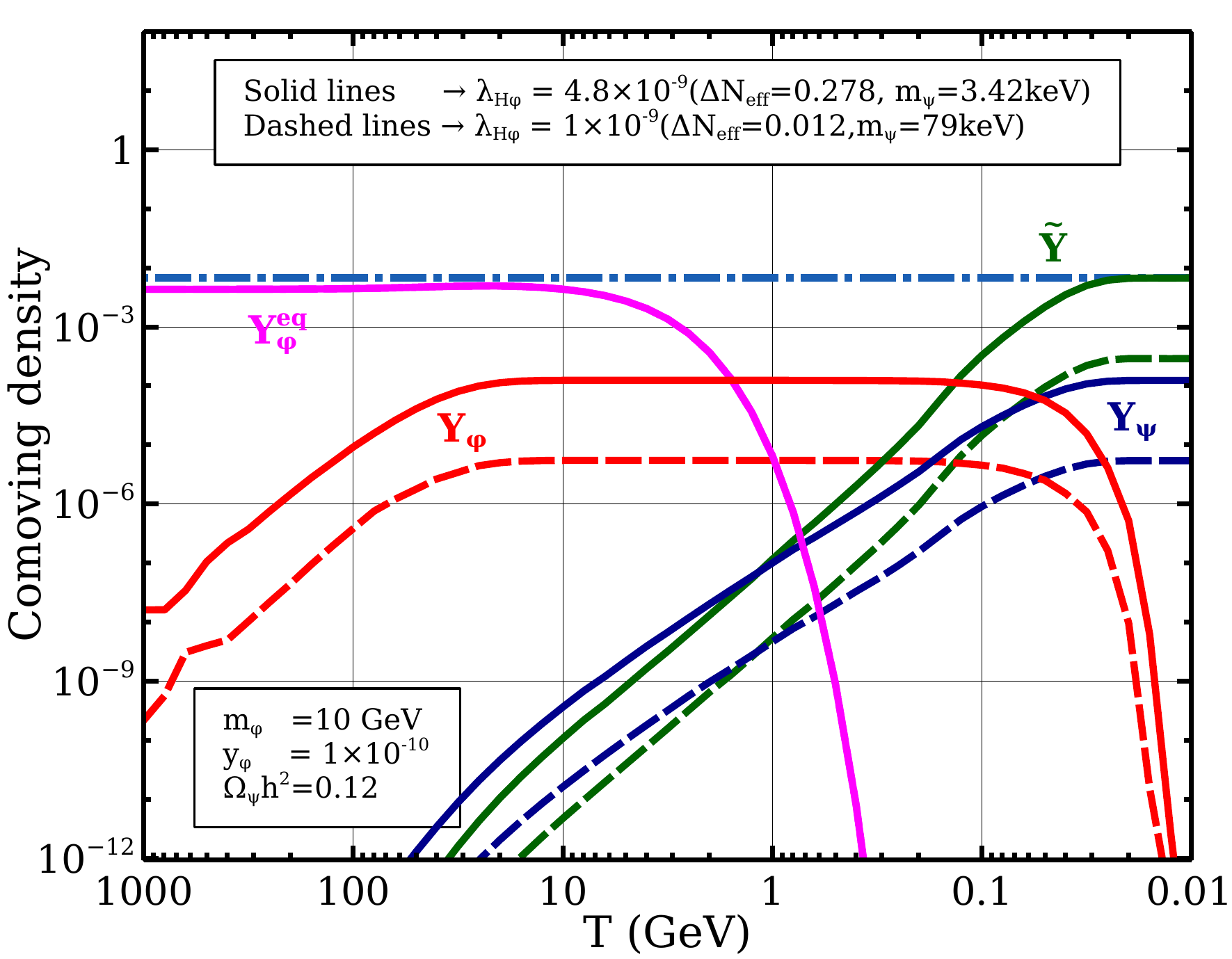}
\includegraphics[height=6cm,width=8.0cm,angle=0]{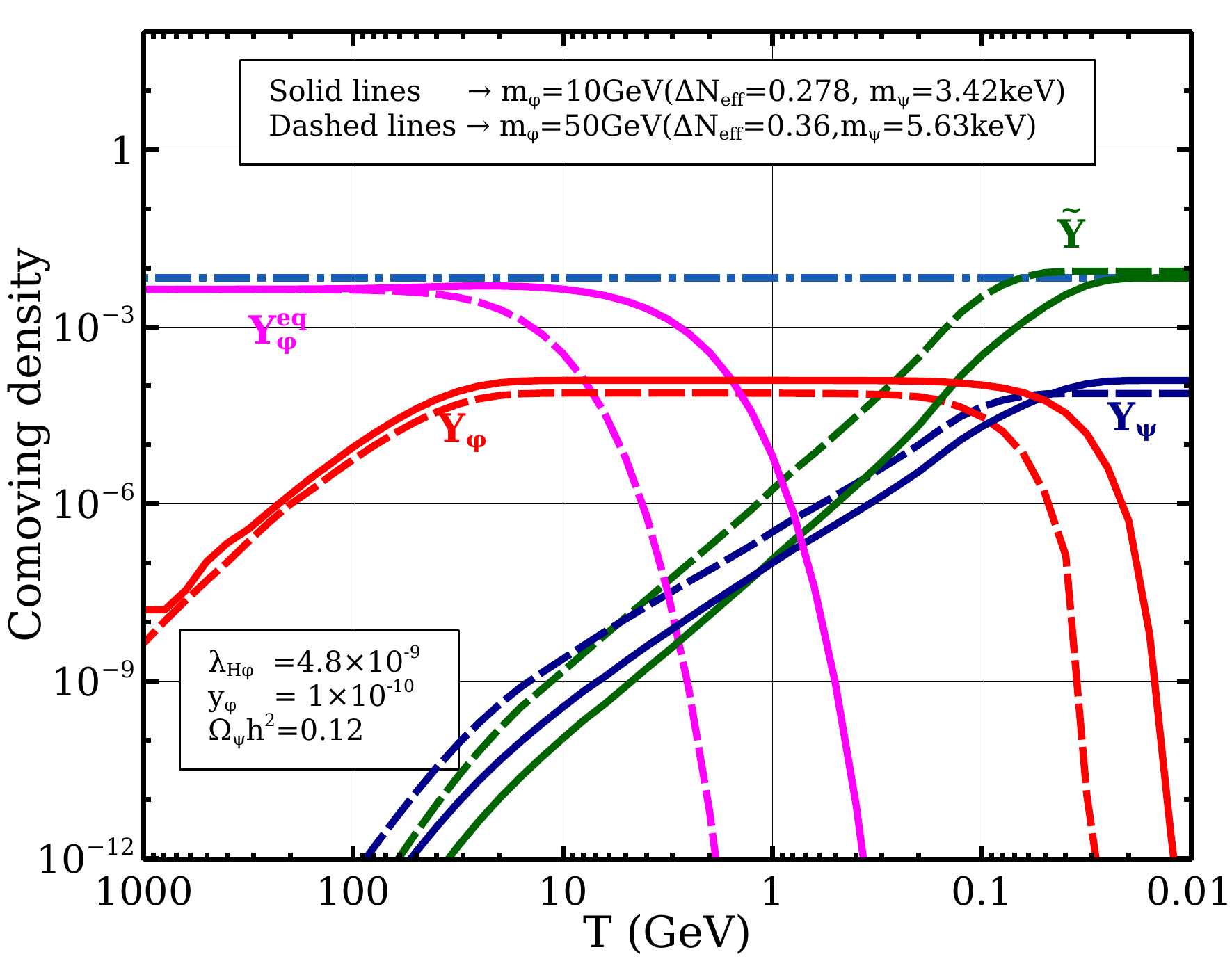}
\caption{Evolution of dark sector particles $(\phi, \psi, \nu_R)$ in case III considering $\phi$ to freeze in from Higgs decay and then decaying into $(\psi, \nu_R)$. Top left, top right and bottom panel plots show the change in evolution for two different choices of $y_\phi, \lambda_{H \phi}, m_\phi$ respectively. Chosen sets of points keep the DM abundance within the Planck limit.}
\label{fig:case3_benchmark}
\end{figure}

In this case, $\phi$ freezes in from Higgs decay and then decays into $\psi$ and $\nu_R$. Similar to earlier cases, we first show the evolution of dark sector particles for suitable choices of model parameters such that both DM abundance as well as $\Delta {\rm N}_{\rm eff}$ remain within Planck $2\sigma$ limits. The corresponding evolution plots are shown in Fig. \ref{fig:case3_benchmark}. We maintain similar colour codes as before namely, magenta, red, blue, green to show the evolution of comoving number densities of $\phi$ (equilibrium), $\phi$ (actual), DM $\psi$, $\nu_R$ respectively. In sharp contrast to case I, II discussed earlier, here we see that the initial abundance of $\phi$ remains negligible and then it slowly freezes in from decay of SM Higgs. In the top left panel of Fig. \ref{fig:case3_benchmark}, we show the differences in these evolution for two different choices of $y_\phi$. As usual, a smaller value of $y_{\phi}$ delays the decay of $\phi$. While final DM density remains same for both the values of $y_\phi$, the smaller value of $y_\phi$ leads to enhancement in $\nu_R$ density. Similar observation was noted in case II as well. In the top right panel of Fig. \ref{fig:case3_benchmark}, we show the variation due to two different choices of Higgs portal coupling $\lambda_{H \phi}$. In sharp contrast to case II, here we get smaller abundance of $\phi$ for smaller value of $\lambda_{H \phi}$ which also highlights the generic difference between freeze-in and freeze-out production mechanisms \cite{Hall:2009bx}. Consequently, smaller $\lambda_{H \phi}$ leads to smaller yields in $\psi, \nu_R$ as clearly seen from the same plot in top right panel. Finally, in the bottom panel plot of Fig. \ref{fig:case3_benchmark}, we show the variation due to two different choices of $m_\phi$. We see a marginal decrease in freeze-in abundance of $\phi$ for larger $m_\phi$ due to the fact that as $m_\phi$ approaches $m_h/2$, the corresponding partial decay width $\Gamma_{h \rightarrow \phi \phi^{\dagger}}$ decreases suppressing the production of $\phi$ slightly. On the other hand, a larger $m_\phi$ corresponds to larger decay width of $\phi$ in the limit $m_\psi \ll m_\phi$ leading to depletion in $\phi$ abundance earlier. The increase in $\phi$ decay width for larger $m_\phi$ also results in increased initial production of $\psi$ and $\nu_R$. While final DM abundance decreases slightly for larger $m_\phi$ due to smaller freeze-in abundance of heavier $\phi$, the abundance of $\nu_R$ gets slightly enhanced for larger $m_\phi$ due to larger decay width. Thus, there exists a competition between two effects: (i) decrease in $\nu_R$ production due to decrease in freeze-in production of $\phi$ for larger $m_\phi$ and (ii) increase in $\nu_R$ production due to increase in $\phi$ decay width for larger $m_\phi$ and the final results will be decided by the dominance of either of these, to be discussed below. In all the plots shown in Fig. \ref{fig:case3_benchmark}, we notice an intermediate plateau region for $\phi$ abundance. This arises when the freeze-in production rate of $\phi$ from Higgs decay and decay rate of $\phi$ into $\psi, \nu_R$ remain comparable.

\begin{figure}[h!]
\includegraphics[scale=0.5]{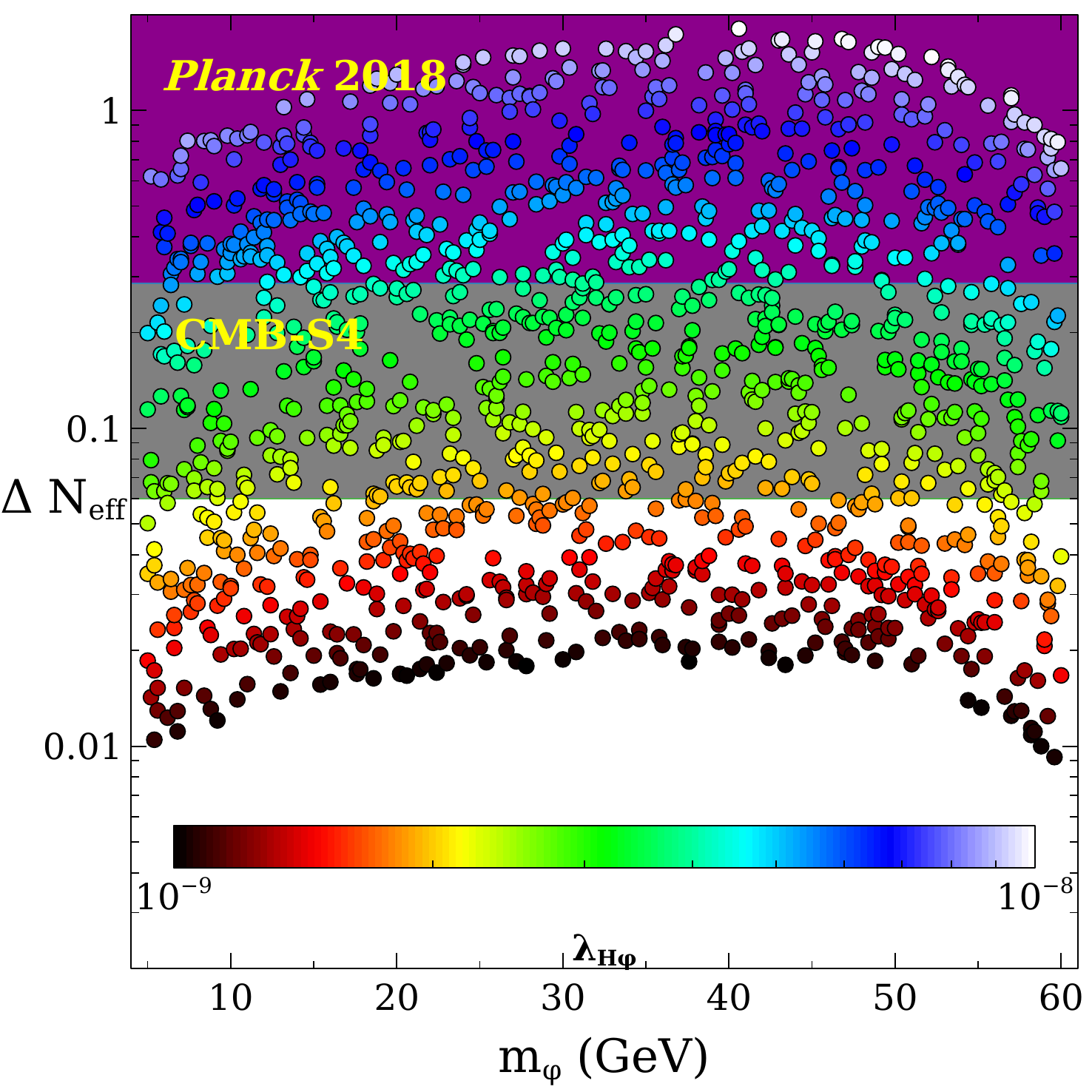}
\includegraphics[scale=0.5]{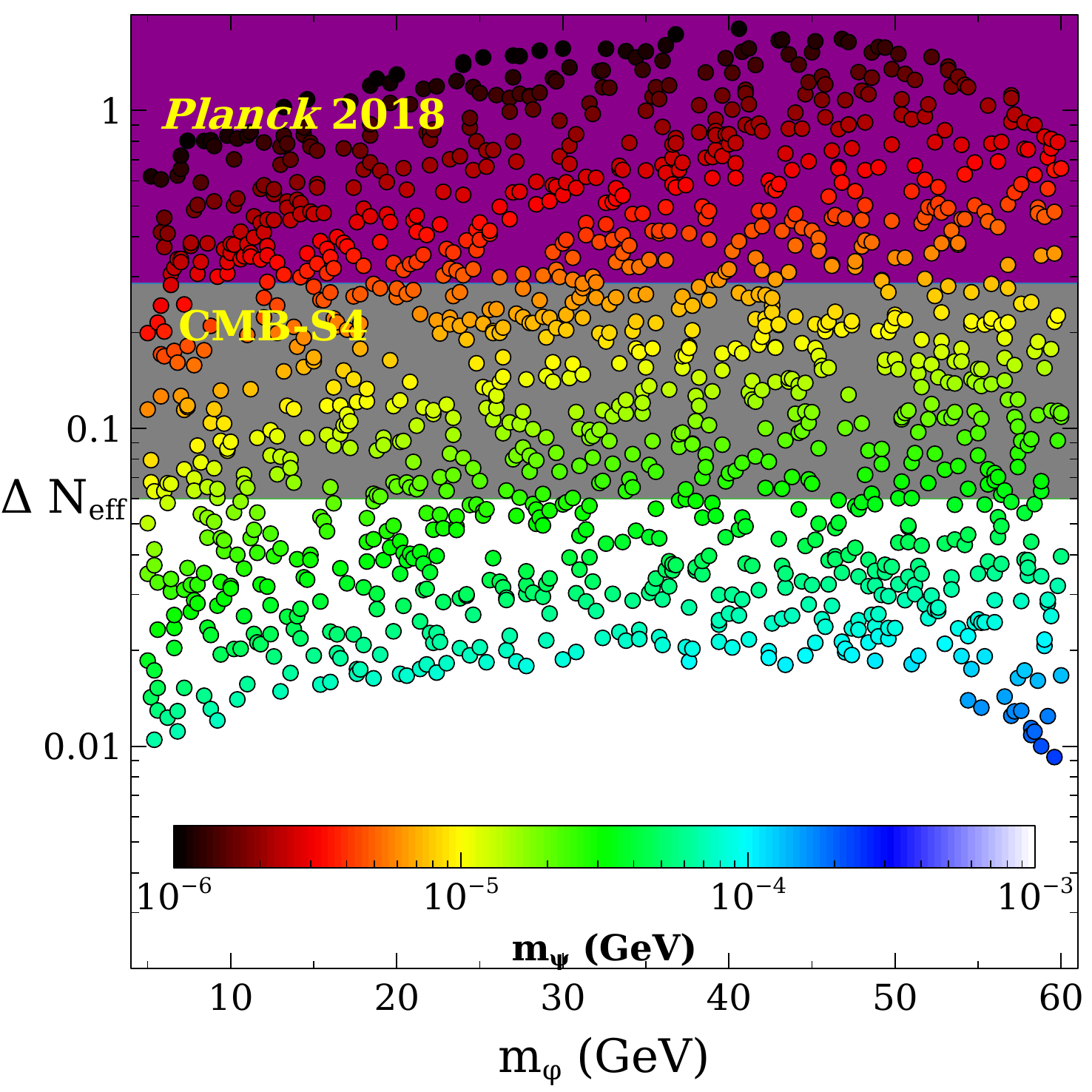}
	\caption{Parameter space plot for case III (considering $\phi$ to freeze in from Higgs decay) obtained from numerical scans, shown in terms of $\Delta {\rm N}_{\rm eff}$ vs $m_{\phi}$ while $\lambda_{H\phi}$ (left panel) and $m_{\psi}$ (right panel) are shown in colour code. The other relevant parameter $y_{\phi}$ is kept fixed at $10^{-10}$. The magenta and grey shaded regions indicate the current and future bound on $\Delta {\rm N}_{\rm eff}$ from Planck 2018 ($2\sigma$) and CMB-S4 respectively.}
	\label{fig:case3_scan}
\end{figure}

We then perform a numerical scan to show the parameter space assuming $\phi$ to be out-of-equilibrium throughout which freezes in only from the SM Higgs decay. In the scan, we vary the relevant parameters in the following range:
\begin{align} 
    5 \, {\rm GeV} & \leq m_{\phi} \leq 60 \, {\rm GeV}, \nonumber \\
    10^{-9} & \leq \lambda_{H\phi} \leq 10^{-8}, \nonumber \\
    1 \, {\rm keV} & \leq m_{\psi} \leq 1 \, {\rm MeV}. \nonumber 
\end{align}
Here also $y_{\phi}$ is kept fixed at $10^{-10}$. The resulting parameter space is shown in $\Delta {\rm N}_{\rm eff}$ vs $m_{\phi}$ plane in Fig. \ref{fig:case3_scan} with the colour bars in left and right panel plots showing the variation in $\lambda_{H\phi}$ and $m_{\psi}$ respectively. Similar to case II, here also the scattered points satisfy the Planck bound on DM relic abundance while the corresponding upper bound (future sensitivity) on $\Delta {\rm N}_{\rm eff}$ is shown by magenta (grey) shaded region. With an increase in $\lambda_{H \phi}$ while keeping $m_\phi$ fixed, we get enhancement in $\Delta {\rm N}_{\rm eff}$ as seen from the left panel plot of Fig. \ref{fig:case3_scan}, in sharp contrast with the corresponding results in case II. As discussed above, this trend is expected as increase in $\lambda_{H \phi}$ leads to increased freeze-in production of $\phi$. Since DM number density also increases from the same $\phi$ decay, we need to choose lighter DM masses for larger $\lambda_{H \phi}$ in order keep $\Omega_{\rm DM}{\rm h}^2$ within observed limits, as seen from the right panel plot of Fig. \ref{fig:case3_scan}. On the other hand, if $\phi$ mass increases for fixed $\lambda_{H \phi}$, we first see an increase in $\Delta {\rm N}_{\rm eff}$ followed by decrease for $m_\phi$ closer to $m_h/2$. The initial rise in $\Delta {\rm N}_{\rm eff}$ can be explained by noting the increase in $\phi$ decay width for larger $m_\phi$. However, if we continue to increase $m_\phi$, taking it closer to $m_h/2$, the partial decay width of the SM Higgs $\Gamma_{h \rightarrow \phi \phi^{\dagger}}$ decreases leading to suppression in freeze-in abundance of $\phi$. Consequently, this leads to decrease in $\nu_R, \psi$ densities. Correct DM abundance can be obtained by choosing heavier DM masses in the high $m_{\phi}$ regime, as seen from the right panel plot of Fig. \ref{fig:case3_scan}. Similar to case II discussed before, here also the bounds on DM mass become more severe, after imposing the structure formation constraints, as we discuss in the next section.

\subsubsection{$m_{\phi}>m_{h}/2$}
\begin{figure}[h!]
\includegraphics[height=6cm,width=8.0cm,angle=0]{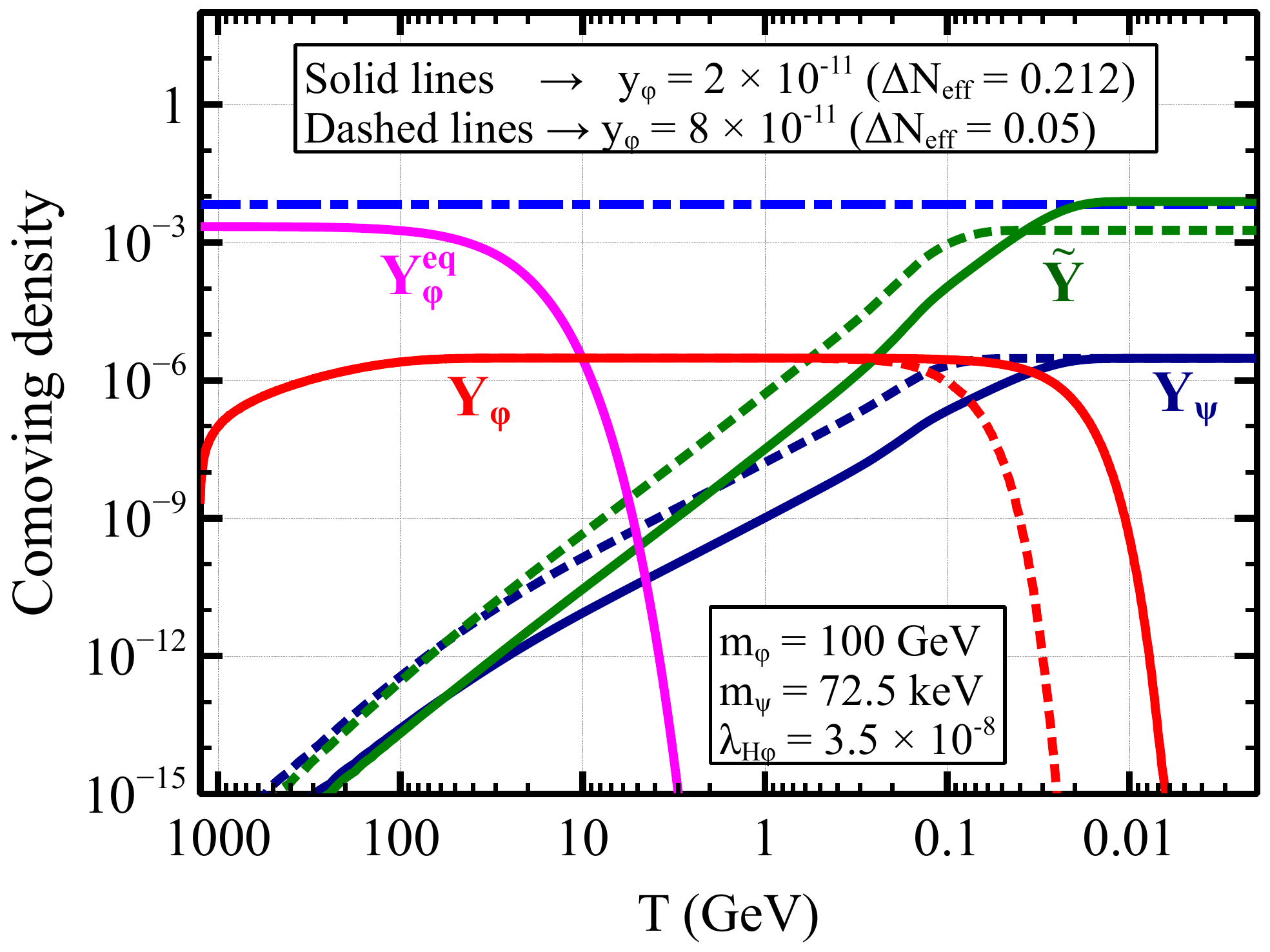}
\includegraphics[height=6cm,width=8.0cm,angle=0]{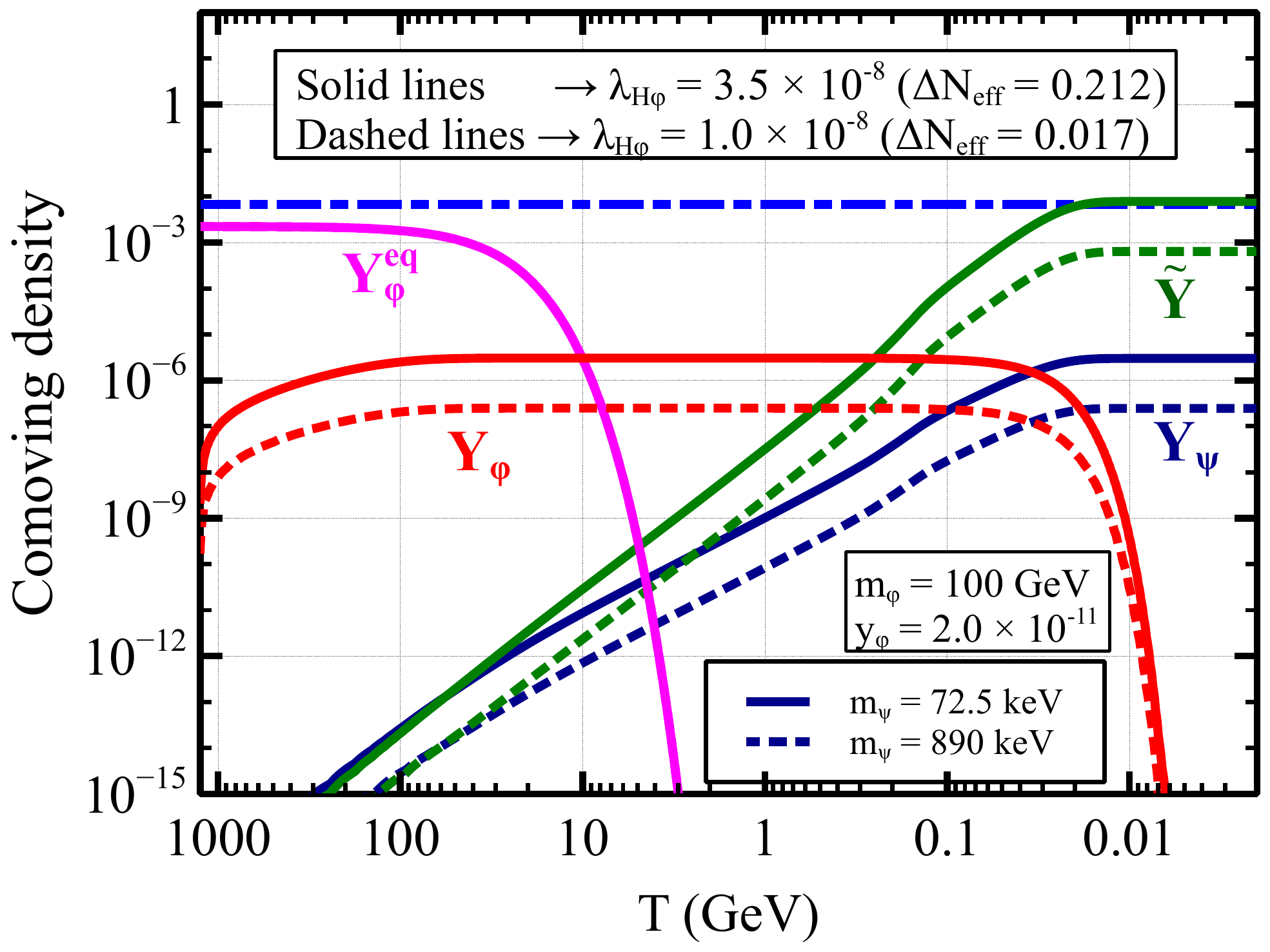}
\includegraphics[height=6cm,width=8.0cm,angle=0]{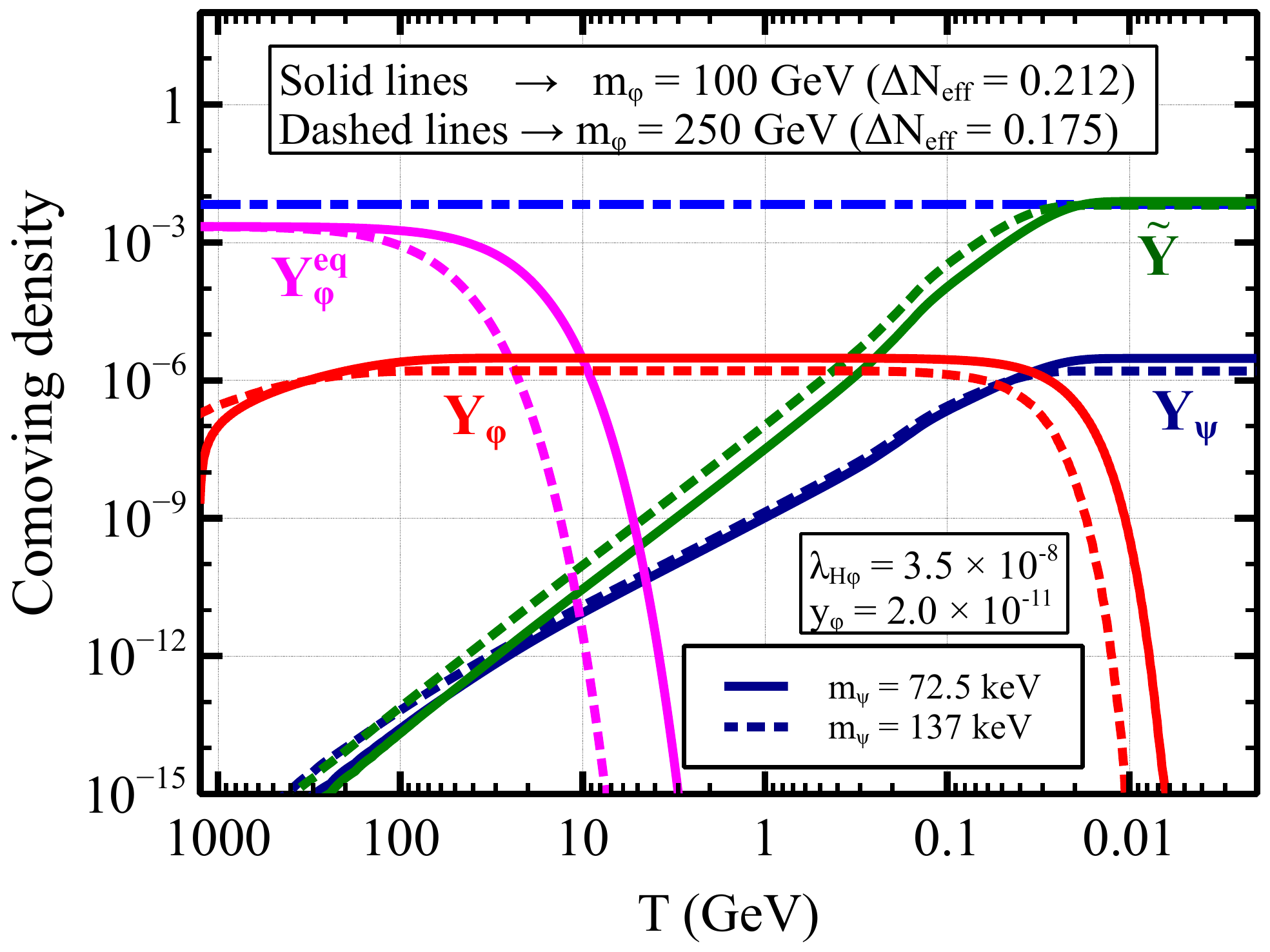}
\caption{Evolution of dark sector particles $(\phi, \psi, \nu_R)$ in case III considering $\phi$ to freeze in from Higgs annihilations and then decaying into $(\psi, \nu_R)$. Top left, top right and bottom panel plots show the change in evolution for two different choices of $y_\phi, \lambda_{H \phi}, m_\phi$ respectively. Chosen sets of points keep the DM abundance within the Planck limit.}
	    \label{fig:case3_annihilation}
\end{figure}

We now briefly discuss the essential features of the non-thermal $\phi$ scenario where its freeze-in production is dominated by annihilations only and decay is forbidden kinematically due to $m_{\phi} > m_{h}/2$. The evolution of dark sector particles in this case are shown in Fig. \ref{fig:case3_annihilation}. Once again, the choice of benchmark parameters is made in such a way that the final DM abundance and $\Delta {\rm N}_{\rm eff}$ remain within Planck 2018 limits. In top left panel of Fig. \ref{fig:case3_annihilation}, we show the variation in evolution for two different choices of $y_\phi$. As expected, this only alters the decay width of $\phi$ and hence the production of $\nu_R, \psi$. While final DM density remains same for both the choices, late production of $\nu_R$ due to smaller $y_\phi$ leads to an enhancement in $\widetilde{Y}$, an observation which was also made in other scenarios discussed above. In top right panel of Fig. \ref{fig:case3_annihilation}, we show the difference in evolution due to variation in Higgs portal coupling $\lambda_{H \phi}$. Naturally, a smaller $\lambda_{H \phi}$ results in smaller freeze-in abundance of $\phi$ from annihilation and hence smaller yields in $\nu_R, \psi$. Variation due to change in $m_\phi$ is shown in the bottom panel plot of Fig. \ref{fig:case3_annihilation}. We do not see much difference between the two values except for the fact that a larger $m_\phi$ increase $\phi$ decay width leading to early depletion. Since the overall features in this case remains similar to the earlier case where $\phi$ is produced from decay only, we expect the parameter space to remain similar. Therefore, we do not perform any numerical scan in this case.

\begin{figure}[h!]
\includegraphics[height=6cm,width=8.0cm,angle=0]{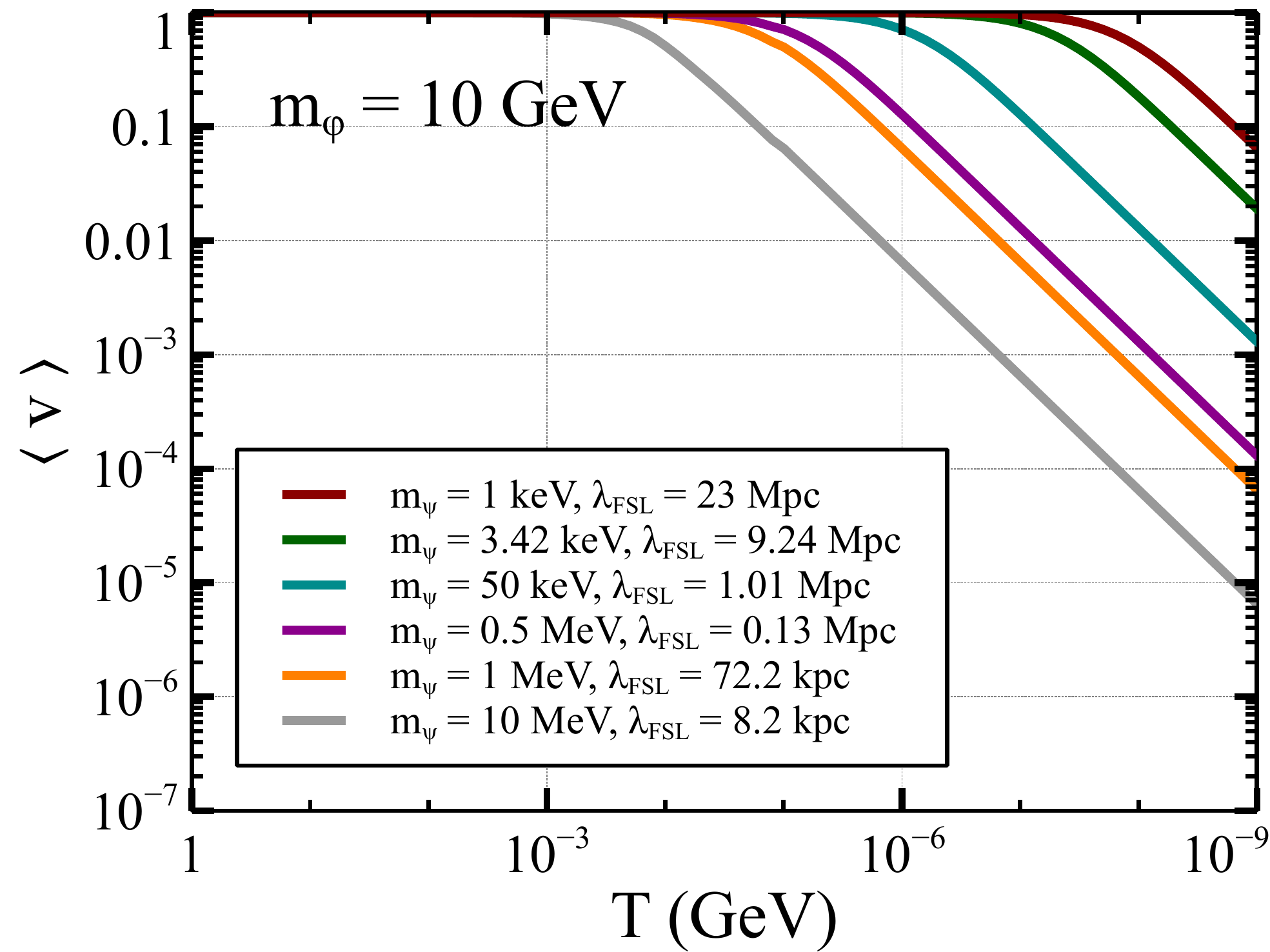}
\includegraphics[height=6cm,width=8.0cm,angle=0]{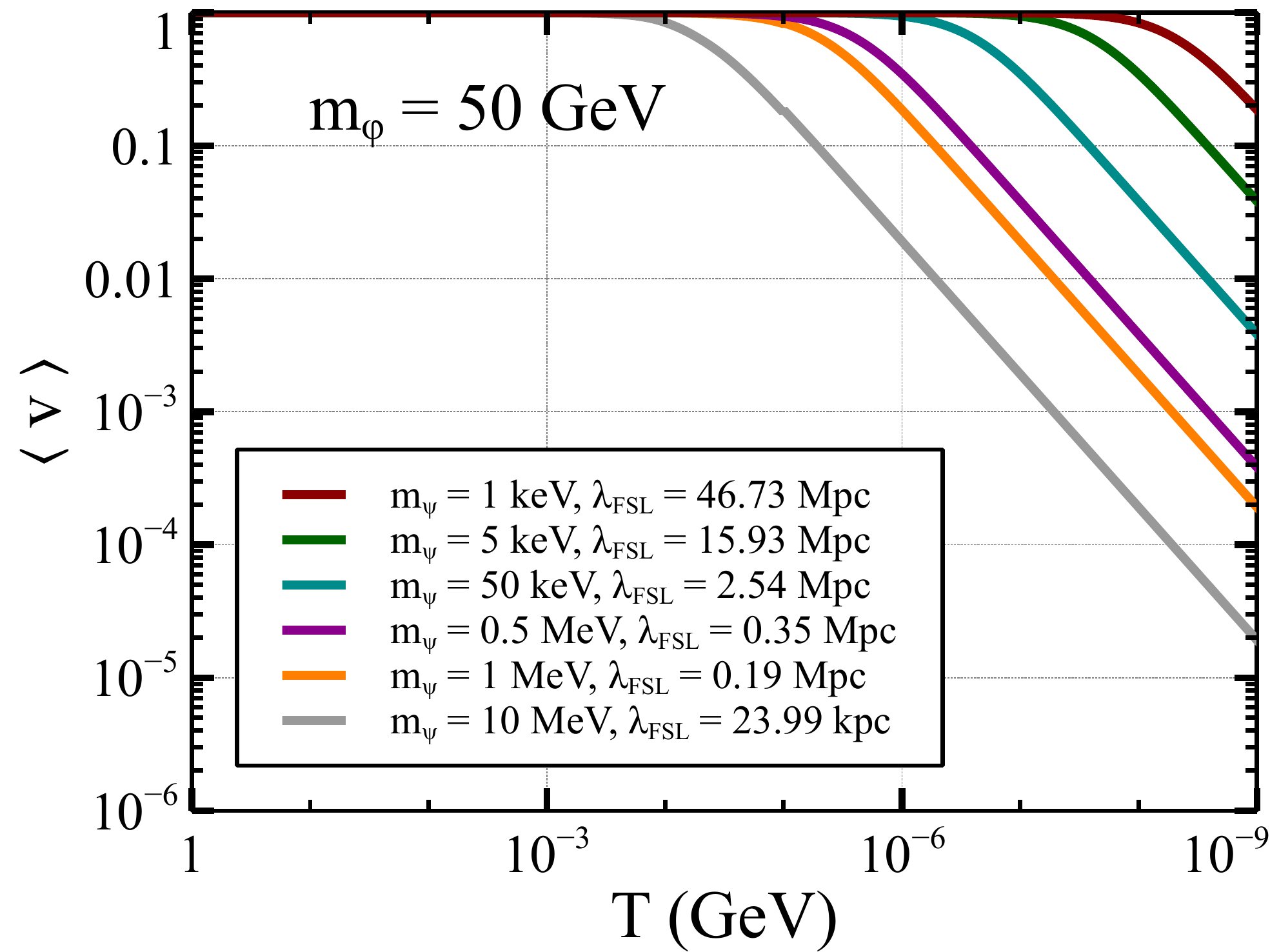}
\caption{Average velocity of DM as a function of temperature in case III for different benchmark combinations of relevant parameters.}
	    \label{fig:case3_FSL}
\end{figure}

\noindent
{\bf Structure formation constraints:} For case III, we have considered the situation when $m_{\phi} < m_{h}/2$. Here, we have considered the same benchmark point as in the bottom plot of Fig. \ref{fig:case3_benchmark} for two different $\phi$ mass, $m_{\phi}=10$ GeV and $m_{\phi}=50$ GeV. The production temperature for both the situation is around $10$ MeV (production temperature of DM for $m_{\phi}=10$ GeV and $m_{\phi}=50$ GeV are about $10$ MeV and $30$ MeV respectively). The Fig. \ref{fig:case3_FSL} shows that the FSL for a particular dark matter mass is more in the right plot where $m_{\phi}$ is $50$ GeV. This is expected as the production temperature is almost same, so an increase in mass of decaying particle injects more energy to the dark matter particles. For the left plot, the DM relic is satisfied when $m_{\psi} = 3.42$ keV and for the right side plot when $m_{\psi} \approx 5$ keV. For both the cases, the FSL when DM mass gives correct DM relic is larger than $0.1$ Mpc making the DM "hot". For these two benchmark points, the $\Delta {\rm N_{eff}}$ is within the current CMB bound. In principle, by increasing the dark sector coupling $y_{\phi}$, the production temperature can be increased making the FSL small.

\begin{table}[]
    \centering
    \caption{Table for case III}
    \begin{tabular}{|c|c|c|c|c|c|c|c|}
    \hline
    \multicolumn{4}{|c|}{Parameters} & \multirow{2}{*}{$\Omega_{\rm DM} {\rm h}^2$}& \multirow{2}{*}{$\Delta {\rm N_{eff}}$} & \multirow{2}{*}{FSL(Mpc)}\\ \cline{1-4}  $m_{\phi}$(GeV)& $\lambda_{H\phi}$ & $y_{\phi}$ & $m_{\psi}$(keV)& \multirow{2}{*}{} & \multirow{2}{*}{} & \multirow{2}{*}{}\\ \hline 
    $10$ & $4.8\times 10^{-9}$ & $10^{-10}$ &  $3.42$ & $0.12$ & $2.7\times10^{-1}$ & $9.42$ \\
    $50$ & $4.8\times 10^{-9}$ & $10^{-10}$ &  $5.63$ & $0.12$ & $3.6\times10^{-1}$ & $15.5$ \\
    \hline
    \end{tabular}
    \label{tab:case_3}
\end{table}

We summarize our FSL results for case III in table \ref{tab:case_3}, by including only those benchmark points from above analysis which satisfy correct DM relic density. Clearly, the constraints on DM mass from FSL criteria can be as severe as $\mathcal{O}(100 \, {\rm keV})$ keeping the $\Delta {\rm N}_{\rm eff} \leq \mathcal{O}(10^{-3})$. In the next section, we briefly comment on possible UV completions which can bring the $\Delta {\rm N}_{\rm eff}$ within CMB-S4 sensitivity while keeping the DM phenomenology similar to above analysis.

\section{Possible UV Completions}
\label{sec:uv}
We have discussed a minimal scenario to illustrate the essential results of freeze-in DM via light Dirac neutrino portal. The minimal nature of this model with only three new BSM fields has led to strong predictions on DM mass as well as $\Delta {\rm N}_{\rm eff}$ allowed from experimental constraints. Possible UV completions of this model can, in principle, give rise to a {\it natural} origin of light Dirac neutrino masses, a gauge symmetric realisation of the discrete $\mathbb{Z}_4$ symmetry while also giving a flexibility to enhance $\Delta {\rm N}_{\rm eff}$ to bring it within future experimental sensitivity.

One simple possibility is to introduce an additional Higgs doublet $H_2$, responsible for generating a light Dirac neutrino mass \cite{Davidson:2009ha}. While the freeze-in contribution to $\Delta {\rm N_{eff}}$ from Dirac Yukawa interaction with the SM Higgs doublet is negligibly small due to tiny Yukawa couplings \cite{Adshead:2020ekg, Luo:2020fdt}, the neutrinophilic Higgs doublet $H_2$ can have a larger Yukawa coupling leading to either thermalised $\nu_R$ or large freeze-in contribution to $\Delta {\rm N_{eff}}$. We can choose the $\mathbb{Z}_4$ charges of SM leptons, $\nu_R, \psi, \phi, H_2$ to be $i, -i, -1, i, -1$ respectively, so that the Yukawa interaction $\overline{L} \tilde{H_2} \nu_R$ is responsible for light Dirac neutrino mass. This charge assignment leaves the dark sector interactions same as in the minimal model. The second Higgs doublet can have a tiny soft-breaking term with the SM Higgs $\mu_{12} H^{\dagger} H_2$ by virtue of which its neutral component can acquire a tiny VEV, leading to a larger Dirac Yukawa. Due to the presence of multiple sources of $\Delta {\rm N_{eff}}$, we can have correct FIMP DM phenomenology while enhancing $\Delta {\rm N_{eff}}$ to remain within the sensitivity of next generation experiments.

Another possibility is to consider a gauge extension of the SM which naturally accommodates three right handed neutrinos required to realise a Dirac neutrino scenario. Perhaps the simplest possibility is to consider the gauged $B-L$ extension of the SM \cite{Davidson:1978pm, Mohapatra:1980qe, Marshak:1979fm, Masiero:1982fi, Mohapatra:1982xz, Buchmuller:1991ce} where three right handed neutrinos arise a minimal possibility to keep the model anomaly free. Depending upon the scalar content, light neutrinos can be purely Dirac in this model \cite{Ma:2015mjd, Reig:2016ewy, Wang:2017mcy,Han:2018zcn,Dasgupta:2019rmf,Nanda:2019nqy, Mahanta:2021plx}. The $B-L$ gauge charges of SM leptons, $\nu_R, \psi, \phi$ are $-1, -1, 0, 1$ to realise the minimal possibility. The fermion singlet DM couples via the same portal $\overline{\psi} \nu_R \phi$ while light Dirac neutrino mass arise from the SM Higgs Yukawa couplings. Although the contribution to $\Delta {\rm N_{eff}}$ from SM Higgs Yukawa interactions remain suppressed, there can be sizeable enhancement to it due to $B-L$ gauge interactions of $\nu_R$. The DM phenomenology will remain similar to the minimal setup except for the fact that $\phi$ can now interact with the SM bath via Higgs as well as $B-L$ gauge portal interactions. Therefore, such non-minimal FIMP DM via light Dirac neutrino portal can lead to observable $\Delta {\rm N_{eff}}$ which can be probed at CMB-S4 as well as other planned experiments like SPT-3G \cite{Benson:2014qhw}, Simons Observatory \cite{Ade:2018sbj}. We leave detailed phenomenological studies of such non-minimal scenarios to future works.


\section{Conclusion}
\label{sec:conclude}
We have studied a minimal scenario where the origin of neutrino mass and dark matter remain connected with interesting observational prospects at CMB experiments. Assuming light neutrinos to be of Dirac nature necessitates the inclusion of right handed neutrinos $\nu_R$ which can also act like a portal to dark sector comprising of a fermion singlet DM and a scalar singlet $\phi$. While the scalar singlet can be directly coupled to the SM bath via Higgs portal coupling, fermion singlet DM can couple only to $\nu_R$ via $\phi$. We have studied in details, the freeze-in production of $\psi$ and $\nu_R$ from $\phi$ decay, by considering three different possibilities with (i) $\phi$ in equilibrium, (ii) $\phi$ undergoing thermal freeze-out and (iii) $\phi$ getting produced via freeze-in. Since $\nu_R$ couples to SM leptons very feebly due to the requirement of generating sub-eV scale Dirac neutrino mass, the corresponding freeze-in production of $\nu_R$ directly from the SM bath remains suppressed. Since the same coupling with $\phi$ leads to freeze-in production of both DM and $\nu_R$ with the latter remaining relativistic throughout, we show the possibility of correlating DM parameter space with effective relativistic degrees of freedom $\Delta {\rm N}_{\rm eff}$. We find that the scenario with $\phi$ in equilibrium throughout leads to tiny enhancement in $\Delta {\rm N}_{\rm eff}$ while being consistent with DM relic criteria. However, for the other two scenarios, due to one additional free parameter in the form of Higgs portal coupling $\lambda_{H\phi}$ at play, we can have correct DM phenomenology while getting a sizeable enhancement in $\Delta {\rm N}_{\rm eff}$ at the same time. Additionally, depending upon the choice of parameters, existing bounds from the Planck experiment can also rule out DM mass up to a few tens of keV. However, structure formation constraints on such non-thermal DM rules out DM masses all the way up to a few hundred keV. Since DM and $\nu_R$ are produced from the same decay in this minimal model, the resulting $\Delta {\rm N}_{\rm eff}$ also gets reduced to $\leq \mathcal{O}(10^{-3})$ to be in agreement with required DM properties. We briefly discuss two possible UV completions which can disentangle the production of DM and $\nu_R$ while still maintaining the light Dirac neutrino portal scenario, such that correct DM properties can be realised even with enhanced $\Delta {\rm N}_{\rm eff}$ within experimental sensitivity.

Since the scalar singlet can be light in these scenarios opening up the possibility of SM Higgs decaying invisibly into a pair of $\phi$, future LHC measurements will be able to constrain the Higgs portal coupling further from measurements of Higgs invisible decay rates. In addition to these specific signatures of our model keeping it very predictive, one can also pursue such neutrino portal dark matter scenarios from the point of view of easing cosmological tensions between early and late universe cosmological observations \cite{Verde:2019ivm}. There have been a few works already in this direction \cite{DiValentino:2017oaw, He:2020zns} which we plan to explore in future works.
\section{Acknowledgements}
One of the authors AB would like to thank Sougata Ganguly for a
useful discussions on non-thermal distribution function
and related computational procedures. 
The research of AB was supported by Basic Science Research Program through the National Research Foundation of Korea (NRF) funded by the Ministry of Education through the Center for Quantum Spacetime (CQUeST) of Sogang University (NRF-2020R1A6A1A03047877). The work of DN is supported by National Research Foundation of Korea (NRF)’s grants, grants no. 2019R1A2C3005009(DN). ND would like to acknowledge Ministry of Education, Government of India for providing financial support for his research via the Prime Minister's Research Fellowship (PMRF) December 2021 scheme.
\appendix
\section{Derivation of Boltzmann equations}
\label{appen1}
The Boltzmann equation in differential form can be written as
\begin{equation}\label{BE}
    \frac{\partial f}{\partial t} - \mathcal{H} p \frac{\partial f}{\partial p} = C[f]
\end{equation}
where $\mathcal{H}$ is the Hubble expansion rate and $C[f]$ is the collision term
for a species with distribution function $f$. In this section, we discuss the derivation of the Boltzmann equations for the relevant species ($\phi, \psi, \nu_R$) in all the cases considered in this work.
\subsection{Case I: $\phi$ in equilibrium}
\subsubsection{For $\psi$ abundance:}

For the process: $\phi(K) \to \psi (P_1) + \overline{\nu_R}(P_2)$

Integrating both sides of Eq.\,\eqref{BE} over the three momentum
$p_1$ of species $\psi$, we get 
\begin{equation}\label{case1dm}
    \int g_{\psi} \frac{d^3 p_1}{(2 \pi)^3}
    \left[\frac{\partial f_{\psi}}{\partial t} - \mathcal{H} p_1 
    \frac{\partial f_{\psi}}{\partial p_1} \right] = 
    \int g_{\psi} \frac{d^3 p_1}{(2 \pi)^3} C[f_{\psi}]\,.
\end{equation}
Using the definition of $n_{\psi}$ and integration by parts method for the term proportional to $\mathcal{H}$, the LHS of Eq. \eqref{case1dm} becomes 
\begin{equation}
\frac{d n_{\psi}}{d t} + 3 \mathcal{H} n_{\psi}\,,
\end{equation}
where,
\begin{equation}
n_{\psi} = \int g_{\psi} \frac{d^3 p_1}{(2 \pi)^3} f_{\psi}\,,
\end{equation}
with $g_{\psi}$ being the internal degree of freedom of $\psi$.
The RHS of Eq. \eqref{case1dm} is
\begin{align}
    \int g_{\psi} \frac{d^3 p_1}{(2 \pi)^3} C[f_{\psi}] & = 
    \int g_{\psi} \frac{d^3 p_1}{(2 \pi)^3} \frac{1}{2 E_1}
    \int g_{\nu_R}\frac{d^3 p_2}{(2 \pi)^3 2 E_2} g_{\phi} 
    \frac{d^3 k}{(2 \pi)^3 2 E_k} \nonumber \\
   & \times (2 \pi)^4 \delta^4(K-P_1-P_2) 
   \lvert \mathcal{M} \rvert^2_{\phi \to \bar{\nu}_R \psi } (
   f_{\phi}^{\rm eq}- f_{\psi}f_{\nu_R}).
\end{align}
We assume that the initial abundances of both $\psi$ and $\nu_{R}$ are negligible, so both $f_{\psi}$ and $f_{\nu_{R}}$ can be set to zero. Thus we can omit the back-reaction term in the above equation. Now using the definition of decay width of $\phi$ in the rest frame of $\phi$ i.e.
\begin{equation}
    \Gamma_{\phi} = \frac{1}{2 m_{\phi}} \int \frac{g_{\psi}d^3 p_1}{(2 \pi)^3 2 E_1} \frac{g_{\nu_{R}}d^3 p_2}{(2 \pi)^3 2 E_2} (2 \pi)^4 \delta^4(K-P_1-P_2) 
    \lvert \mathcal{M} \rvert^2_{\phi \to \bar{\nu}_R \psi},
\end{equation}
we get
\begin{equation}\label{case1dm6}
    {\rm RHS} =  g_{\phi}\int \frac{d^3 k}{(2 \pi)^3}
    \frac{2 m_{\phi}}{2 E_k} \Gamma_{\phi} f_{\phi}^{\rm eq}.
\end{equation}
Here, the decay width $\Gamma_{\phi}$ is given by
\begin{eqnarray}
    &&\Gamma_{\phi} = \frac{g_{\psi} g_{\nu_R}}{16 \pi m_{\phi}} \lvert \mathcal{M} \rvert^2_{\phi \to \bar{\nu}_R \psi } \left(1 - \frac{m_{\psi}^2}{m_{\phi}^2} \right) \nonumber \\
    &&\hspace{-3cm}{\rm and} \nonumber \\
    &&\lvert \mathcal{M} \rvert^2_{\phi \to \bar{\nu}_R \psi } =  \frac{1}{g_{\phi}g_{\psi}g_{\nu_R}} y_{\phi}^2 (m_{\phi}^2-m_{\psi}^2).
\end{eqnarray}
Using $f_{\phi}^{\rm eq} = e^{-E_k/T}$, the Maxwell-Boltzmann distribution, we get,
\begin{eqnarray}\label{case1dm7}
    {\rm RHS} &=&  g_{\phi} \Gamma_{\phi} 
    \int \frac{d^3 k}{(2 \pi)^3} \frac{2 m_{\phi}}{2 E_k} e^{-E_k/T} \nonumber \\
    &=&  g_{\phi} \Gamma_{\phi} \frac{T}{2 \pi^2} m_{\phi}^2 K_1(m_{\phi}/T).
\end{eqnarray}
Putting $n_{\phi}^{\rm eq} = \dfrac{g_{\phi}}{2 \pi^2}
m_{\phi}^2 T K_2(m_{\phi}/T)$, the RHS becomes 
\begin{equation}
    {\rm RHS} =  \Gamma_{\phi} 
    \frac{K_1(m_{\phi}/T)}{K_2(m_{\phi}/T)} n_{\phi}^{\rm eq}.
\end{equation}
Finally, after equating LHS and RHS of Eq. \eqref{case1dm}, the Boltzmann equation for $n_{\psi}$ becomes
\begin{equation}\label{case1dm9}
    \frac{d n_{\psi}}{d t} + 3 \mathcal{H} n_{\psi} =  \Gamma_{\phi} \frac{K_1(m_{\phi}/T)}{K_2(m_{\phi}/T)} n_{\phi}^{\rm eq}.
\end{equation}
Now,instead of $n_{\psi}$, we can write the equation in terms of a new variable $Y_{\psi} = n_{\psi}/s$, known as comoving number density. Using the fact
that $sa^3 = constant$ with $s, a$ being the entropy density,  cosmic scale factor of the FLRW metric respectively,
the LHS of Eq. \eqref{case1dm9} becomes
\begin{eqnarray}
    s \frac{dY_{\psi}}{dt} &=& \frac{d n_{\psi}}{d t} + 3 \mathcal{H} n_{\psi}
\nonumber \\
\implies \frac{dY_{\psi}}{dT} &=&  - \frac{1}{3 \mathcal{H} s} \left[\frac{3}{T} + \frac{d g_{s}/dT}{g_s}\right] \left(\frac{d n_{\psi}}{d t} + 3 \mathcal{H} n_{\psi}\right) \nonumber \\
&=&  - \frac{1}{3 \mathcal{H}} \left[\frac{3}{T} + \frac{d g_{s}/dT}{g_s}\right] g_{\psi}  g_{\nu_R} g_{\phi} \Gamma_{\phi} \frac{K_1(m_{\phi}/T)}{K_2(m_{\phi}/T)} Y_{\phi}^{\rm eq}\nonumber \\
&=&  - \frac{ 1 }{ \mathcal{H} T} \left[1 + \frac{T d g_{s}/dT}{3 g_s}\right] \Gamma_{\phi} \frac{K_1(m_{\phi}/T)}{K_2(m_{\phi}/T)} Y_{\phi}^{\rm eq}.
\end{eqnarray}
Now defining $x=m_{\phi}/T$, we can write the above equation in terms of dimensionless variables $x$
\begin{equation} \label{case1_psi}
    \frac{dY_{\psi}}{dx} = \frac{ \beta}{x \mathcal{H} } \Gamma_{\phi} \frac{K_1(x)}{K_2(x)} Y_{\phi}^{\rm eq},
\end{equation}
where,
\begin{equation}
    \beta = \left[1 + \frac{T d g_{s}/dT}{3 g_s}\right].
\end{equation}

\subsubsection{For $\nu_R$ energy density:}

Let us start with the differential Boltzmann equation for $\nu_R$
\begin{equation}
    \frac{\partial f_{\nu_R}}{\partial t} - 
    \mathcal{H} p_2 \frac{\partial f_{\nu_R}}{\partial p_2} = C[f_{\nu_R}].
\end{equation}
Integrating both side with $\int g_{\nu R} E_2 \frac{d^3 p_2}{(2 \pi)^3}$, we get 
\begin{equation} \label{case1nu}
   \int g_{\nu_R} E_2 \frac{d^3 p_2}{(2 \pi)^3} \left(\frac{\partial f_{\nu_R}}{\partial t} - \mathcal{H} p_2 
   \frac{\partial f_{\nu_R}}{\partial p_2} \right) 
   = \int g_{\nu_R} E_2 \frac{d^3 p_2}{(2 \pi)^3} C[f_{\nu_R}].
\end{equation}
The LHS, after simplification becomes -
\begin{equation}
    \int g_{\nu_R} E_2 \frac{d^3 p_2}{(2 \pi)^3} \left(\frac{\partial f_{\nu_R}}{\partial t} - \mathcal{H} p_2 \frac{\partial f_{\nu_R}}{\partial p_2} \right) = \frac{d \rho_{\nu_R}}{dt} + 4 \mathcal{H} \rho_{\nu_R},
\end{equation}
where,
\begin{equation}
    \rho_{\nu_R} = \int g_{\nu_R} \frac{d^3p_2}{(2 \pi)^3} E_2 f_{\nu_R}.
\end{equation}
Expanding the collision term, the RHS becomes
\begin{align}
    \int g_{\nu_R} E_2 \frac{d^3 p_2}{(2 \pi)^3} C[f_{\nu_R}] & = g_{\nu_R} \int \frac{d^3 p_2}{(2 \pi)^3} \frac{1}{2 E_2} \int g_{\psi} \frac{d^3 p_1}{(2 \pi)^3 2 E_1} g_{\phi}\frac{d^3k}{(2 \pi)^3 2 E_k} \nonumber \\
    & \times E_2 (2 \pi)^4 \delta^4(K-P_1-P_2) \lvert \mathcal{M}
    \rvert^2_{\phi \to \bar{\nu}_R \psi }\,f_{\phi}^{\rm eq}.
    \label{eq:rhs_rho_nuR}
\end{align}
Let us do the following integral first.
\begin{align}
   I = & \int \frac{d^3 p_1}{(2 \pi)^3 2 E_1} 
   \frac{d^3p_2}{(2 \pi)^3 2 E_2} E_2 (2 \pi)^4 \delta^4(K-P_1-P_2) \lvert \mathcal{M} 
   \rvert^2_{\phi \to \bar{\nu}_R \psi } \nonumber \\
    = & \frac{1}{4 (2 \pi)^2} \int \frac{d^3 p_1}{E_1}
    d^3p_2 \delta^4(K-P_1-P_2) \lvert \mathcal{M}
    \rvert^2_{\phi \to \bar{\nu}_R \psi}\,.
\end{align}
We first do the integration over $\vec{p_2}$ using the Dirac delta function,
\begin{align}
    I = & \frac{1}{4 (2 \pi)^2} \int \frac{d^3 p_1}{E_1} \delta(E_k-E_1-E_{k-1}) \lvert \mathcal{M} \rvert^2_{\phi \to \bar{\nu}_R \psi} \nonumber \\
    = & \frac{2 \pi}{4 (2 \pi)^2} \int \frac{p^2_1 dp_1 d(\cos \theta)}{E_1} \delta(f(\theta)) \lvert \mathcal{M} \rvert^2_{\phi \to \bar{\nu}_R \psi }.
\end{align}
Here, $\theta$ is the angle between $\Vec{k}$ and $\Vec{p_1}$ and
$f(\theta) = E_k-E_1-E_{k-1}$ with $E_{k-1}=\sqrt{(\vec{k}-\vec{p_1})^2+m^2_{\nu}}$.
Now to find the root of $f(\theta)$, we set -
\begin{eqnarray}
    f(\theta)  &=& 0 \nonumber \\
    \implies E_k-E_1-E_{k-1} &=& 0 \nonumber \\
    \implies \cos\theta &=& \frac{2 E_k E_1 - (m_{\phi}^2 + m_{\psi}^2 - 
    m_{\nu}^2)}{2 \lvert {\Vec{k}} \rvert \lvert {\Vec{p}_1} \rvert } \equiv \cos\theta_0.
\end{eqnarray}
Also, 
\begin{equation}
   \left.\frac{df}{d \cos\theta}\right\vert_{\cos\theta = \cos\theta_0} = \frac{ \lvert {\Vec{k}} \rvert \lvert {\Vec{p}_1} \rvert }{E_k - E_1}.
\end{equation}
Thus, the integral $I$ reduces to
\begin{eqnarray}
    I &=& \frac{1}{4 (2 \pi)} \int \frac{p_1^2 dp_1}{E_1} \int d(\cos\theta)\frac{\delta(\cos\theta - \cos \theta_0)}{\left\vert\frac{df}{d \cos\theta}\right\vert_{\theta = \theta_0}} \lvert \mathcal{M} \rvert^2_{\phi \to \bar{\nu}_R \psi } \nonumber \\
     &=& \frac{\lvert \mathcal{M} \rvert'^2_{\phi \to \bar{\nu}_R \psi }}{8 \pi} \int \frac{p_1^2 dp_1}{E_1} \frac{E_k - E_1}{\lvert {\Vec{k}} \rvert \lvert {\Vec{p}_1} \rvert} \nonumber \\
   &=& \frac{\lvert \mathcal{M} \rvert'^2_{\phi \to \bar{\nu}_R \psi }}{8 \pi \lvert {\Vec{k}} \rvert } \int_{E_1^{\rm min}}^{E_1^{\rm max}} dE_1 (E_k - E_1).
\end{eqnarray}
In the above, $\lvert \mathcal{M} \rvert'$ implies $ \lvert \mathcal{M} \rvert$ at $\theta=\theta_0$. The limits of the integration will come from the condition 
\begin{equation}
    -1 \leq \cos{\theta_0} \leq 1.
\end{equation}
Working through it, we get 
\begin{align}
    E_1^{\rm min} = \frac{E_k(m_{\phi}^2 + m_{\psi}^2 - m_{\nu}^2) - \sqrt{E_k^2(m_{\phi}^2 + m_{\psi}^2 - m_{\nu}^2)^2 - m_{\phi}^2(\Lambda + 4E_k^2 m_{\psi}^2)}}{2 m_{\phi}^2} \equiv g_1(E_k) \nonumber \\
    E_1^{\rm max} = \frac{E_k(m_{\phi}^2 + m_{\psi}^2 - m_{\nu}^2) + \sqrt{E_k^2(m_{\phi}^2 + m_{\psi}^2 - m_{\nu}^2)^2 - m_{\phi}^2(\Lambda + 4E_k^2 m_{\psi}^2)}}{2 m_{\phi}^2} \equiv g_2(E_k),
\end{align}
where,
\begin{equation}
    \Lambda = (m_{\phi}^2 + m_{\psi}^2 - m_{\nu}^2)^2 - 4 m_{\phi}^2 m_{\psi}^2.
\end{equation}
Hence, $I$ becomes
\begin{align}
    I = & \frac{g_2(E_k) - g_1(E_k)}{8 \pi \lvert {\Vec{k}} \rvert} \lvert \mathcal{M} \rvert'^2_{\phi \to \bar{\nu}_R \psi } \left(E_k - \frac{g_2(E_k) + g_1(E_k)}{2}\right) \nonumber \\
    = & \frac{\sqrt{E_k^2(m_{\phi}^2 + m_{\psi}^2 - m_{\nu}^2)^2 - m_{\phi}^2(\Lambda + 4E_k^2 m_{\psi}^2)}}{8 \pi \lvert {\Vec{k}} \rvert m_{\phi}^2} \lvert \mathcal{M} \rvert'^2_{\phi \to \bar{\nu}_R \psi } \left(E_k - \frac{E_k(m_{\phi}^2 + m_{\psi}^2 - m_{\nu}^2)}{2m_{\phi}^2}\right) \nonumber \\
    = & \lvert \mathcal{M} \rvert'^2_{\phi \to \bar{\nu}_R \psi } \frac{\sqrt{E_k^2(m_{\phi}^2 + m_{\psi}^2 - m_{\nu}^2)^2 - m_{\phi}^2(\Lambda + 4E_k^2 m_{\psi}^2)}}{8 \pi \lvert {\Vec{k}} \rvert m_{\phi}^2} E_k \left(\frac{m_{\phi}^2 - m_{\psi}^2 + m_{\nu}^2}{2m_{\phi}^2}\right).
\end{align}
Finally, the RHS becomes -
\begin{align}
    {\rm RHS} & = g_{\phi} g_{\psi} g_{\nu_R} \frac{\lvert \mathcal{M} \rvert'^2_{\phi \to \bar{\nu}_R \psi }}{32 \pi^3} \frac{(m_{\phi}^2 - m_{\psi}^2 + m_{\nu}^2)}{2 m_{\phi}^4} \nonumber \\
    & \times \int_{m_{\phi}}^{\infty} E_k f_{\phi}^{\rm eq} \sqrt{E_k^2(m_{\phi}^2 + m_{\psi}^2 - m_{\nu}^2)^2 - m_{\phi}^2(\Lambda + 4E_k^2 m_{\psi}^2)}\,\,dE_k \nonumber \\
    &=  g_{\phi} g_{\psi} g_{\nu_R} \frac{\lvert \mathcal{M} \rvert'^2_{\phi \to \bar{\nu}_R \psi }}{32 \pi^3} \frac{(m_{\phi}^2 - m_{\psi}^2)^2}{2 m_{\phi}^4} \int_{m_{\phi}}^{\infty} E_k f_{\phi}^{\rm eq} \sqrt{E_k^2 - m_{\phi}^2}\,\,dE_k \qquad 
    (\because m_{\nu} \simeq 0)
    \label{RHS_mnu_eq_zero}\\
    &=  g_{\phi} g_{\psi} g_{\nu_R} \frac{\lvert \mathcal{M} \rvert'^2_{\phi \to \bar{\nu}_R \psi }}{32 \pi^3} \frac{(m_{\phi}^2 - m_{\psi}^2)^2}{2 m_{\phi}^4} \int_{m_{\phi}}^{\infty} E_k e^{-E_k/T} \sqrt{E_k^2 - m_{\phi}^2}\,\,dE_k \nonumber \\
    & = g_{\phi} g_{\psi} g_{\nu_R} \frac{\lvert \mathcal{M} \rvert'^2_{\phi \to \bar{\nu}_R \psi }}{32 \pi^3} \frac{(m_{\phi}^2 - m_{\psi}^2)^2}{2 m_{\phi}^4} m_{\phi}^2 T K_2(m_{\phi}/T) \nonumber \\
    & = \langle E\Gamma\rangle n_{\phi}^{\rm eq},
\end{align}
where
\begin{equation}
    \left<E\Gamma\right> = g_{\psi} g_{\nu_R} \frac{\lvert \mathcal{M} \rvert'^2_{\phi \to \bar{\nu}_R \psi }}{32 \pi} \frac{(m_{\phi}^2 - m_{\psi}^2)^2}{ m_{\phi}^4}.
\end{equation}
So, the final form of the evolution equation of $\rho_{\nu_R}$ is 
\begin{align}
    \frac{d \rho_{\nu_R}}{dt} & + 4 \mathcal{H} \rho_{\nu_R} =  \langle E\Gamma\rangle n_{\phi}^{\rm eq} \nonumber\\
    \implies \frac{d\widetilde{Y}}{dT} = & - \frac{\beta}{\mathcal{H} T s^{4/3}} \left<E\Gamma\right> n_{\phi}^{\rm eq}
    \qquad ({\rm where}\,\,\widetilde{Y} = \dfrac{\rho_{\nu_R}}{s^{4/3}}) \,\,.
\end{align}
In terms of $x = m_{\phi}/T$, the above equation becomes 
\begin{equation} \label{case1_nuR}
    \frac{d \widetilde{Y}}{dx} =    \frac{\beta}{\mathcal{H} s^{1/3} x} \langle E\Gamma \rangle Y_{\phi}^{\rm eq}.
\end{equation}

\subsection{Case II}
In this case, $\phi$ is not in equilibrium always. It is produced in equilibrium and at some epoch it goes out of equilibrium due to
thermal freeze-out. 
\subsubsection{For $\psi$ abundance:}\label{case2psi}
The procedure to obtain the Boltzmann equation for $\psi$ in this case
is same as the above case from Eq. \eqref{case1dm} to Eq. \eqref{case1dm6} except that $f_{\phi}^{\rm eq}$ is now replaced by
$f_{\phi}$. Thus, the Boltzmann equation for $\psi$ is 
\begin{equation}
    \frac{d n_{\psi}}{d t} + 3 \mathcal{H} n_{\psi} =  g_{\phi} 
    \int \frac{d^3 k}{(2 \pi)^3} \frac{2 m_{\phi}}{2 E_k} \Gamma_{\phi} f_{\phi}\,.
\label{BEpsi_CII}    
\end{equation}
Since $\phi$ was in equilibrium earlier and goes
out of equilibrium after freeze-out, we can write
the general form of the Maxwell-Boltzmann distribution function
for $\phi$ with a chemical potential that is nonzero
only after the freeze-out of $\phi$
i.e. $f_{\phi} = e^{\mu/T}e^{-E_k/T}$. The chemical
potential $\mu$ is defined as
$\mu = T\ln\left(\dfrac{n_{\phi}(T)}{n_{\phi}^{\rm eq}(T)}\right)$.
Substituting $f_{\phi}$ in Eq.\,\eqref{BEpsi_CII},
the Boltzmann equation becomes
\begin{eqnarray}
    \frac{d n_{\psi}}{d t} + 3 \mathcal{H} n_{\psi} =  g_{\phi} e^{\mu/T} \int \frac{d^3 k}{(2 \pi)^3} \frac{2 m_{\phi}}{2 E_k} \Gamma_{\phi} e^{-E_k/T}.
\end{eqnarray}
The RHS of the equation is same as Eq.\,\eqref{case1dm7}
in case I except for the $e^{\mu/T}$ factor. Hence, following
the same procedure as Eq.\,\eqref{case1dm7} to Eq.\,\eqref{case1dm9}
and replacing $\mu$ by number density, we get
\begin{eqnarray}
    \frac{d n_{\psi}}{d t} + 3 \mathcal{H} n_{\psi} &=&  e^{\mu/T} \Gamma_{\phi} \frac{K_1(m_{\phi}/T)}
    {K_2(m_{\phi}/T)} n_{\phi}^{\rm eq}\,, \nonumber \\
    &=& \Gamma_{\phi} \frac{K_1(m_{\phi}/T)}
    {K_2(m_{\phi}/T)} n_{\phi}\,.
\label{eq:BE_npsi_for_nonEq_phi}
\end{eqnarray}
We can write the above equation in terms of
$Y_{\psi} = n_{\psi}/s$, $Y_{\phi} = n_{\phi}/s$ and
$x = m_{\phi}/T$. In terms of these dimensionless
quantities the above equation takes the following form
\begin{equation}
     \frac{dY_{\psi}}{dx} = \frac{ \beta }{x \mathcal{H} } 
     \Gamma_{\phi} \frac{K_1(x)}{K_2(x)} Y_{\phi}\,.
\label{BE_psi_case2}     
\end{equation}
\subsubsection{For $\nu_R$ energy density:}
To find the energy density of $\nu_R$ in this case, we will follow
the same procedure as in the previous case, the only difference
will be that now $f_{\phi}^{eq}$ will be replaced by $f_{\phi}=e^{\mu/T}e^{E_k/T}$. Hence starting from
Eq.\,\eqref{RHS_mnu_eq_zero}, the R.H.S. of the Boltzmann
equation for $\rho_{\nu_R}$ can be written as
\begin{align}
    \frac{d \rho_{\nu R}}{dt} + 4 \mathcal{H} \rho_{\nu R} = & g_{\phi} g_{\psi} g_{\nu_R} \frac{\lvert \mathcal{M} \rvert'^2_{\phi \to \bar{\nu}_R \psi }}{32 \pi^3} \frac{(m_{\phi}^2 - m_{\psi}^2)^2}{2 m_{\phi}^4} \int_{m_{\phi}}^{\infty} E_k f_{\phi} \sqrt{E_k^2 - m_{\phi}^2} dE_k\,,\nonumber\\
    = & g_{\phi} g_{\psi} g_{\nu_R} \frac{\lvert \mathcal{M} \rvert'^2_{\phi \to \bar{\nu}_R \psi }}{32 \pi^3} \frac{(m_{\phi}^2 - m_{\psi}^2)^2}{2 m_{\phi}^4} \int_{m_{\phi}}^{\infty} E_k e^{\mu/T}e^{E_k/T} \sqrt{E_k^2 - m_{\phi}^2}\,, \nonumber\\
    = & g_{\phi} g_{\psi} g_{\nu_R} \frac{\lvert \mathcal{M} \rvert'^2_{\phi \to \bar{\nu}_R \psi }}{32 \pi^3} \frac{(m_{\phi}^2 - m_{\psi}^2)^2}{2 m_{\phi}^4} e^{\mu/T} \int_{m_{\phi}}^{\infty} E_k e^{E_k/T} \sqrt{E_k^2 - m_{\phi}^2}\,, \nonumber \\
    = & g_{\phi} g_{\psi} g_{\nu_R} \frac{\lvert \mathcal{M} \rvert'^2_{\phi \to \bar{\nu}_R \psi }}{32 \pi^3} \frac{(m_{\phi}^2 - m_{\psi}^2)^2}{2 m_{\phi}^4} e^{\mu/T} m_{\phi}^2 T K_2(m_{\phi}/T)\,, \nonumber \\
    = &  \langle E\Gamma\rangle  e^{\mu/T} n_{\phi}^{\rm eq}\,, \nonumber\\
    \implies \frac{d \rho_{\nu R}}{dt} + 4 H \rho_{\nu R}
    = &  \langle E\Gamma\rangle n_{\phi}.
\end{align}
Now expressing $\rho_{\nu_R}$ by the comoving energy density,
$\widetilde{Y}$, the above equation in terms of $T$ and $x=m_{\phi}/T$ are given by
\begin{align} \label{case2_nuR}
    \frac{d\widetilde{Y}}{dT} = & -  \frac{\beta}{\mathcal{H} T s^{1/3}} \langle E\Gamma \rangle Y_{\phi}\,, \nonumber\\
    \frac{d \widetilde{Y}}{dx} = &  \frac{\beta}{\mathcal{H} s^{1/3} x} \langle E\Gamma \rangle Y_{\phi}\,.
\end{align}
\subsubsection{For comoving number density of non-thermal $\phi$:}
The calculation of the number density of $\phi$ will involve two
processes : $X(K'_1)+\bar{X}(K'_2)\rightarrow \phi(K_1)+
\phi^{\dagger}(K_2)$ and $\phi(K_1)
\rightarrow \psi(P_1)+\overline{\nu_R}(P_2)$. Hence, the
differential form of the Boltzmann equation is
($X$ is any SM particle) -
\begin{align}
    \frac{\partial f_{\phi}}{\partial t} 
    - \mathcal{H}\,k_1 \frac{\partial f_{\phi}}{\partial k_1} = & 
    C^{X\bar{X}\rightarrow\phi\phi^{\dagger}}[f_{\phi}] - 
    C^{\phi\rightarrow \psi \nu_R}[f_{\phi}] \nonumber \\
    \int g_{\phi} \frac{d^3k_1}{(2\pi)^3}
    \left(\frac{\partial f_{\phi}}{\partial t} 
    - \mathcal{H}\,k_1 \frac{\partial f_{\phi}}{\partial k_1}\right)
    = &  \int g_{\phi} \frac{d^3k_1}{(2\pi)^3}
    \left(C^{X\bar{X}\rightarrow\phi\phi^{\dagger}}[f_{\phi}] 
    - C^{\phi\rightarrow \psi \nu_R}[f_{\phi}]\right)\,.
\end{align}
The LHS is
\begin{align}
    \int g_{\phi} \frac{d^3k_1}{(2\pi)^3}
    \left(\frac{\partial f_{\phi}}{\partial t} 
    - \mathcal{H}\,k_1 \frac{\partial f_{\phi}}
    {\partial k_1}\right) = 
    \frac{d n_{\phi}}{d t} + 3 \mathcal{H} n_{\phi}.
\end{align}
The first term of RHS is
\begin{align}
    & \int g_{\phi} \frac{d^3k_1}{(2\pi)^3}
    C^{X\bar{X}\rightarrow\phi\phi^{\dagger}}[f_{\phi}]\,, \nonumber\\
    = & \int g_{\phi} \frac{d^3k_1}{(2\pi)^3} \frac{1}{2E_{k_1}} \int g_X \frac{ d^3k'_1}{(2\pi)^3 2E_{k'_1}} \int g_X \frac{ d^3k'_2}{(2\pi)^3 2E_{k'_2}} \int g_{\phi} \frac{ d^3k_2}{(2\pi)^3 2E_{k_2}} (2 \pi)^4 \delta^4(K'_1 + K'_2 - K_1 - K_2) \nonumber \\ 
    & \qquad \times \lvert \mathcal{M} \rvert^2_{X\bar{X}\rightarrow
    \phi\phi^{\dagger}} (f_{k'_1}f_{k'_2} - f_{k_1}f_{k_2})\,, \nonumber \\
    = &  (n_{\phi}^{\rm eq})^2 \left<\sigma v\right>_{\phi \phi^{\dagger{}} \to X \bar{X}} \left(\left(\frac{n_{X}}{n_{X}^{\rm eq}}\right)^2 - \left(\frac{n_{\phi}}{n_{\phi}^{\rm eq}}\right)^2\right)\,, \qquad \left(\because f_{i} = e^{\mu_i/T}e^{-E_{i}/T} = 
    \dfrac{n_{i}}{n_{i}^{\rm eq}}\right)\,,\nonumber\\
    = & \left<\sigma v\right>_{\phi \phi^{\dagger{}} \to X \bar{X}} \left((n_{\phi}^{\rm eq})^2 - 
    (n_{\phi})^2 \right)\,, \qquad (\because n_{X}^{\rm eq} = n_{X})
\end{align}
where,
\begin{align}
    \langle \sigma v \rangle &= \frac{1}{(n_{\phi}^{\rm eq})^2} \int g_{\phi} \frac{d^3k_1}{(2\pi)^3} \frac{1}{2E_{k_1}} \int g_{X} \frac{ d^3k'_1}{(2\pi)^3 2E_{k'_1}} \int g_{X} \frac{ d^3k'_2}{(2\pi)^3 2E_{k'_2}} \int g_{\phi} \frac{ d^3k_2}{(2\pi)^3 2E_{k_2}} \nonumber \\
    & \times (2 \pi)^4 \delta^4(K'_1 + K'_2 - K_1 - K_2)
    \lvert \mathcal{M} \rvert^2_{X\bar{X}\rightarrow
    \phi\phi^{\dagger}} e^{-(E_{k_1}+E_{k_2})/T} \,,\nonumber \\
    & = \frac{1}{(n_{\phi}^{\rm eq})^2} \int g_{\phi} \frac{d^3k_1}{(2\pi)^3 } \int g_{\phi} \frac{ d^3k_2}{(2\pi)^3} \frac{1}{4 E_{k_1} E_{k_2}} \int g_{X} \frac{ d^3k'_1}{(2\pi)^3 2E_{k'_1}} \int g_{X} \frac{ d^3k'_2}{(2\pi)^3 2E_{k'_2}} \nonumber \\
    & \times (2 \pi)^4 \delta^4(K'_1 + K'_2 - K_1 - K_2) 
    \lvert \mathcal{M} \rvert^2_{\phi\phi^{\dagger}\rightarrow X\bar{X}} e^{-(E_{k_1}+E_{k_2})/T}\,, \quad (\because \lvert M\rvert^2_{X\bar{X}\rightarrow \phi\phi}= \lvert M \rvert^2_{\phi\phi\rightarrow X\bar{X}}) \nonumber \\ 
    & = \frac{g_{\phi}^2}{(n_{\phi}^{\rm eq})^2} \int  \frac{d^3k_1}{(2\pi)^3 } \frac{ d^3k_2}{(2\pi)^3} (\sigma v)_{\phi\phi^{\dagger}\rightarrow X\bar{X}} e^{-(E_{k_1}+E_{k_2})/T}\,,  \nonumber \\
    & = \frac{\int  \frac{d^3k_1}{(2\pi)^3 } \int  \frac{ d^3k_2}{(2\pi)^3} (\sigma v)_{\phi\phi^{\dagger}\rightarrow X\bar{X}} e^{-(E_{k_1}+E_{k_2})/T}}{\int  \frac{d^3k_1}{(2\pi)^3 } \int  \frac{ d^3k_2}{(2\pi)^3} e^{-(E_{k_1}+E_{k_2})/T}}\,, \nonumber \\
    & =  \dfrac{1}{8 m_{\phi}^4 T K_2^2(m_{\phi}/T)} \int_{4 m_{\phi}^2}^{\infty} (\sigma)_{\phi\phi^{\dagger}\rightarrow X\bar{X}} (s-4m_{\phi}^2)\sqrt{s}K_1(\sqrt{s}/T)ds.
\end{align}
We have obtained the last expression following the prescription
given in \cite{Gondolo:1990dk}. Now the second term in the RHS is
\begin{align}
    \int g_{\phi} \frac{d^3k_1}{(2\pi)^3}
    C^{\phi\rightarrow \psi \nu_R}[f_{\phi}] = 
    \int g_{\phi} \frac{d^3k_1}{(2\pi)^3} \frac{1}{2 E_{k_1}} 
    \int g_{\psi} \frac{d^3p_1}{(2\pi)^3 2 E_1} 
    \int g_{\nu R} \frac{d^3p_2}{(2\pi)^3 2 E_2}
    (2\pi)^4 \delta^4(K_1-P_1-P_2) f_{\phi}.
\label{RHS_phi_decay}    
\end{align}
Here due to non-thermal nature of $\psi$ and $\nu_R$, we have omitted
the back reaction term which otherwise will be there in
Eq.\,\eqref{RHS_phi_decay} and is proportional to $f_{\psi} f_{\nu_R}$.
This is the same decay process that we have worked through
the section \ref{case2psi} when $\phi$ is not in equilibrium.
Therefore, from Eq.\,\eqref{eq:BE_npsi_for_nonEq_phi} we obtain
\begin{align}
    \int g_{\phi} \frac{d^3k_1}{(2\pi)^3}
    C^{\phi\rightarrow \psi \nu_R}[f_{\phi}] = 
    \Gamma_{\phi} \frac{K_1(m_{\phi}/T)}{K_2(m_{\phi}/T)} n_{\phi}\,.
\end{align}
Finally, the full equation for the evolution of $n_{\phi}$
is 
\begin{align}
    \frac{d n_{\phi}}{d t} + 3 \mathcal{H} n_{\phi} = 
    & - \left<\sigma v\right>_{\phi \phi^{\dagger{}} \to X \bar{X}}
    \left((n_{\phi})^2 - (n_{\phi}^{eq})^2 \right)
    -  \Gamma_{\phi} \frac{K_1(m_{\phi}/T)}{K_2(m_{\phi}/T)} n_{\phi}\,.
\end{align}
In terms of comoving number density $Y_{\phi}$,
\begin{align}\label{case2_phi}
    \dfrac{dY_{\phi}}{dT} = & -\dfrac{\beta s}{\mathcal{H} T}
    \left(- \left<\sigma v\right>_{\phi \phi^{\dagger{}} \to X \bar{X}} \left((Y_{\phi})^2 -  (Y_{\phi}^{\rm eq})^2 \right) - \dfrac{\Gamma_{\phi}}{s}
    \dfrac{K_1(m_{\phi}/T)}{K_2(m_{\phi}/T)} Y_{\phi}
    \right)\,, \nonumber \\
\implies \frac{dY_{\phi}}{dx} = &\,\,\,\, \dfrac{\beta s}{\mathcal{H} x}  \left(-\left<\sigma v\right>_{\phi \phi^{\dagger{}} \to X \bar{X}}
\left((Y_{\phi})^2 -  (Y_{\phi}^{\rm eq})^2\right) 
- \dfrac{\Gamma_{\phi}}{s}
\dfrac{K_1(m_{\phi}/T)}{K_2(m_{\phi}/T)} Y_{\phi} 
\right)\,.
\end{align}
\subsection{Case III}
\subsubsection{Distribution function of $\phi$ :}
The case III, where $\phi$ never attains thermal equilibrium with
the SM bath, has the same forms of Boltzmann equations for
$n_{\phi}$ and $\rho_{\nu_R}$ as those are in case I except here
we need to replace the thermal distribution function of $\phi$
by the non-thermal distribution function. The differential form of
the Boltzmann equation to find the distribution function of $\phi$,
$f_{\phi}$ is given by \cite{Konig:2016dzg, Biswas:2016iyh} 
\begin{equation}
    \frac{\partial f_{\phi}}{\partial t} - \mathcal{H} p_1 \frac{\partial f_{\phi}}{\partial p_1} = C^{h\to \phi \phi^{\dagger}} + C^{h h \to \phi \phi ^{\dagger} } + C^{\phi \to \bar{\nu}_R \psi}.
\label{eq:BE_dis_fun}    
\end{equation}
Here $C^{h \to \phi \phi^{\dagger}}$ is the collision term
for production of $\phi\phi^\dagger$ pair from the decay
of the SM Higgs boson $h(K) \to \phi(P_1) + \phi^{\dagger}(P_2)$.
The expression of $C^{h \to \phi \phi^{\dagger}}$ is given by
\begin{align}
    C^{h\to \phi \phi^{\dagger}} = \frac{1}{2E_{p_1}} &\int \frac{d^3p_2}{2E_{p_2}(2\pi)^3}\frac{d^3k}{2E_{k}(2\pi)^3} (2\pi)^4 \delta^4(K-P_1-P_2) \nonumber \\
    &\times  \lvert \mathcal{M} \rvert^2_{h\to \phi \phi^{\dagger}} \left(f_h^{\rm eq}(k)-f_{\phi}(p_1)f_{\phi^{\dagger}}(p_2)\right)\,,
    \nonumber \\
    =  \frac{1}{2E_{p_1}(2\pi)^2} &\int \frac{d^3p_2}{4E_{p_2}E_{p_1 + p_2}} \delta(E_{p_1 + p_2}-E_{p_1}-E_{p_2}) \nonumber \\
    &\times \lvert \mathcal{M} \rvert^2_{h\to \phi \phi^{\dagger}}
    \left(f_h^{\rm eq}(k)-f_{\phi}(p_1)f_{\phi^{\dagger}}(p_2)\right)\,.
\end{align}
Now we can write $d^3p_2 = p_2^2 dp_2 d(\cos{\theta})d\phi$,
where $\theta$ is the angle between $\vec{p_1}$ and $\vec{p_2}$.
Therefore, the Dirac delta function $\delta(E_{p_1 + p_2}-E_{p_1}-E_{p_2})$ actually fixes the angle $\theta$. So, from the condition $E_{p_1+p_2} = E_{p_1} + E_{p_2}$, we will get
\begin{align}
    \cos{\theta} = \frac{2m^2_{\phi}-m^2_{h}+
    2E_{p_1}E_{p_2}}{2p_1 p_2} \equiv \cos{\theta_{0}}.
\end{align}
Therefore,
\begin{align}
     C^{h\to \phi \phi^{\dagger}} = \frac{1}{2E_{p_1}(2\pi)^2} \int \frac{p^2_2 dp_2 (2\pi)}{4E_{p_2}} \int_{-1}^{1} \frac{d(\cos{\theta})\delta(\cos{\theta}-\cos{\theta_{0}})}{E_{p_1+p_2}\lvert \frac{df}{d\cos{\theta}}\rvert_{\theta=\theta_{0}}} \nonumber \\
     \times \lvert \mathcal{M} \rvert^2_{h\to \phi \phi^{\dagger}}
    \left(f_h^{\rm eq}(E_{p_1+p_2})-f_{\phi}(p_1)f_{\phi^{\dagger}}(p_2)
    \right),
\end{align}
where $f(\cos{\theta}) = E_{p_1 + p_2} - E_{p_1} - E_{p_2} $ with
$E_{p_1 + p_2} = \sqrt{|\vec{p_1} + \vec{p_2}|^2 + m^2_h}$ and 
\begin{eqnarray}
    \dfrac{df}{d\cos\theta}\biggm\lvert_{\theta = \theta_0} =
    \dfrac{p_1p_2}{E_{p_1} + E_{p_2}} \,,\\
    E_{p_1+p_2} \Bigm\lvert_{\theta = \theta_0} = 
    E_{p_1} + E_{p_2}\,.
\end{eqnarray}
After some simplification, the collision term takes the following form 
\begin{align}
    C^{h\to \phi \phi^{\dagger}} = \frac{1}{16 \pi E_{p_1}p_1} 
    \int^{p^{\rm max}_2}_{p^{\rm min}_2}
    \frac{p_2 dp_2}{E_{p_2}} 
    \lvert \mathcal{M} \rvert^2_{h\to \phi \phi^{\dagger}}
    \left(f^{\rm eq}_{\phi}(E_{p_1})f^{\rm eq}_{\phi^{\dagger}}
    (E_{p_2}) - f_{\phi}(p_1)f_{\phi^{\dagger}}(p_2)\right).
\label{eq:chphiphi}    
\end{align}
The limits of the integration is obtained from the condition $-1 \leq \cos{\theta_{0}} \leq 1$. This condition translates to -
\begin{align}
   p_2^{\rm min} = & \bigg \lvert \frac{p_1 (m_h^2 - 2 m_{\phi}^2) - m_h \sqrt{(m_h ^2 - 4 m_{\phi}^2)(p_1^2 + m_{\phi}^2)}}{2 m_{\phi}^2} \bigg \rvert\,,  \nonumber \\
    p_2^{\rm max} = & \frac{p_1 (m_h^2 - 2 m_{\phi}^2) + m_h \sqrt{(m_h ^2 - 4 m_{\phi}^2)(p_1^2 + m_{\phi}^2)}}{2 m_{\phi}^2}.
\end{align}
Here we have neglected the inverse decay term in Eq.\,\eqref{eq:chphiphi}
as it is substantially smaller compared to the decay term
as long as $\phi$ is non-thermal. Therefore, the collision term
$C^{h\to \phi \phi^{\dagger}}$ becomes 
\begin{align}
    C^{h\to \phi \phi^{\dagger}} = & \frac{1}{16 \pi E_{p_1}p_1} \int_{p_2^{min}}^{p_2^{max}} \frac{p_2 dp_2}{E_{p_2}} \lvert \mathcal{M} \rvert^2_{h\to \phi \phi^{\dagger}} e^{-E_{p_1}/T}e^{-E_{p_2}/T} \,,\nonumber \\
    = &  \frac{\lvert \mathcal{M} \rvert^2_{h\to \phi \phi^{\dagger}} T e^{-E_{p_1}/T}}{16 \pi E_{p_1}p_1}\left( e^{-E_{p_2}^{\rm min}/T} - e^{-E_{p_2}^{\rm max}/T}   \right)\,,
    \label{eq:chphiphi_final}
\end{align}
and $E^{\rm max(min)}_{p_2} = \sqrt{\left(p^{\rm max(min)}_2\right)^2
+ m^2_{\phi}}$. 

Now, we will briefly discuss the derivation of the 
collision term $C^{h h \to \phi \phi^{\dagger}}$
for the production of $\phi\phi^\dagger$ pair due the
scattering of the Higgs boson
$h(K_1) + h(K_2) \to \phi(P_1) + \phi^{\dagger}(P_2)$.
\begin{align}
    C^{h h \to \phi \phi^{\dagger}} = \frac{1}{2 E_{p_1}} &\int \frac{d^3k_1}{2E_{k_1}(2\pi)^3} \frac{d^3k_2}{2E_{k_2}(2\pi)^3} \frac{d^3p_2}{2E_{p_2}(2\pi)^3} (2\pi)^4 \delta^4(K_1 + K_2 - P_1 - P_2) \nonumber \\
    & \lvert \mathcal{M} \rvert^2_{hh\to \phi \phi^{\dagger}} \left(f_{h}(k_1)f_{h}(k_2) - f_{\phi}(p_1)f_{\phi^{\dagger}}(p_2)\right)\,, \nonumber \\
    = \frac{1}{2 E_{p_1}} &\int \frac{d^3p_2}{2E_{p_2}(2\pi)^3} \left[\int \frac{d^3k_1}{2E_{k_1}(2\pi)^3} \frac{d^3k_2}{2E_{k_2}(2\pi)^3} (2\pi)^4 \delta^4(K_1 + K_2 - P_1 - P_2) \right] \nonumber \\
    & \lvert \mathcal{M} \rvert^2_{hh\to \phi \phi^{\dagger}} \left(f_{h}(k_1)f_{h}(k_2) - f_{\phi}(p_1)f_{\phi^{\dagger}}(p_2)\right).
\end{align}
The term inside the square bracket is Lorentz invariant, 
and we can do that integration easily in the centre of momentum
frame. Here, for the calculational simplification, we assume
that the matrix amplitude square
$\lvert \mathcal{M} \rvert^2_{hh\to \phi \phi^{\dagger}}$
depends only on the Mandelstam variable $s$ which is true
for $s$-channel scatterings and contact interactions. 
For a general matrix amplitude square depending on all three
Mandelstam variables one can use the prescription given
in \cite{Hannestad:1995rs}. 
\begin{align}
    I = \int \frac{d^3k_1}{2E_{k_1}(2\pi)^3}
    \frac{d^3k_2}{2E_{k_2}(2\pi)^3} (2\pi)^4
    \delta^4(K_1 + K_2 - P_1 - P_2). 
\end{align}
This will give -
\begin{align}
    I = \frac{1}{8 \pi} \sqrt{1- \frac{4 m^2_h}{s}},
\end{align}
Now, since $I$ is a Lorentz invariant quantity, we can use 
this result in any inertial frame of reference with proper definition
of $s$.\,\,In any arbitrary reference frame, the Mandelstam
variable $s(p_1,p_2,\cos{\alpha}) = (P_1 + P_2)^2 =
2 m_{\phi}^2 + 2 E_{p_1} E_{p_2} - 
2 \lvert \vec{p_1} \rvert \lvert \vec{p_2} \rvert 
\cos{\alpha}$, $\alpha$ is the angle between 
$\vec{p}_1$ and $\vec{p}_2$ which is $\pi$ in
the centre of momentum frame.
Hence, the collision term in an arbitrary
inertial frame of reference is given by
\begin{align}
    C^{h h \to \phi \phi^{\dagger}} = \frac{1}{16 \pi E_{p_1}} &\int \frac{d^3p_2}{2E_{p_2}(2\pi)^3}\sqrt{1- \frac{4 m^2_h}{s(p_1,p_2,\cos{\alpha})}} \nonumber \\
    & \times \lvert \mathcal{M} \rvert^2_{hh\to \phi \phi^{\dagger}}(s) \left(f_{h}(k_1)f_{h}(k_2) - f_{\phi}(p_1)f_{\phi^{\dagger}}(p_2)\right)\,, \nonumber \\
    = \frac{2\pi}{16 \pi E_{p_1} 2 (2\pi)^3} &\int \frac{p^2_2 dp_2 d(\cos{\alpha})}{E_{p_2}} \sqrt{1- \frac{4 m^2_h}{s(p_1,p_2,\cos{\alpha})}} \nonumber \\
    & \times \lvert \mathcal{M} \rvert^2_{hh\to \phi \phi^{\dagger}}(s) f_{h}(k_1)f_{h}(k_2)\,,
\end{align}    
where, in the last step we have neglected the back scattering term. Now
using the Maxwell-Boltzmann distribution function for the SM Higgs
boson and $f_{h}(k_1)f_{h}(k_2) = e^{-\left(E_{k_1}+E_{k_2}\right)/T}=
e^{-\left(E_{p_1}+E_{p_2}\right)/T}$, we obtain
\begin{align}
 C^{h h \to \phi \phi^{\dagger}} 
    &= \frac{e^{-E_{p_1}/T}}{16 E_{p_1} (2\pi)^3} \int_{0}^{\infty} \frac{p^2_2 dp_2}{\sqrt{p^2_2 + m^2_{\phi}}} e^{-E_{p_2}/T} \nonumber \\
    & \times \int_{-1}^{\cos{\alpha}_{max}} d(\cos{\alpha} ) \sqrt{1- \frac{4 m^2_h}{s(p_1,p_2,\cos{\alpha})}}
    \lvert \mathcal{M} \rvert^2_{hh\to \phi \phi^{\dagger}}(s).
\end{align}
The limit on $\cos{\alpha}$ will come from the condition that
$\sqrt{1- \frac{4 m^2_h}{s(p_1,p_2,\cos{\alpha})}}$ is real.
This is possible only when $s\geq 4m^2_h$ and therefore
\begin{eqnarray}
\cos \alpha \leq 
\frac{2 m_{\phi}^2 - 4 m_h^2 + 2 E_{p1}E_{p2}}
{2 \lvert \vec{p_1} \rvert \lvert \vec{p_2} \rvert} \equiv
\cos \alpha_{0}\,.
\end{eqnarray}
Thus the upper limit of the integration is   
\begin{align}
    \cos{\alpha}_{\rm max} = & {\rm Min}
    \left[{\rm Max}\left[\cos\alpha_0,-1\right],1\right]\,.
\end{align} 

And, lastly, the collision term $C^{\phi \to \bar{\nu_R} \psi}$
is for the decay of $\phi$ into $\overline{\nu_R}$ and
$\psi$ ($\phi(P_1) \to \overline{\nu_R}(q) + \psi(q')$)
and it has the following expression \cite{Biswas:2016iyh} 
\begin{align}
    C^{\phi \to \bar{\nu_R} \psi} = - f_{\phi} \frac{m_{\phi}}{\sqrt{p_1^2 + m_{\phi}^2}}\Gamma_{\phi \to \bar{\nu}_R \psi}\,.
\end{align}
The LHS of Eq.\,\eqref{eq:BE_dis_fun}, can be greatly simplified
in we transform the variables from $p_1$ and $T$ to new
variables $r=m_0/T$ and $\xi = \left(\frac{g_s(T_0)}{g_s(T)}\right)^{1/3} \frac{p_1}{T}$ where $m_0$ is any arbitrary mass scale.
In terms of the two new variables, the LHS of
Eq.\,\eqref{eq:BE_dis_fun} depends only on $r$
\cite{Konig:2016dzg, Biswas:2016iyh} 
\begin{align}
    \frac{\partial f_{\phi}}{\partial t} - \mathcal{H} p_1 \frac{\partial f_{\phi}}{\partial p_1} = r \mathcal{H} \left( 1 + \frac{T g'_s(T)}{3 g_s(T)} \right)^{-1} \frac{\partial f_{\phi}}{\partial r}.
\end{align}

Therefore, the full Boltzmann equation for $f_{\phi}$ is 
\begin{align}
    \frac{\partial  f_{\phi}(\xi,r)}{\partial r} = 
    \dfrac{ \left(1 - \frac{r}{3 g_s(r)}\frac{dg_s(r)}{dr}
    \right)}{r H} (C^{h\to \phi \phi^{\dagger}}(\xi,r) +C^{h h \to \phi \phi ^{\dagger}}(\xi,r) + C^{\phi \to \bar{\nu}_R \psi}(\xi,r))\,.
    \label{eq:BE_dis_func_final}
\end{align}
Now, the number density of $\phi$ can be written as 
\begin{align}
    n_{\phi}(r) = \frac{g_{\phi}}{2 \pi^2} \mathcal{A}(r)^3 \left(\frac{m_0}{r}\right)^3 \int d\xi\,\xi^2\,f_{\phi}(\xi,r),
\end{align}
where 
\begin{align}
    \mathcal{A}(r) = \left(\frac{g_s(m_0/r)}{g_s(m_0/T_0)}\right)^{1/3}.
\end{align}
After solving the Eq.\,\eqref{eq:BE_dis_func_final} for
the non-thermal distribution function $f_{\phi}(\xi,r)$,
we can now calculate comoving number density of $\psi$
and $\widetilde{Y}$ using the following Boltzmann equations
\begin{align}
    \frac{dY_{\psi}}{dr} = \frac{g_{\phi}\beta}{r \mathcal{H} s} \frac{\Gamma_{\phi} m_{\phi}}{2 \pi^2} \int_{0}^{\infty} \frac{\left( \mathcal{A} \frac{m_0}{r} \right)^3 \xi^2\,f_{\phi}(\xi,r)}{\sqrt{\left(\xi \mathcal{A} \frac{m_0}{r}\right)^2 + m_{\phi}^2}}d\xi \,, \nonumber \\
    \frac{d\widetilde{Y}}{dr} = \frac{g_{\phi}\beta}{rH s^{4/3}} \left<E\Gamma\right> \frac{1}{2 \pi^2} \int_{0}^{\infty} \left(\mathcal{A}\frac{m_0}{r}\right)^3 \xi^2 f_{\phi}(\xi,r)\, d\xi .
\end{align}
\section{Equations for $\Omega_{\rm DM}{\rm h}^2$ and 
$\Delta {\rm N}_{\rm eff}$ }
\label{appen2}
The effective number of relativistic degrees of
freedom ${\rm N}_{\rm eff}$ can be defined as 
\begin{equation}
    {\rm N}_{\rm eff} = \frac{8}{7} \left ( \frac{11}{4} \right)^{4/3} \left ( \frac{\rho_{\rm rad}-\rho_{\gamma}}{\rho_{\gamma}} \right )
\end{equation}
where $\rho_{\rm rad}, \rho_{\gamma}$ denote total radiation and photon densities respectively. The change in ${\rm N}_{\rm eff}$ is defined as $\Delta {\rm N}_{\rm eff}={\rm N}_{\rm eff}-{\rm N}^{\rm SM}_{\rm eff} $. While the expected value in the SM is close to 3 due to three left
handed neutrinos, in our scenario this can increase due to the
presence of three right handed neutrinos $\nu_R$ which are
relativistic. Thus, taking $\rho_{\nu_R}$ to be part of
$\rho_{\rm rad}$, we can write $\Delta {\rm N}_{\rm eff}$ as
\begin{align} \label{Neff}
    \Delta {\rm N}_{\rm eff} = & 2 \times 3 \left( \frac{\rho_{\nu_R}}{\rho_{\nu_L}}\right)_{\rm CMB} \nonumber \\
    = & 2 \times 3 \left( \frac{\rho_{\nu_R}}{\rho_{\nu_L}}\right)_{10 \, \rm MeV} (\because \rho_{\nu_L} \propto \frac{1}{a^4}; \rho_{\nu_L} \propto \frac{1}{a^4}) \nonumber \\
    = & 2 \times 3 \left( \frac{s^{4/3} \widetilde{Y}}{\rho_{\nu_L}}\right)_{10 \, \rm MeV},
\end{align}
where in the second step, we equate the ratio $\rho_{\nu_R}/\rho_{\nu_L}$ at the scale of recombination or CMB to that of BBN $\sim \mathcal{O}(10)$ MeV. This is possible as we ensure the production of $\nu_R$ is complete before the BBN epoch.

Similarly, final DM abundance $\Omega_{\rm DM}{\rm h}^2$ can be written in terms of corresponding comoving number density as
\begin{align} \label{Abun}
    \Omega_{\rm DM} {\rm h}^2 = & 2 \times \frac{\rho_{\psi}^0}{\rho_{c}^0} {\rm h}^2 = 2 \times \frac{m_{\psi} s^0 Y_{\psi}^0 }{\rho_{c}^0} {\rm h}^2 = 2 \times \frac{m_{\psi} s^0 (Y_{\psi})_{10}}{\rho_{c}^0} {\rm h}^2. 
\end{align}
Since we have taken $g_{\phi} = 1$ throughout (the value of $g_{\psi}$ and $g_{\nu_R}$ are taken as 2), this implies that we are considering either the equations for $\phi$ or $\phi^{\dagger}$. Hence, $Y_{\psi}$ and $\widetilde{Y}$ are only for either particles or anti-particles. So, in the expressions for $\Delta {\rm N}_{\rm eff}$ and $\Omega_{\rm DM}{\rm h}^2$ above, we have included a factor of 2 to incorporate both particles and antiparticles. Also a factor of 3 is included in $\Delta {\rm N}_{\rm eff} $ for three flavours of $\nu_R$.

\section{Approximate analytical solutions for case I and case II :}
\label{appen3}

\subsection{Case I}
 The Eqs.\,\,\eqref{case1_psi_1} and $\eqref{case1_nuR_1}$ for case I can be solved analytically neglecting the variation of $g_{s}$ and $g_{\rho}$. The expressions of $Y_{\psi}$ and $\widetilde{Y}$ after freeze-in are 
        \begin{eqnarray} \label{Appen_C_1}
    Y_{\psi} &=& \frac{135\,g_{\phi}}{1.66 \times 
    8 \pi^3 g_{s} \sqrt{g_{\rho}}}
   \frac{M_{pl} \Gamma_{\phi}}{m_{\phi}^2}\,,
   \nonumber \\   
    \widetilde{Y} &=& \frac{675\,g_{\phi}}{1.66\times 8\pi^3 g_s\sqrt{g_{\rho}}} \left(\frac{45}{2\pi^2\,g_s}\right)^{1/3}
    \frac{M_{pl}\langle E\Gamma\rangle}{m^3_{\phi}}\,,
    \end{eqnarray}
    where $g_s$ and $g_{\rho}$ are effective number of degrees of freedoms at the freeze-in temperature $T\sim m_{\phi}$ 
    and 
    \begin{eqnarray} \label{Appen_C_2}
    \langle E\Gamma \rangle = \frac{m_{\phi}}{2}
    \left(1-\frac{m^2_{\psi}}{m^2_{\phi}}\right)\Gamma_{\phi}\,.
    \end{eqnarray}
    With this, the ratio of $\widetilde{Y}$ to $Y_{\psi}$ in the limit $m_{\phi}>>m_{\psi}$
    is given by
    \begin{eqnarray} \label{Appen_C_3}
    \frac{\widetilde{Y}}{Y_{\psi}} = 
    \frac{675}{270} \left(\frac{45}{2\pi^2}\right)^{1/3} \frac{1}{g^{1/3}_s}\,\,.
    \end{eqnarray}
    Using this ratio, we can easily establish a relation between $\Delta{\rm {N}_{eff}}$ and $\Omega_{\rm DM} {\rm h}^{2}$ as
\begin{eqnarray} \label{Appen_C_4}
\Delta{\rm N}_{\rm eff} = 3.29\frac{C_2}{C_1\,m_{\psi}}
\frac{\Omega_{\rm DM}{\rm h}^2}{g^{1/3}_s}
\end{eqnarray}    
where $C_1 = 2\times 2.755\times 10^{8}$ GeV$^{-1}$ and
$C_2 = 3\times 1.16\times (43/4)^{4/3}$ are constants.
\subsection{Case II}
For case II, we have solved the Eq.\,\,\eqref{case2_phi_2} neglecting its 1st term i.e. after the freeze out of $\phi$. This gives
    \begin{eqnarray} \label{Appen_C_5}
    Y_{\phi} = Y^{fo}_{\phi} e^{-\frac{\Gamma_{\phi} M_{pl}}{1.66\times \sqrt{g^{\rho}_{*}} m^2_{\phi}} \left(\frac{x^2}{2} - \frac{(x^{f})^2}{2}\right)}\,.
    \end{eqnarray}
    Now this expression can be used to solve Eqs.\,\,\eqref{case2_psi_2} and \eqref{case2_nur_2} analytically (once again we are neglecting the temperature dependence of $g_{s}$ and $g_{\rho}$)
    \begin{eqnarray} \label{Appen_C_6}
    Y_{\psi} &\approx& Y^{fo}_{\phi}\,, \nonumber \\
    \widetilde{Y} &\approx& \frac{Y^{fo}_{\phi}}{g^{1/3}_sg^{1/2}_{\rho}} \frac{M_{pl}\langle E\Gamma\rangle}{m^3_{\phi}} f_{1} \frac{e^{\frac{f_{2}}{2}(x^{f})^2}}{f^{3/2}_{2}},
    \label{ytilde2}
    \end{eqnarray}
    where 
    \begin{eqnarray} \label{Appen_C_7}
    f_{1} &=& \frac{1}{1.66}
    \sqrt{\frac{{\pi}}{2}}
    \left(\frac{45}{2\pi^2}\right)^{1/3}
     \nonumber \\
    f_{2} &=& \frac{\Gamma_{\phi} M_{pl}}{1.66\,\sqrt{g_{\rho}}\,m^2_{\phi}}
    = \frac{\Gamma_{\phi}}{H(m_{\phi})}\,\,.
    \end{eqnarray}
    Here $Y_{\phi}^{fo}$ is the abundance of $\phi$
    just after freeze-out. The expression for $\widetilde{Y}$ given in Eq.\,\,\eqref{ytilde2} is valid as long as the product $f_2\,(x^f)^2<<1$. Now
    in the limit $m_{\phi}>>m_{\psi}$ the ratio of $\widetilde{Y}$
    to $Y_{\psi}$ is given by
    \begin{eqnarray} \label{Appen_C_8}
    \frac{\widetilde{Y}}{Y_{\psi}} \approx
    \frac{1}{g^{1/3}_sg^{1/2}_{\rho}} \frac{M_{pl}\Gamma_{\phi}}{2\,m^2_{\phi}} f_{1} 
    \frac{e^{\frac{f_{2}}{2}(x^{f})^2}}{f^{3/2}_{2}}\,,
    \end{eqnarray}
    and finally,
    \begin{eqnarray} \label{Appen_C_9}
    \Delta{\rm N}_{\rm eff} \approx
    \frac{M_{pl}\Gamma_{\phi}}{2\,m^2_{\phi}} f_1\,
    \frac{e^{\frac{f_2}{2}(x_f)^2}}{f^{3/2}_2}
    \frac{C_2}{C_1\,m_{\psi}}
    \frac{\Omega_{\rm DM}{\rm h^2}}
    {g^{1/3}_sg^{1/2}_{\rho}}\,\,.
    \end{eqnarray}

\providecommand{\href}[2]{#2}\begingroup\raggedright\endgroup

\end{document}